\documentclass[aps,reprint,unsortedaddress,superscriptaddress]{revtex4-1}
\usepackage{bm}
\usepackage{amsfonts,amssymb,amsmath}
\usepackage{subfigure}
\usepackage{graphicx}
\usepackage{color}
\usepackage{colortbl}
\usepackage{hyperref}
\usepackage{textcmds}
\usepackage{ulem} 
\usepackage{url}

\providecommand{\apj}[0]{Astrophys. J.}
\providecommand{\apjl}[0]{Astrophys. J. Lett.}
\providecommand{\apjs}[0]{Astrophys. J. Supp. Ser. }

\providecommand{\aap}[0]{Astron. Astrophys. }

\providecommand{\araa}[0]{Ann.\ Rev. Astron. Astroph. }
\providecommand{\physrep}[0]{Phys. Rep. }
\providecommand{\mnras}[0]{Mon. Not. Roy. Astron. Soc. }

\providecommand{\prl}[0]{Phys. Rev. Lett.}
\providecommand{\prd}{Phys. Rev. D.}

\providecommand{\nar}[0]{New Astron. Rev.}

\def\imbh#1{intermediate mass black hole#1(IMBH#1)\gdef\imbh{IMBH}}
\def\smbh#1{supermassive black hole#1(SMBH#1)\gdef\smbh{SMBH}}
\def\bbh#1{binary black hole#1 (BBH#1)\gdef\bbh{BBH}}
\def\bh#1{black hole#1 (BH#1)\gdef\bh{BH}}
\def\ns#1{neutron star#1 (NS#1)\gdef\ns{NS}}
\def\gw#1{gravitational wave#1 (GW#1)\gdef\gw{GW}}
\def\pnw#1{post-Newtonian#1 (PN#1)\gdef\pnw{PN}}
\def\eos#1{equation of state#1 (EOS#1)\gdef\eos{EOS}}
\def\grb#1{gamma-ray burst#1 (GRB#1)\gdef\grb{GRB}}
\def\amr#1{adaptive mesh refinement#1 (AMR#1)\gdef\amr{AMR}}
\def\isco#1{innermost stable circular orbit#1 (ISCO#1)\gdef\isco{ISCO}}

\begin{document}
\title{The Transient Gravitational-Wave Sky}
\date{\today}
\author{Nils	Andersson} \affiliation{	School of Mathematics, University of Southampton, Southampton, SO17 1BJ, UK	}
\author{John	Baker} \affiliation{	Gravitational Physics Lab, NASA GSFC, Greenbelt, MD 20771, USA	}
\author{Kris	Belczynski} \affiliation{	Astronomical Observatory, University of Warsaw, Al. Ujazdowskie 4, 00-478 Warsaw, Poland	} \affiliation{Center for Gravitational Wave Astronomy, University of Texas at Brownsville, Brownsville, TX 78520, USA}
\author{Sebastiano	Bernuzzi} \affiliation{	Theoretical Physics Institute, University of Jena,  07743 Jena, Germany	}
\author{Emanuele	Berti} \affiliation{Department of Physics and Astronomy, The University of Mississippi, University, MS 38677, USA} \affiliation{Theoretical Astrophysics 350-17, California Institute of Technology, Pasadena, CA 91125, USA}
\author{Laura	Cadonati} \affiliation{	Physics Department, University of Massachusetts, Amherst, MA 01003, USA	}
\author{Pablo	Cerd\'a-Dur\'an} \affiliation{	Departamento de Astronom\'ia y Astrof\'isica, Universidad de Valencia, Dr. Moliner 50, 46100 Burjassot, Spain	}
\author{James	Clark} \affiliation{	Physics Department, University of Massachusetts, Amherst, MA 01003, USA	}
\author{Marc	Favata} \affiliation{	Montclair State University, 1 Normal Ave, Montclair NJ 07043, USA	}
\author{Lee Samuel	Finn} \affiliation{	Department of Physics \& Department of Astronomy \& Astrophysics, The Pennsylvania State University, University Park, PA 16802, USA}	
\author{Chris	Fryer} \affiliation{	CCS-2, Los Alamos National Laboratory, Los Alamos, NM 87545, USA	}
\author{Bruno	Giacomazzo} \affiliation{	JILA, University of Colorado and National Institute of Standards and Technology, Boulder, CO 80309, USA	}
\author{Jose Antonio Gonz\'alez} \affiliation{	Instituto de F\'isica y Matem\'aticas, Universidad Michoacana de San Nicol\'as de Hidalgo. Edificio C-3 Ciudad Universitaria 58040, Morelia, Michoac\'an, M\'exico	}
\author{Martin	Hendry} \affiliation{	SUPA, School of Physics and Astronomy, University of Glasgow, G12 8QQ, UK	}
\author{Ik Siong	Heng} \affiliation{	SUPA, School of Physics and Astronomy, University of Glasgow, G12 8QQ, UK	}
\author{Stefan	Hild} \affiliation{	Institute for Gravitational Research, University of Glasgow, G12 8QQ, UK	}
\author{Nathan	Johnson-McDaniel} \affiliation{	Theoretical Physics Institute, University of Jena,  07743 Jena, Germany	}
\author{Peter	Kalmus}
\affiliation{LIGO Laboratory, California Institute of Technology, Pasadena, CA 91125, USA}
\author{Sergei	Klimenko} \affiliation{	University of Florida, PO Box 118440, Gainesville, Florida 32611, USA	}
\author{Shiho	Kobayashi} \affiliation{	Astrophysics Research Institute, Liverpool John Moores University,  Birkenhead, CH41 1LD, UK	}
\author{Kostas	Kokkotas} \affiliation{	Theoretical Astrophysics, IAAT, Eberhard Karls University of Tuebingen, Tuebingen 72076, Germany	}
\author{Pablo	Laguna} \affiliation{	School of Physics, Georgia Institute of Technology, Atlanta, Georgia 30332, USA	}
\author{Luis	Lehner}
\affiliation{Perimeter Institute for Theoretical Physics, Waterloo, Ontario, Canada}
\author{Janna	Levin} \affiliation{	Department of Physics and Astronomy, Barnard College of Columbia University, New York, NY 10027, USA	}
\author{Steve	Liebling} \affiliation{	Long Island University, Brookville, NY 11548, USA	}
\author{Andrew	 MacFadyen} \affiliation{	Physics Department, New York University, New York, NY 10003, USA	}
\author{Ilya	Mandel} \affiliation{	School of Physics and Astronomy, University of Birmingham, Edgbaston, B15 2TT, UK	}
\author{Szabolcs	Marka}
\affiliation{Department of Physics, Columbia University, New York, New York 10027, USA}	
\author{Zsuzsa	Marka}
\affiliation{Columbia Astrophysics Laboratory, Columbia University, New York, New York 10027, USA}
\author{David	Neilsen} \affiliation{	Department of Physics \& Astronomy, Brigham Young University, Provo, UT 84602, USA}	
\author{Paul	O'Brien} \affiliation{	Department of Physics \& Astronomy, University of Leicester, University Road, Leicester, LE1 7RH, UK	}
\author{Rosalba	Perna} \affiliation{	JILA, University of Colorado and National Institute of Standards and Technology, Boulder, CO 80309, USA	}
\author{Harald	Pfeiffer} \affiliation{	Canadian Institute for Theoretical Astrophysics, University of Toronto, Toronto, Ontario M5S 3H8, Canada}	
\author{Jocelyn	 Read} \affiliation{	Department of Physics, California State University Fullerton, Fullerton, CA 92831, USA	}
\author{Christian	Reisswig} \affiliation{Theoretical Astrophysics 350-17, California Institute of Technology, Pasadena, CA 91125, USA}
\author{Carl	Rodriguez} \affiliation{	Department of Physics and Astronomy, Northwestern University, Evanston, IL 60208, USA	}
\author{Max	Ruffert} \affiliation{	School of Mathematics and Maxwell Institute, King's Buildings, University of Edinburgh, Edinburgh EH16 5JN, Scotland, UK}	
\author{Erik	Schnetter}
\affiliation{Perimeter Institute for Theoretical Physics, Waterloo, Ontario, Canada}
\affiliation{Department of Physics, University of Guelph, Guelph, Ontario, Canada}
\affiliation{Center for Computation \& Technology, Louisiana State University, Louisiana, USA}
\author{Antony	Searle}
\affiliation{LIGO Laboratory, California Institute of Technology, Pasadena, CA 91125, USA}
\author{Peter	Shawhan} \affiliation{	University of Maryland, College Park, MD 20742, USA	}
\author{Deirdre	Shoemaker} \affiliation{	School of Physics, Georgia Institute of Technology, Atlanta, Georgia 30332, USA	}
\author{Alicia	Soderberg} \affiliation{	Harvard-Smithsonian Center for Astrophysics, 60 Garden st. Cambridge, MA 02138, USA	}
\author{Ulrich	Sperhake}
\affiliation{DAMTP, Center for Mathematical Sciences, Cambridge CB3 0WA, UK}
\affiliation{Theoretical Astrophysics 350-17, California Institute of Technology, Pasadena, CA 91125, USA}
\affiliation{CENTRA-IST, Lisbon, Portugal}
\author{Patrick	Sutton} \affiliation{	School of Physics and Astronomy, Cardiff University, Cardiff, CF24 3AA, UK	}
\author{Nial	Tanvir} \affiliation{	Department of Physics \& Astronomy, University of Leicester, University Road, Leicester, LE1 7RH, UK	}
\author{Michal	Was}
\affiliation{Albert-Einstein-Institut, Max-Planck-Institut f\"ur Gravitationsphysik, D-30167 Hannover, Germany}
\author{Stan	Whitcomb}
\affiliation{LIGO Laboratory, California Institute of Technology, Pasadena, CA 91125, USA}

\begin{abstract}
  Interferometric detectors will very soon give us an unprecedented
  view of the gravitational-wave sky, and in particular of the
  explosive and transient Universe. Now is the time to challenge our
  theoretical understanding of short-duration gravitational-wave
  signatures from cataclysmic events, their connection to more
  traditional electromagnetic and particle astrophysics, and the data
  analysis techniques that will make the observations a reality.  This
  paper summarizes the state of the art, future science opportunities,
  and current challenges in understanding gravitational-wave transients.
\end{abstract}

\maketitle

\section{Introduction} 
\label{sec:intro}

The \gw{} sky is an unexplored frontier which holds a great potential
for discovery and a promise for understanding one of the most
mysterious interactions of nature: gravity.
Predicted by Einstein's theory of general relativity, \gw{s} are
ripples in the fabric of space-time, produced by the accelerated
motion of masses.  
They carry information from the bulk, coherent motion of matter,
complementary to the multi-wavelength electromagnetic spectrum of
traditional astronomy and to the neutrinos and cosmic rays of particle
astrophysics.  Their observation will play a transformative role in
our understanding of the Universe. 

The existence of \gw{s} was indirectly proven by over three decades of
measurements of the orbit of the binary pulsar PSR1913+16, which has
steadily been evolving due to the emission of gravitational radiation
in agreement with the predictions of general
relativity~\cite{Weisberg:2004hi}.  However, the direct measurement
of \gw{s} remains a challenge, due to their tiny amplitude once they
reach Earth.

New and upgraded \gw{} detectors are pursuing their first detection,
which will transition gravitational physics to an observation-driven
field and usher in a new \gw{} astronomy.  The instrumental landscape
includes a new generation of ground-based laser interferometers \cite{advLIGO,advVirgo,GEOHF}, pulsar-timing arrays
\cite{janssen:633,Jenet:2009hk,Hobbs:2008yn} and future detectors on the ground and in space
\cite{LIGO:India,KAGRA,ET,ET:2011,LISA,AmaroSeoane:2012km}.  In particular, the
first generation of interferometric detectors has achieved design
sensitivity \cite{LIGO,Virgo,GEO600:2010}, and next-generation ground-based interferometers are
expected to be taking data within a few years.

The ``perfect storm'' in the transient sky of short-duration
cataclysmic events is about to arrive, with \gw{} observations from
stellar core collapse, gamma-ray burst engines, 
rapidly rotating \ns{s} and mergers of compact object
binaries. The storm will be fueled by \gw{} observations from Advanced
LIGO and Advanced Virgo, but the success in interpreting observations
will hinge on our ability to model the complex physics at the heart of
these transient astrophysical sources. This includes \gw{} emission
mechanisms that, in general, are not yet fully understood. The
importance of this enterprise was highlighted in \emph{New Worlds, New
Horizons in Astronomy and Astrophysics}, a Decadal Survey of Astronomy
and Astrophysics by the National Research Council
\cite{national2010New}.
 
This paper summarizes the current state of the art, future science
opportunities and open challenges for \gw{} transients. In
$\S$\ref{sec:SourceScience}, we review the main sources of the \gw{}
transient sky: mergers of compact objects, core collapse supernovae,
and \ns{} oscillations.  For each burst source, we discuss its chances
of detection and its connection with electromagnetic observations.  In
$\S$\ref{sec:instrumentation}, we describe \gw{} antennas as well as
traditional astronomical observatories available today and in the near
future. We elaborate on the detection challenges and open questions of
the \gw{} transient universe in $\S$\ref{sec:challenges}.  Finally,
$\S$\ref{sec:concl} summarizes our conclusions.

This review was conceived as the peroration of the
\emph{Gravitational Wave Bursts} workshop series. The first of these workshops was
held in 
Chichen-Itza, Mexico on December 9-11, 2009 and the second in
Tobermory, Scotland, on May 29-31, 2012. 
These meetings provided a forum for astrophysicists, \gw{} data
analysts and numerical relativists to explore transformative views of
the \gw{} transient sky, focused on: i) critical examination of
current methodologies to model, detect and characterize transients,
ii) current understanding of the physics behind burst sources, iii)
requirements on detector technology and data analysis, and iv)
imagining the future of \gw{} transient science.

\section{Sources of Gravitational Wave Transients} 
\label{sec:SourceScience}

Compact objects such as \ns{s} and \bh{s} will likely be protagonists
in most of the astrophysical events detectable by the next generation
of ground-based \gw{} interferometers. Their role could start at birth
(core collapse supernovae) or later in their life, either as members
of binary systems or as isolated objects. This section provides an
overview of what we know, observationally and theoretically, about
compact objects as potential sources of transient \gw{s}, including
their predicted energetics and rates.

\subsection{Compact Object Binaries and Short Gamma-ray Bursts}
\label{sec:nsbinaries}

Binaries of coalescing compact objects are the
main target of ground-based \gw{} astronomy.  In many instances, their
gravitational waveform is expected to contain many cycles in the
sensitive band of the detectors.  Therefore, with the aid of well
understood models (and assuming that Einstein's general relativity
provides the correct description of gravity in dynamical, strong-field
systems) they are ideal candidates for discovery via matched
filtering.  Given what we know about these sources and their merger
rates deduced from known \ns{} systems~\cite{2008LRR....11....8L},
binaries with either two \ns{s} or a \ns{} and a \bh{} (also called
{\em mixed binaries}) are expected to be bread-and-butter sources for
ground-based detectors, even though recent population synthesis
calculations suggest that \bh{}-\bh{} binaries may be more numerous
than initially expected \cite{2010ApJ...715L.138B,Dominik:2012}.
Unlike binary \bh{s}, binaries containing at least one \ns{} have an
additional appeal: their potential to act as the central engine of
short \grb{s} (and possibly other observable electromagnetic
signals). Short \grb{s} are bursts with duration shorter than $\sim
2$~seconds, with most of them lasting a few hundreds of
milliseconds. The expected properties of a population of short GRBs
from double compact object mergers have been estimated by several
studies~\cite{2002ApJ...570..252P,2006ApJ...648.1110B,2008ApJ...675..566O}.
To date, only rough comparisons can be made between theory and
observations~\cite{2012arXiv1202.2179C,2012arXiv1205.4621K}.  However,
the theoretical predictions of compact object merger rates are in
general consistent with the observed short \grb{}
rates~\cite{2008ApJ...675..566O,Fong:2012}.

There are several pieces of evidence supporting the association of
short \grb{s} with the merger of double \ns{} or mixed
binaries~\cite{Berger2009}. One is that short \grb{s} do not seem to
be associated with supernovae, and some of them explode in ``dead''
elliptical galaxies (i.e., galaxies with negligible ongoing
star-formation).  Long \grb{s} have distinctly different host galaxies
from short \grb{s}, and they mainly occur in star-forming galaxies.
Short \grb{s} appear more closely correlated with the rest-frame
optical light (old stars) than the UV light (young massive
stars). Furthermore, offsets of short \grb{s} relative to their host
galaxy centers are significantly larger than for
long \grb{s} \cite{Fong:2009bd}.

The observations listed above are promising, but several challenges
remain to establish an unequivocal association between short \grb{s}
and compact binary mergers.
As we will see below, the answer to these questions for now can only
be obtained via sophisticated numerical models including the relevant
microphysics. For example, a problem that can only be resolved via
numerical methods is whether mergers can produce accretion disks
massive enough to power the observed electromagnetic emission. Other
demanding aspects of the bursts are precursors~\cite{Troja2010} and
extended emission phases~\cite{Margutti2011} that happen on timescales
larger than 10 seconds: these timescales are beyond the reach of
current simulations, which presently last less than a second. Looking
forward, a smoking gun for the association between short \grb{s} and
binary mergers will be the coincident detection of electromagnetic
signals and \gw{s}. Such a coincident detection is one of the most
exciting multi-messenger observations that could occur in the advanced
detector era.

\subsubsection{Neutron Star-Neutron Star Binaries}

The last few years have witnessed remarkable progress in fully
general relativistic simulations of compact object
binaries~\cite{Faber2012}.  The first simulations of \ns{} binaries
were performed by Shibata and collaborators~\cite{Shibata2000}, but
only recently have simulations been extended from the late inspiral up
to the coalescence and eventual formation of a \bh{} surrounded by a
massive torus. Also recent is the inclusion of more sophisticated
physics -- realistic \eos{}, magnetic fields and neutrino radiation --
as well as the implementation of advanced numerical algorithms, such
as \amr{} techniques.

\begin{figure*}[!hbt]
\begin{center}
\begin{tabular}{cc}
\includegraphics[width=0.45\textwidth]{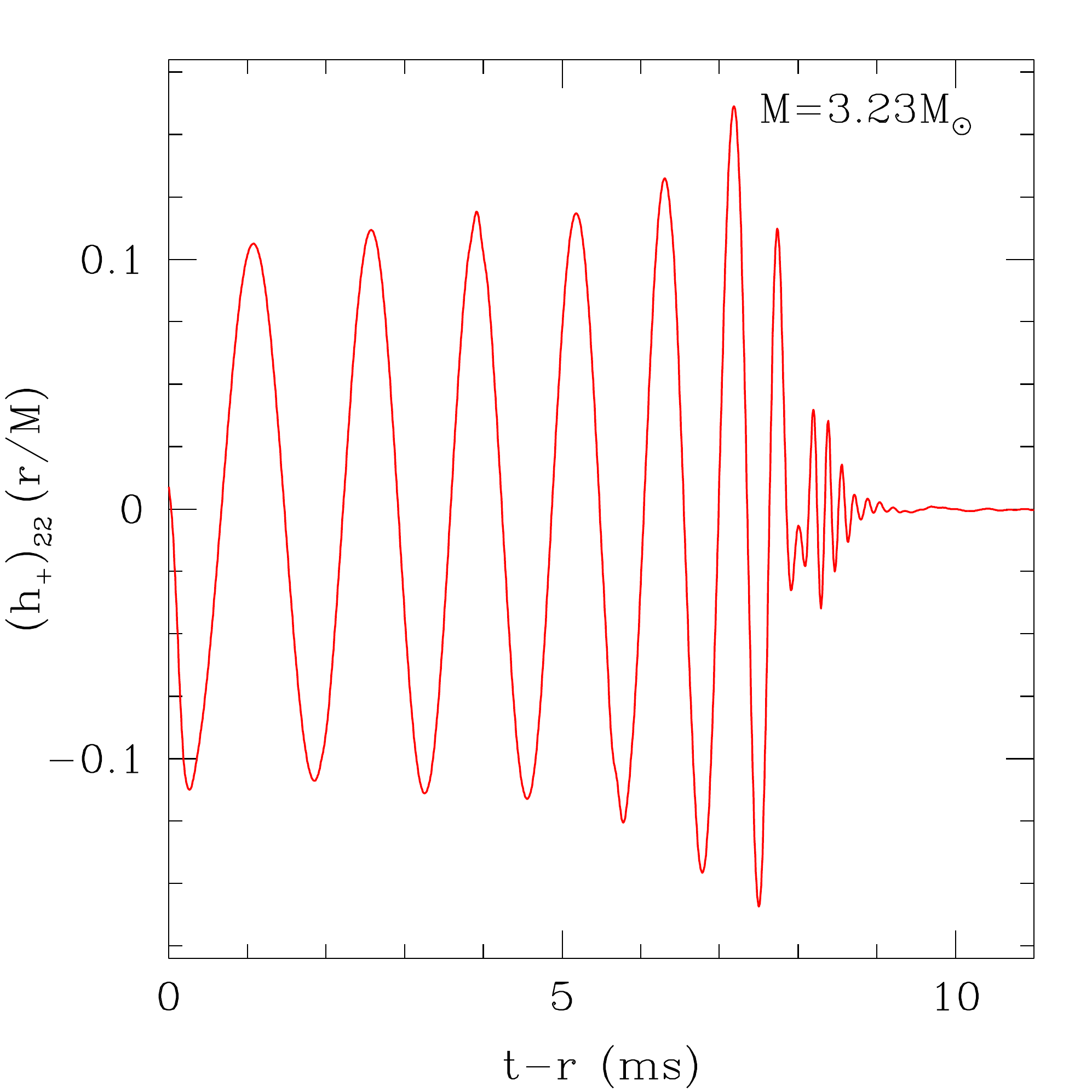} &
\includegraphics[width=0.45\textwidth]{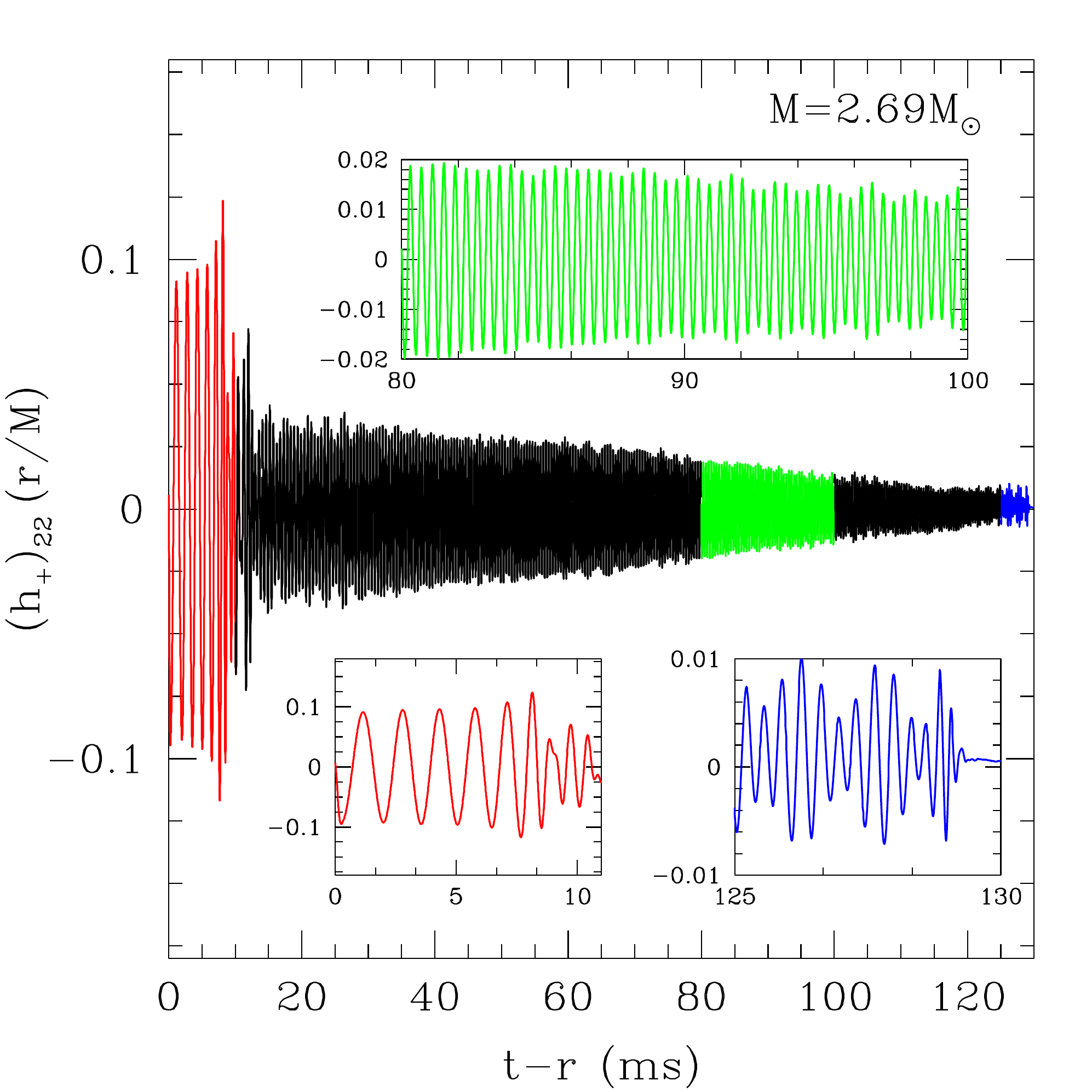}
\end{tabular}
\end{center}
\caption{\gw{} signal from the merger of equal-mass binary \ns{s}. The
left panel refers to a ``high-mass'' case with a total gravitational
mass of $3.23 M_{\odot}$; the right panel, to a ``low-mass'' binary
with a gravitational mass of $2.69 M_{\odot}$. In the ``high-mass''
case the total mass of the system is large enough to produce a prompt
collapse to \bh{} soon after the merger, and the \gw{} signal is
characterized only by inspiral, merger and \bh{} ringdown. In the
right panel, on the other hand, the mass of the system is lower; a
hypermassive \ns{} is formed and survives for $\sim 120$ ms before
collapsing to a \bh{} (note the very different time scale between the
two panels). In the right panel, different insets show zoom-ins on
different parts of the signal. The bottom-left inset shows the
inspiral and merger on the same time scale as the left panel. The top
inset shows the signal emitted by the newborn hypermassive \ns{},
which has a peak frequency at $\sim 2.5$ kHz. The bottom-right inset
shows the final collapse to a \bh{}. The figure was produced using
data from~\cite{Rezzolla2010}.}
\label{fig_bns_gws}
\end{figure*}

New physics and \amr{} have been essential ingredients to accurately
predict \gw{} emission and to establish a possible connection with
short \grb{s}.  In particular, having \amr{} in place in the numerical
codes has allowed various groups to model the inspiral and merger of
\ns{s}~\cite{Anderson2008,Baiotti2008,Yamamoto2008,Baiotti2009} and
to study the formation of hydrodynamic instabilities in the presence
of magnetic fields~\cite{Anderson2008,Liu2008MHD,Giacomazzo2009}.
Similarly, the use of a realistic \eos{} has led to the
suggestion~\cite{Shibata2005,Shibata2006} that there are two broad
classes of \ns{} binary mergers. Binaries with initial total masses
above $\sim 2.8-3.2 M_{\odot}$ (depending on the EOS) promptly form a
BH soon after the merger. On the other hand, binaries with lower
masses yield a metastable, hypermassive \ns{} before collapsing to
a \bh{}. This difference has very important observational
consequences. In the case of massive binaries, the \gw{} signal shows
a quick transition from ``chirping'' during the inspiral to the
characteristic quasinormal ringdown of the
final \bh{} \cite{2009CQGra..26p3001B}. On the other hand, if the
merger yields a hypermassive \ns{,} the chirp in the \gw{s} is
followed by quasi-periodic oscillations at frequencies $2-4$ kHz: see
e.g. Figure~\ref{fig_bns_gws}, adapted from~\cite{Rezzolla2010}.
Unfortunately this quasi-periodic signal will only be detectable by
Advanced LIGO and Virgo interferometers if the merger takes place
within $\sim 20$ Mpc~\cite{Sekiguchi2011b, Kiuchi2012}, but these
systems could be an interesting target for future detectors.

Recent studies~\cite{Kiuchi2009,Kiuchi2010} have also addressed the
association of binary \ns{} mergers with short GRBs in the more
general case of unequal-mass binaries. These investigations showed
that the formation of a massive torus, which could lead to the
emission of gamma rays, leaves a characteristic signature in the
\gw{} spectrum, consisting of an exponential decay followed by a hump in the $\sim
2-7$ kHz range.  The cut-off frequency and the frequency and amplitude
of the hump can be directly associated to the \ns{} EOS and to the
mass of the disk.  While Advanced LIGO and Virgo may not be
sufficiently sensitive at these frequencies, a third-generation
detector (such as the Einstein Telescope) would be sensitive to
sources within 100 Mpc.
Moreover, simulations have shown that unequal-mass binaries can
produce tori with masses up to $\sim 0.35 M_{\odot}$, which would be
more than sufficient to power short \grb{s}~\cite{Rezzolla2010}.
Notice however that simple equipartition arguments would suggest that
even tori with masses as low as $\sim 0.01 M_{\odot}$ may be
sufficient to power short \grb{s} (see e.g. \cite{enrico} for a
discussion).

Present estimates suggest that observations of lowest-order tidal
perturbations in \ns{} mergers with third-generation detectors would
yield measurements of \ns{} radii with $\sim$~1~km
precision~\cite{Read2009,Hinderer2009}, while observations of
strong-field tidal effects during mergers might allow similar
accuracies even with second-generation detectors if sufficiently
accurate models are available~\cite{Markakis:2011}.  Such \gw{}
observations, which probe the bulk properties of neutron stars and
could make it possible to constrain the EOS of matter in the \ns{}
core, will be complementary to electromagnetic
observations~\cite{Ozel2010,2012ApJ...748....5O}, which are sensitive
to surface properties.  This enterprise will require a large number
of \gw{} templates. To reduce the computational burden, there have
been successful attempts to construct semi-analytical \gw{} templates
(e.g. using the effective one-body approach) that include the
influence of tidal
deformation~\cite{Baiotti2010,Baiotti2011,Bernuzzi2012}. More recent
studies in the conformally-flat
approximation~\cite{Bauswein:2011tp,Bauswein:2012ya} have suggested
that the \gw{} signal emitted by the hypermassive \ns{} formed after
merger may also be used to constrain the \ns{} EOS. If the result of
the merger is a long-lived hypermassive \ns{}, the \gw{} signal may be
used to measure the \ns{} radius with an accuracy of up to $\sim
200 \mathrm{m}$ already with Advanced
LIGO~\cite{Bauswein:2011tp,Bauswein:2012ya}. While this is a very
interesting result, a more accurate treatment of general relativistic
effects and magnetic fields may affect the evolution of the
hypermassive \ns{}~\cite{Anderson2008,Giacomazzo2011} and make such
measurements more difficult.

Our understanding of the potential role of magnetic fields in the
dynamics of \ns{} binaries has significantly improved in recent
times~\cite{Anderson2008MHD,Liu2008MHD,Giacomazzo2009,Giacomazzo2011,Rezzolla2011,2012PhRvD..86j4035L,2013arXiv1301.7074P}.
These studies have provided additional support to the view that the
merger of magnetized \ns{s} can provide the central engine for short
GRBs; furthermore, one could have potentially observable emissions
even in cases where a SGRB is not
realized \cite{Rezzolla2011,2012PhRvD..86j4035L,2013arXiv1301.7074P}.
Unfortunately, the studies also showed that the effects of magnetic
fields on the \gw{} signal are most appreciable at frequencies larger
than $\sim 1$~kHz, where Advanced LIGO and Virgo are less sensitive.

Simulations of \ns{} mergers that include the effects of neutrino
cooling are in their infancy. Pioneering work in this
direction \cite{Sekiguchi2011, Sekiguchi2011b, Kiuchi2012} shows that
for a stiff, finite-temperature \eos{} and with neutrino cooling, a
hypermassive \ns{} is the canonical outcome of the merger of \ns{}
binaries with masses smaller than $3.2 M_\odot$, and that thermal
pressure support may be important.
The neutrino luminosity associated with these events could potentially
be detected for mergers within 5~Mpc by
hyper-Kamiokande~\cite{Kiuchi2012}.  Finally, studies of the \gw{}
emission from \ns{} mergers on eccentric orbits have also
begun~\cite{Gold2011,2012arXiv1212.0837E}. These systems may be formed
via dynamical capture in dense stellar environments, and they may
account for a fraction of \ns{} binaries~\cite{Lee2010}.

\subsubsection{Neutron Star-Black Hole Binaries}

In the last few years, the interest in modeling mixed binary systems
comprised of a \bh{} and a \ns{} has also intensified
(see \cite{ShibataLRR} for a recent review).  The first simulations
tracked the merger of a \ns{} with a
non-spinning \bh{}~\cite{Shibata06d,ShibataUryu2007,ShibataTaniguchi2008}. These
early simulations showed that the merger produced a torus with mass
$\sim 0.2 M_{\odot}$ when the \bh{} and the \ns{} have comparable
masses. Subsequent studies with higher numerical accuracy excluded the
possibility of short GRBs in mixed binary mergers if the \bh{} is
non-spinning~\cite{Etienne2008,Duez2008,Yamamoto2008,Giacomazzo2013},
even in the case of equal-mass systems~\cite{Duez2008}.

If the \bh{} is spinning, the merger can potentially power a short
GRB.  Simulations show that a disk with mass $\sim 0.2 M_{\odot}$ can
be formed for binary mass ratios of $1/3$ and \bh{} spins of
$a/M_h=0.75$~\cite{Etienne2009}, where $a=J/M$ is the Kerr spin
parameter (in geometrical units $G=c=1$). Recently, the case of \ns{s}
merging with $10 M_{\odot}$ \bh{s} was explored
in~\cite{Foucart2012a}. If the
\bh{} is rapidly rotating ($a/M_h \sim 0.7-0.9$) the merger can result in
accretion disks massive enough to power a
short \grb{}~\cite{Foucart2012a}; however, we remark once again that
even less massive tori may be able to power short \grb{s} in the
presence of instabilities \cite{enrico}. Furthermore, neutron-rich
ejecta from BH-NS systems are possible, and it has been shown that the
calculated flux and the time to return to the central engine are
consistent with models for extended emissions from the
r-process \cite{2010PhRvL.105k1101C}.

Analytic models have also been developed to compute the mass of the
disk that can be formed after merger~\cite{Pannarale2011,Foucart2012},
as well as other features of the merger
remnant \cite{Pannarale:2012ux}. These studies have the main advantage
that they allow us to explore a larger portion of the parameter
space. They confirmed that massive tori can be formed even at low mass
ratios ($q \sim 0.1$) if the \bh{} is rapidly
spinning~\cite{Pannarale2011,Foucart2012}, and therefore that these
binaries could in principle power short \grb{s}.

\gw{s} emitted from mixed binary systems can be grouped in three
broad classes depending on their behavior near the \isco{}~\cite{Shibata2009}:

\begin{itemize}

\item Type I: the \ns{} is disrupted outside the
\isco{.}

\item Type II: the mass transfer from the \ns{} to the \bh{} takes
place close to the \isco{.} 

\item Type III: the \ns{} is not disrupted
outside the \isco{}, and all of the matter falls immediately into the
\bh{.} 

\end{itemize}

While gravitational waveforms from type-III mergers are difficult to
distinguish from \bh{} binary systems, type I and II will be
sufficiently distinct that they could provide information about the
mass ratio and the \ns{} compactness. The likely frequency of these
distinctive signatures is above 2 kHz, i.e. within the reach of
third-generation detectors, such as the Einstein Telescope. In the
case of non-spinning \bh{s} and for large mass ratios between $\sim
0.3$ and $1$ (i.e., type-I/type-II mergers), tidal deformations
induced in the \ns{} during the inspiral may also allow for Advanced
LIGO measurements of the \ns{} radius (and hence for constraints on
the \eos{}) with accuracy $\sim 10\%-50\%$ for sources located at a
distance of $100$ Mpc~\cite{Pannarale2011b,Lackey2012,Lackey2013}.

While early simulations of mixed binary mergers adopted a simple
ideal-fluid \eos{,} recent work accounts for more
realistic \eos{s}~\cite{Duez2010,Kyutoku2010,Kyutoku2011}, the effects
of the orientation of the \bh{} spin on the formation of the torus
and \gw{} signatures~\cite{Foucart2011}, and the presence of magnetic
fields~\cite{Chawla2010,Etienne2012}. Magnetic fields do not seem to
have a detectable \gw{} signature in the case of mixed binaries, but
they could provide a mechanism for the production of relativistic jets
when a torus is formed after the merger, and various models predict
that they could induce possible observable electromagnetic
counterparts \cite{2001MNRAS.322..695H,2010PhRvD..82d4045P,2011PNAS..10812641N,2011ApJ...742...90M}.

Finally, recent studies considered eccentric orbits in mixed binary
mergers~\cite{Stephens2011,East2012}. In particular, these simulations
addressed the effect of eccentricity on the formation of the massive
torus and on \gw{} emission. While binaries in quasi-circular orbits
emit a periodic signal during their inspiral phase, eccentric binaries
emit a series of quasi-periodic \gw{} bursts, detectable by Advanced
LIGO and Virgo up to distances of 300~Mpc.

\subsubsection{Event Rates}
\label{subs:rates}

Techniques for estimating rates of compact binary mergers have been
recently summarized in \cite{MandelOShaughnessy:2010}.  The rate
of \ns{}-\ns{} binary mergers in our Galaxy can be estimated from
measured parameters of known binary
pulsars \cite{2003ApJ...584..985K}.  Given the lack of direct
observations of \ns{}-\bh{} systems, their rates must be estimated by
modeling, typically involving population-synthesis studies of large
numbers of simulated binaries \cite{OShaughnessy:2007}, or
attempts to predict the future evolution of specific observed systems
that represent possible progenitors of compact
binaries \cite{CygnusX3:2012}.

Unfortunately, both methods currently suffer from significant
uncertainties.  Population-synthesis results depend critically on
assumptions about common-envelope evolution, typical supernova kicks,
mass-loss rates, metallicity, and other astrophysical
parameters \cite{2010ApJ...715L.138B,2012ApJ...759...52D}.
Extrapolations from binary pulsar observations have fewer free
parameters, but their accuracy is limited by small-number statistics
due to the paucity of observed binary \ns{} systems in the Galaxy, the
imperfect understanding of selection effects in pulsar surveys, and
uncertain knowledge of the pulsar luminosity function.

According to the compilation \cite{Abadie:2010cf}, \ns{}-\ns{} merger
rates plausibly range from $1$ to $1000$ mergers per Milky Way Galaxy
per million years.  This range is also consistent with extrapolations
from the observed rate of short GRBs 
assuming a relatively high correction for beaming \cite{Fong:2012}.
Meanwhile, \ns{}-\bh{} merger rates fall in the range $0.05$ to $100$
per million years in the Galaxy.

The conversion of merger rates to detection rates depends on the
assumed detector sensitivities, data quality, and details of the
search pipelines.  The uncertainties in these factors are typically
small relative to the uncertainties considered above, but additional
astrophysical uncertainties encountered when scaling up from the
Galaxy, such as including the contribution of low-metallicity
environments or elliptical galaxies, could be more
significant \cite{OShaughnessy:2009}.  The merger rate ranges
quoted above correspond to detection rates of $0.04$ to $400$ per year
for \ns{}-\ns{} binaries and $0.2$ to $300$ per year for \ns{}-\bh{}
binaries in the era of advanced \gw{} detectors operating at full
sensitivity \cite{Abadie:2010cf}.

\subsection{Binary Black Hole Mergers}
\label{sec:bbh}
The defining characteristic of \bh{s} is their event horizon: a
``surface of no return,'' from within which not even light can escape.
Until now, we have been able to infer the existence of \bh{s} only
indirectly, in particular by modeling phenomena associated with the
neighborhood of the putative \bh{} horizons \cite{lrr-2008-9}.  The
evidence gathered so far is from the behavior of astrophysical
objects, matter or fields that cannot be explained by other means than
by appealing to the presence of a \bh{}, under the assumption that
Einstein's theory of gravity is correct.  Examples are the emission
from active galactic nuclei, the orbits of stars at the center of our
Galaxy, X-ray sources and tidal disruptions of stars, to name a
few. One rather extreme point of view is that no electromagnetic
observation will ever provide conclusive proof of the existence
of \bh{s}
\cite{Abramowicz:2002vt}.
On the other hand, there is general consensus that the detection and
characterization of \gw{s} from the merger of two \bh{s} will offer
compelling evidence for their existence: see
e.g. \cite{2009CQGra..26p3001B,Sperhake:2011xk} and references
therein.
%

Binary stellar mass \bh{s} can be formed either through the evolution
of isolated binaries in galactic fields, or through dynamical
formation scenarios in dense stellar environments, such as globular
clusters or galactic nuclear clusters
(see \cite{MandelOShaughnessy:2010,Abadie:2010cf} and references
therein).  Due to the lack of direct observations of any
binary \bh{s}, predictions about rates and mass distributions of these
systems must rely on simulations.

In Section \ref{subs:rates} we referred to population-synthesis models
for the evolution of isolated compact-object binaries. These models
have particularly large uncertainties in the case of binary \bh{s},
including the effects of metallicity (lower metallicity tends to
decrease mass loss through stellar winds and increase the number of
merging binary \bh{s}), the uncertainty in supernova birth kicks
for \bh{s} (higher kicks may disrupt binaries), and the uncertain
future of binaries that enter the common envelope as they are going
through the Hertzsprung gap (such binaries may merge directly, without
a \gw{} signature).  According to the compilation of predictions
in \cite{Abadie:2010cf}, advanced detectors may observe \gw{s} from
merging binary \bh{s} at a rate between one detection in a few years
and a thousand detections per year.  More recent simulations
considering a wider range of updated models are presented
in \cite{2012ApJ...759...52D}. These simulations indicate that rate
uncertainties still span several orders of magnitude, but also that
rates appear more promising than in the past.  Attempts to model the
future evolution of \bh--Wolf-Rayet binaries IC 10 X-1 and NGC 300 X-1
also indicate that advanced detectors may observe hundreds of events
per year \cite{Bulik:2008}, even though the precise modeling of the
evolution of such systems is still a difficult task.

Another possible channel to produce observable binary \bh{} mergers
consists of dynamical interactions in globular clusters and nuclear
star clusters. In these systems, \bh{s} are likely to sink to the
center through mass segregation and replace other members of existing
binary systems via three-body encounters, leading eventually to
binary \bh{} mergers.  Several simulations (see e.g.~\cite{Sadowski,
MillerLauburg:2008, OLeary:2008, Banerjee:2010}) have indicated that,
although many \bh{s} could be ejected from globular clusters during
three-body encounters, dynamically formed \bh-\bh{ } binaries could
make significant contributions to the overall rates of detected
systems.

Most \bh{} binaries produced in population synthesis models or via
three-body encounters in globular clusters and nuclear star clusters
have masses such that the binaries will inspiral in the band of
interest for Earth-based \gw{} detectors; some of them may also
produce detectable ringdown signals in band. Post-Newtonian
approximations and progress in numerical relativity since the
2005-2006
breakthroughs~\cite{Pretorius:2005gq,Campanelli:2005dd,2006PhRvL..96k1102B}
are supplying accurate knowledge of the type of signal emitted in the
coalescence, by providing waveforms with enough cycles to cover the
inspiral, merger and ringdown (see
e.g. \cite{Pretorius:2007nq,Sperhake:2011xk,Pfeiffer:2012pc} for
recent reviews).  Knowledge of the waveform allows for
matched-filtering searches for these signatures in \gw{} data. Efforts
to create analytical and/or phenomenological models calibrated to
numerical-relativity data are ongoing: see
e.g. \cite{Buonanno:2007pf,Ajith:2009bn,Santamaria:2010yb,Sturani:2010yv,Taracchini:2012ig,2012arXiv1212.4357D}.

Beyond the mass range of stellar-mass \bh{s}, binaries involving
intermediate-mass black holes (IMBHs) could represent exciting \gw{}
sources.  Detectable binaries in  advanced LIGO and Virgo 
would have total mass in the range between $\sim 100$ and $\sim 500$
solar masses.  Both theoretical formation scenarios and observational
evidence for IMBHs are topics of active research and
debate \cite{MillerColbert:2004, Miller:2009}.  If IMBHs have a
non-negligible occupation number in globular clusters, they could
capture stellar-mass \ns{s} or \bh{s}, with \gw{s} from the
ensuing intermediate mass ratio inspirals being detectable at rates of
up to tens per year \cite{Brown:2007, Mandel:2007rates}.  It may also
be possible to detect mergers of two
IMBHs \cite{imbhlisa-2006,Amaro:2006imbh}.  The highly uncertain rates
of these processes are summarized in \cite{Abadie:2010cf}.


The characteristic chirp-like signal from binaries in a quasi-circular
inspiral could be radically modified if the \bh{s} merge in a highly
eccentric, precessing orbit. Such orbits could be the result of
scattering events expected in the environment of dense galactic
cores. The resulting \gw{s} will appear as bursts of radiation near
periastron, followed by quiescent phases while the \bh{s} travel to
and return from apoastron. The time elapsed between subsequent bursts
decreases as the binary hardens and the eccentricity
decreases~\cite{2002PhRvD..66d4002G,2008PhRvD..77j3005L,
2009PhRvD..79d3016L,2009CQGra..26w5010L,2009PhRvL.103m1101H,2007CQGra..24...83P,Gold:2009hr}.
Depending on the masses of the \bh{s}, \gw{s} from highly eccentric
binaries should be visible by both ground- and space-based
interferometers~\cite{2006ApJ...648..411K,2009MNRAS.395.2127O}.

\subsection{Core-Collapse Supernovae and Long Gamma-ray Bursts}
\label{sec:core}

Core collapse supernovae (CCSNe) and long-duration gamma-ray bursts
(LGRBs) share a common origin: the collapse of stars with masses
$\gtrsim 8~\rm M_{\odot}$.  Observational evidence comes from the
positional and temporal association of several nearby LGRBs with
supernovae of Type Ic: see \cite{wb06,2012grbu.book..169H} for
reviews.  These SNe associated with
\grb{s} are distinguished from other core-collapse events based on (i)
the absence of hydrogen and helium in their optical spectra
\cite{fil97}, characterized predominantly by broad
absorption features of intermediate mass elements, and (ii) a strong
non-thermal afterglow component best studied at radio wavelengths
\cite{kfw+98}.  Long-term monitoring of the afterglow reveals
that the LGRB is characterized by a bi-polar relativistic outflow, and
jet opening angles $\lesssim 30^\circ$ are commonly inferred
\cite{piran04}.  Intriguingly, comparing the collimation-corrected
rates of LGRBs and Type Ic SNe, we find that {\it most} LGRBs are
associated with a SN, but less than 1\% of SN Ic are associated with a
LGRB.  This is confirmed through detailed radio studies of local Type
Ic SNe, which indicate that only $\sim 0.7\%$ of SNe Ic drive
relativistic outflows, some of which do not give rise to detectable
($E_{\gamma}\gtrsim 10^{48}~\rm erg$; 25-150 keV) gamma-ray emission
\cite{skn+06,scp+10}.

While the progenitors of SNe Ic and/or LGRBs have yet to be directly
detected in pre-explosion imaging \cite{smartt09}, theoretical
considerations point to massive stars that have been stripped of their
hydrogen envelope prior to explosion, either by their own strong
radiation-driven stellar winds \cite{wlw95} or through the interaction
with a close binary companion \cite{pjh92}.  The critical ingredient
that enables only 1\% of SNe to produce GRBs remains unclear, but a
key component is probably low metallicity ($Z \lesssim
0.5~Z_{\odot}$), allowing the stellar core to retain angular momentum
by suppressing the line-driven winds
\cite{wh06}.  With the advent of new wide-field surveys 
(e.g., Pan-STARRS, Palomar Transient Factory), the rate of CCSN
discoveries in metal-poor galaxies is growing. Combined with radio
follow-up observations, the metallicity dependence of relativistic
outflows in CCSNe can be directly tested.

GRB-associated SNe are not the only explosions
with evidence for asphericity.  Late-time spectroscopy of CCSNe in the
nebular phase \cite{tvb+09}, polarization measurements of local CCSNe
\cite{cfl+10}, and detailed high spatial resolution studies of
Galactic SN remnants such as Cas A \cite{fhm+06} all suggest that
ejecta asymmetries are in fact common place.  Thus, it remains unclear
what distinguishes the progenitors of ordinary CCSNe from GRBs, and it
could be that a significant fraction of CCSNe (not just those
associated with {\it detected} gamma-ray bursts) could experience
exotic explosion mechanisms.

\subsubsection{Theoretical Modeling}

At the end stage of stellar evolution, the core of a massive star is
supported against gravity by the pressure of relativistically
degenerate electrons. Collapse is initiated when the core exceeds its
effective Chandrasekhar mass and continues until the inner core
reaches nuclear density. There, the nuclear EOS stiffens, leading to
core bounce and the formation of the hydrodynamic bounce shock. The
shock runs into the supersonically collapsing outer core, losing its
energy to the break-up of infalling heavy nuclei into nucleons and to
neutrinos that are made by electron capture in the region behind the
shock, and stream out freely as the shock reaches regions of low
neutrino optical depth. The shock stalls, turns into an accretion
shock, and must be revived by the \emph{core-collapse supernova
  mechanism} to drive a core-collapse supernova explosion. This is the
basic picture that has been established since Bethe's authoritative
1990 review \cite{bethe:90}.

A variety of {mechanisms} have been proposed in the literature (see,
e.g., \cite{janka:07} for a recent review) and all leading
candidates involve GW-emitting aspherical dynamics. The \emph{neutrino
  mechanism} relies on the net deposition of energy by charged-current
neutrino absorption in the region immediately behind the stalled
shock.  While the neutrino mechanism fails to blow up ordinary massive
stars in spherical symmetry, the multi-dimensional phenomena of
convection, turbulence, and the standing-accretion shock instability
(SASI, an advective-acoustic instability of the stalled shock) very
likely enhance the neutrino mechanism's efficacy
\cite{fryerwarren:04,marek:09,yakunin:10,nordhaus:10,murphy:08,hanke:11,takiwaki:11b}.

The \emph{magnetorotational mechanism} relies on rapid rotation and
magnetic field amplification due to flux compression in collapse,
rotational winding, and the magnetorotational energy that converts
free energy of differential rotation into magnetic field
\cite{burrows:07b,bh:91,obergaulinger:09,cerda:08,takiwaki:11a}.
These processes may lead to magnetic fields with strengths of order
$10^{15}\,\mathrm{G}$, which would be sufficient to launch bipolar
magnetohydrodynamic jets, leading to an energetic, strongly aspherical
explosion \cite{couch:09}. The key ingredient required for the
magnetorotational mechanism is rapid progenitor star rotation, but
most massive stars, perhaps up to 99\%, are presently expected to be
slow rotators \cite{heger05,ott:06spin}.

A third potential way of driving core-collapse supernova explosions is
the \emph{acoustic mechanism} proposed by
\cite{burrows:06,burrows:07a}. In their simulations, SASI-modulated
turbulence and accretion downstreams hitting the protoneutron star
(PNS) excited pulsations (predominantly of $l=1$ spatial character) of
the latter, that grew to non-linear amplitudes and dissipated in sound
waves. Propagating along the density gradient behind the stalled
shock, the sound waves steepened to secondary shocks, injecting
additional heat into the postshock region, eventually leading to
explosion. This mechanism is robust, but there are many unresolved
issues with it. Explosions occur quite late and would imply \ns{}
masses and nucleosynthetic yields that are likely inconsistent with
observations.  Only one group and code have produced this mechanism to
date.  Though many see the necessary ingredients for this mechanism,
including excitation of PNS pulsational modes, it is not clear whether
the amplitudes obtained by \cite{burrows:06,burrows:07a} are produced
in nature.  Furthermore, \cite{weinberg:08} showed that nonlinear
parametric instabilities may limit the oscillation amplitudes of the
PNS by funneling oscillation power into daughter modes.

The details of the long GRB central engine may be as uncertain as the
core-collapse supernova mechanism, but the relativistic beamed
outflows observed from GRBs strongly suggest that rapid rotation plays
a major role in the central engine.

In the collapsar scenario, outlined first by \cite{woosley:93}, a
rotating core-collapse supernova fails to explode or explodes weakly
or very aspherically, leading to \bh{} formation before (type-I
collapsar) or after (type-II collapsar) an explosion by fallback
accretion.  Eventually, typically seconds after BH formation
\cite{oconnor:11}, an accretion disk is expected to form near
the BH. Accretion energy or extracted BH spin energy, mediated via MHD
processes \cite{nagataki:07,komissarov:09} and/or neutrino pair
annihilation \cite{birkl:07,harikae:10}, may then drive the
relativistic GRB outflow, while MHD disk winds and viscous heating may
power a GRB-accompanying energetic core-collapse supernova
explosion \cite{macfadyen:01,lindner:11}.

The rapid progenitor rotation required in the collapsar scenario may
lead to an energetic magnetorotational explosion, preventing BH
formation \cite{dessart:08a}. This possibility gives rise to the
competing millisecond protomagnetar model for the long GRB central
engine \cite{thompson:04,bucciantini:08,metzger:11}. In this model, a
magnetorotational core-collapse supernova explosion excavates the
polar regions, allowing the driving of an ultra-relativistic wind by
the spin-down of the strongly magnetized neutrino cooling protoneutron
star (the protomagnetar). The protomagnetar model is able to explain
the prompt GRB emission, and prolonged magnetar activity may explain
long-duration X-ray afterglow observed in long GRBs.

\subsubsection{Gravitational Wave Emission}
\label{subs:SNGWs}

The ubiquituous aspherical dynamics in stellar collapse, core-collapse
supernovae and long GRBs gives rise to bursts of GWs with typical
durations from milliseconds to seconds, whose waveforms are impossible
to predict precisely by simulations. The reason is that much of the GW
emission is influenced or dominated by stochastic dynamics
(i.e. turbulence), and also that much of the input physics (e.g., the
nuclear EOS) and the initial conditions are complicated and impossible
to know exactly.

Aspherical stellar collapse was early on considered as a source of
detectable GWs \cite{weber:66,ruffini:71} and has been studied
extensively: see \cite{fryernew:11,ott:09,kotake:06a} for recent
reviews. In nonrotating or only slowly rotating core-collapse
supernovae, the GW emission is dominated by convective overturn in the
PNS and in the region behind the shock, modulated by the SASI and
enhanced by fast accretion downstreams that are decelerated in the
stably stratified outer layers of the PNS.  The emitted GW signal has
random polarization, a broad spectrum with power at
$\sim$$100 - 1000\,\mathrm{Hz}$, and dimensionless strain amplitudes
of order $10^{-22}$ at a source distance of $10\,\mathrm{kpc}$
\cite{yakunin:10,marek:09b,murphy:09,mueller:11,kotake:11}.
Contributions at low frequencies ($\lesssim 30\,\mathrm{Hz}$) come
from anisotropic emission of
neutrinos \cite{epstein:78,marek:09b,kotake:11,mueller:11} and
explosion asphericities \cite{murphy:09,yakunin:10,mueller:11}.

If the neutrino mechanism lacks efficacy and the explosion is delayed
to late time, the strong PNS oscillations associated with the acoustic
mechanism may be excited. Their quadrupole components emit GWs at
momentarily fixed (secularly changing due to changes in the PNS
structure) frequencies of $\sim$$600 - 1000\,\mathrm{Hz}$, with strain
amplitudes of $10^{-21} - 10^{-20}$ at 10
kpc \cite{ott:06prl,ott:09}. So far, all simulations of these
pulsations have been axisymmetric, predicting linearly polarized
signals, but in 3D correlated emission in the second GW polarization
can be expected.

Rapid rotation, if present, will lead to a characteristic burst of GWs
emitted at core bounce, when the inner core undergoes the greatest
acceleration. This signal has been shown to be linearly polarized
(i.e., the dynamics is axisymmetric) and increases in amplitude with
increasing initial inner core angular velocity, up to the point where
centrifugal forces become dominant and decelerate the bounce dynamics
\cite{dimmelmeier:08,dimmelmeier:07,kotake:03,scheidegger:10b,ott:07prl}. 
Typical signal amplitudes are of order $10^{-21}$ at 10 kpc for cores
with precollapse central spin periods of $2-4\,\mathrm{s}$; the GW
emission peaks around $700 - 800\,\mathrm{Hz}$, decreasing to below
$\sim$$200\,\mathrm{Hz}$ for very rapid rotation. Simulations that
take into account magnetic fields found that extreme precollapse iron
core fields in excess of $10^{12}\,\mathrm{G}$ would be necessary to
modify the bounce dynamics and GW signal
\cite{kotake:04,obergaulinger:06a,obergaulinger:06b,takiwaki:11a,shibata:06}.
More moderate initial fields can be amplified in the postbounce phase
and will modify the postbounce dynamics and GW signal.

Subsequent to core bounce, nonaxisymmetric rotational instabilities
may develop. These require rapid spin and/or differential rotation.  A
ratio of rotational kinetic to gravitational energy $T/|W|$ above
$\sim$27\% is required for a classical high-$T/|W|$ dynamical
instability, that leads to an $m=2$ deformation of the PNS. A secular
instability (driven by GW radiation reaction or viscosity) may set in
at $T/|W|\gtrsim 14\%$ \cite{stergioulas:03}. Typical rapidly spinning
cores lead to PNS with $T/|W| \lesssim
10\%$ \cite{dimmelmeier:08}. Core collapse naturally produces a nearly
uniformly spinning PNS core with a strongly differentially rotating
outer mantle \cite{ott:06spin}. This differential rotation can drive a
rotational shear instability leading to angular momentum
redistribution, nonaxisymmetric deformation and GW
emission \cite{centrella:01,saijo:03,watts:05,rotinst:05,saijo:06,ott:07prl,scheidegger:08,scheidegger:10b,corvino:10}. Typical
signal characteristics are strain amplitudes of order $10^{-21}$
at 10~kpc and quasi-periodic emission at twice the frequency of the
unstable mode -- typically $\sim$$800 - 1000\,\mathrm{Hz}$
\cite{ott:07prl,scheidegger:08,scheidegger:10b} --
for a duration of $10-\mathrm{few}\,100\,\mathrm{ms}$.

After the onset of an explosion, the GW signal emitted by dynamics
between the PNS core and the shock will subside quickly, leaving the
more gradually decaying GW signal from PNS convection and,
potentially, nonaxisymmetric rotational dynamics behind. If the
explosion fails, a BH forms after $\sim$$1-3\,\mathrm{s}$ (the exact
time is determined by the nuclear EOS and the progenitor
structure). If the PNS is spinning, this will give rise to a second
pronounced peak in the GW signal, with strain of order $10^{-20}$ at
$10\,\mathrm{kpc}$ and most GW power at frequencies above $1-2\,\mathrm{Khz}$ \cite{ott:11a}.

In a collapsar-type long GRB, the GW emission will be very similar to
a rapidly spinning core-collapse supernova up to BH formation. The
latter will be followed by a multi-second GW-silent phase after which
instabilities in the inner accretion disk and/or outer accretion torus
may give rise to GW emission lasting, possibly, for the duration of
the GRB \cite{kiuchi:11,piro:07,korobkin:11}. Detailed waveforms of such
instabilities have yet to be predicted by simulations.

In a millisecond-protomagnetar long GRB, the signal from BH formation
and the GW-silent phase would be absent. The GW emission due to
nonaxisymmetric rotational dynamics of the protomagnetar may continue
for the duration of the GRB and, if the instability is secular, possibly
throughout the early afterglow phase~\cite{corsi:09,piro:11}, and may be
detectable by advanced LIGO out to $\sim$100~Mpc~\cite{corsi:09}.

\subsection{Isolated Neutron Stars}
\label{sec:lonelyNS}

The \ns{} menagerie offers a rich variety of electromagnetic
phenomenology and possibilities for \gw{s}.  In addition to the binary
NS coalescences discussed in Section \ref{sec:nsbinaries}, the birth
of a \ns{} following the death throes of a medium sized star in a
supernova explosion may also be visible with advanced detectors.  Each
scenario may (depending on the maximum
\ns{} mass) lead to a hot, possibly rapidly spinning, remnant
with violent dynamics. Initially this remnant is opaque to neutrinos,
but after a few tens of seconds \cite{1986ApJ...307..178B,
1988PhR...163...51B,1999ApJ...513..780P} it becomes transparent and
cools, the thermal pressure drops and a \ns{} is formed. During
the initial phase, the \gw{} signature of the new-born \ns{}
may evolve considerably \cite{2011PhRvD..84d4017B}, due to changes in
thermal gradients and interior composition. This evolution, in fact,
continues for the first few months of the \ns{} life, as the
crust freezes and the various superfluid/superconducting components
establish themselves. At the end of the process, a mature \ns{}
has a complex structure, the modeling of which requires an
understanding of much extreme physics.

In this section, we will focus on \gw{} burst signals from isolated
\ns{s}. We take the, possibly simplistic, view that the related
phenomena can be understood in terms of the star's oscillation
modes. This is certainly the case for many of the mechanisms that have
been discussed in the literature, ranging from various
mode-instabilities to magnetar flares and radio pulsar glitches. The
interest in instabilities is natural, since they provide an
explanation for the excitation of the modes and the associated \gw{}
signal. Similarly, it is reasonable to consider scenarios associated
with known electromagnetic phenomena, like magnetar flares and pulsar
glitches. There are, however, issues with each of these scenarios. In
the first case, the presence of an instability will not guarantee a
detectable \gw{} signal. The relevance of the mechanism depends on the
physics that counteracts an unstable, growing mode. This involves both
the mechanics of the problem -- whether non-linear hydrodynamics
saturates the
instability \cite{2002PhRvD..66d1303G,2007PhRvD..76f4019B,2009PhRvD..79j4003B,2010PhRvD..82j4036K}
-- and microphysics, as encoded in the relevant viscous damping
channels \cite{2001IJMPD..10..381A}. While we have made good progress
on understanding such issues in the last few years, it is clear that
many challenges remain. In the second case, most estimates are based
on simple energetics. However, the link between the observed
electromagnetic signal and any \gw{s} that may be generated by the
underlying mechanism is not at all clear. This is rather obvious,
since the detailed mechanisms leading to observed flares and glitches
remain rather poorly modelled. Most current estimates are based on
plausibility arguments. To make progress we need a better
understanding of these enigmatic events. That this is a challenge is
clear from the fact that the pulsar glitch mechanism and high-energy
emissions from pulsars are not well understood, despite four decades of
study \cite{2011MNRAS.414.1679E}.

\ns{s} may radiate \gw{s} through a  range of mechanisms. For traditional reasons, the associated signals tend to be categorized either as ``bursts'' or ``periodic'' signals (mainly because of the different data analysis strategies used to look for the signals). However, this division is rather arbitrary. For example, rotating \ns{s} with quadrupolar deformations (linked to the geological history of the elastic crust or the structure of the magnetic field)  would produce continuous quasi-sinusoidal GW signals. Such signals have already  have been the subject of GW searches \cite{2010ApJ...713..671A,2008ApJ...683L..45A,2011ApJ...737...93A}.
However, it is by no means clear that the signal will remain unchanged
over a long term observation lasting months to years. In fact, one may
argue that \ns{} ``mountains'' ought to be transient, evolving due to
plastic flow \cite{2010MNRAS.407L..54C}. However, as the timescale and
detailed behavior of the evolution is essentially unknown, this
possibility has not been considered so far. A closely related problem
concerns \ns{s} that interact with their environment, as in the case
of accretion from a binary partner in a Low-Mass X-ray Binary. While
it is natural to assume that the accretion of material leads to some
level of quadrupole deformation and \gw{}
emission \cite{1998ApJ...501L..89B}, it is far from clear how such
``mountains'' are established and to what extent they evolve as the
accretion rate changes. This is a key question that needs to be
addressed if we are to search for signals from such systems. The
problem is also linked to that of instabilities evolving on a secular
timescale, as in the case of f- and r-modes discussed below.
Instabilities may trigger a violent behaviour, but they may also be
rather subtle, leading to the system simmering on the threshold of
stability. The latter behaviour is, in fact, what is expected for the
instability associated with inertial
r-modes \cite{2002MNRAS.337.1224A,2007PhRvD..76f4019B,2009PhRvD..79j4003B}.

While searches for GWs from \ns{s} have already been performed on data
from first-generation
detectors~\cite{2007PhRvD..76f2003A,2010ApJ...713..671A,2008ApJ...683L..45A,2011ApJ...737...93A,2011ApJ...734L..35A,2011PhRvD..83d2001A,2010ApJ...722.1504A,2008PhRvL.101u1102A},
it is generally expected that such signals will require
third-generation detectors like the Einstein Telescope. A review of
relevant sources of \gw{s}, including isolated \ns{s}, for the
Einstein Telescope can be found in the design
study~\cite{2010CQGra..27s4002P} as well as recent review
articles~\cite{2008RvMA...20..140K,2011GReGr..43..409A}.  Search techniques for long-duration transients have been proposed in \cite[e.g.,][]{Prix:2011,Thrane:2011}.
Here, we
will present a brief overview of the main mechanisms for ``burst''
emission from isolated \ns{s} touching on the science returns for
observing such signals, and highlighting issues that require further
attention.

\subsubsection{Instabilities}

\ns{s} may suffer various instabilities as they evolve from hot
remnants to cold mature objects. Some of these instabilities may be
efficient \gw{} emitters and so are of obvious interest for \gw{}
astronomy. The relevant instabilities can be broadly divided into two
classes. Dynamical instabilities tend to grow rapidly, and do not
require additional ``physics'' for their existence.  The most commonly
considered such instability is the bar-mode instability associated
with the star's fundamental (f-) mode.  Secular instabilities, on the
other hand, are relatively subtle. They owe their existence to
dissipative mechanisms and tend to grow on the associated dissipation
timescale. The most important such instabilities (in the present
context) are driven by \gw{} emission via the so-called
Chandrasekhar-Friedman-Schutz
mechanism \cite{1970ApJ...161..561C,1978ApJ...222..281F} and are
associated with the f-mode and the inertial r-mode in a rotating star.
The basic picture is that an instability sets in at some threshold,
say above a critical rotation rate, leading to the growth of a
non-axisymmetric perturbation. The existence of an instability and the
early phase of its evolution are relatively easy to establish, since
they can be studied within linear perturbation theory. Once the
growing mode reaches a sizeable amplitude the situation becomes less
clear, as one must account for the full nonlinear dynamics. At some
point one would expect the instability to saturate. Understanding the
mechanism for, and level of, saturation is key if we want reliable
estimates of the emerging \gw{s}.

The dynamical bar-mode instability has been studied in some detail via
numerical simulations. As anticipated in Section \ref{subs:SNGWs},
this instability sets in once the ratio of rotational kinetic energy
($T$) to gravitational binding energy ($|W|$) exceeds a certain
threshold. At that point, the f-mode grows and deforms the star into a
(rotating) bar-shape. This would be a very efficient configuration for
emitting \gw{s}. However, it is not clear that real \ns{s} exhibit
this instability. The main problem is that the critical threshold
requires a significant amount of differential rotation. Uniformly
rotating \ns{s} can not get near the critical value of $T/|W|$, because
they reach the break-up limit before this happens.  There are, of
course, situations where differential rotation is expected to develop,
most notably a few milliseconds after core-bounce in a supernova or
tenths of seconds later, as the \ns{} contracts due to cooling by
neutrino emission. Present
simulations \cite{2001ApJ...548..919S,2007PhRvD..75d4023B,2009ApJ...707.1610C}
suggest that this may lead to $T/|W|$ becoming large enough to trigger
the bar-mode instability, but unfortunately the system evolves away
from this regime rather quickly.  Recent simulations also cast doubt
on the notion that the unstable bar-mode would last for many
rotations \cite{2007PhRvD..75d4023B}, as required to make the
effective \gw{} amplitude detectable from sources outside our
galaxy. A closely related, somewhat more subtle instability may be
more important.  There is evidence that instabilities may be triggered
at much lower values of $T/|W|$, provided that the system exhibits
significant differential
rotation \cite{2001ApJ...550L.193C,2002MNRAS.334L..27S,2005ApJ...618L..37W,2007CoPhC.177..288C,
2010CQGra..27k4104C}.  This class of instabilities is much less well
understood at the moment. Most importantly, we need to establish
whether real astrophysical systems may evolve into the relevant part
of parameter space.

In the last few years, the main focus has been on secular (\gw{}
driven) instabilities. This kind of instability sets in when the
pattern speed of a given modes changes from counter- to co-rotating
with respect to an inertial frame. This effectively means that the
system radiates positive angular momentum, drawn from a negative
angular momentum reservoir, leading to a runaway process.

Early work on this mechanism focused on the instability of the f-mode
in Newtonian \ns{}
models \cite{1983PhRvL..51..718F,1991ApJ...373..213I,1995ApJ...444..804L,1995ApJ...442..259L}.
The results suggest (perhaps somewhat optimistically) that the
unstable modes could be observable from sources beyond our
Galaxy. However, the f-mode instability only operates near the mass
shedding limit, so \ns{s} would have to be born rapidly
spinning for the mechanism to kick in during their early life. Current
observations suggest that the subset of \ns{s} born spinning
sufficiently fast may be rather small \cite{Noutsos:2013}, but this is certainly not well
understood at the present time. Another problem for the f-mode
instability is that it may be completely quenched by dissipation. In
particular, the so-called mutual friction associated with superfluid
vortices may suppress the instability once the star cools below the
threshold for
superfluidity \cite{1995ApJ...444..804L,2009PhRvD..79j3009A}. This
means that the f-mode instability is unlikely to operate in mature
\ns{s}, e.g. ones spun up by accretion in a Low-Mass X-ray
Binary. Our understanding of the f-mode instability has improved
significantly recently, with accurate numerical simulations both at
the linear and nonlinear level. In particular, we now have a clearer
picture of the instability for relativistic
stars \cite{2008PhRvD..78f4063G,PhysRevD.81.084055}. These results
indicate an enhancement of the f-mode instability due to relativistic
effects, renewing interest in the mechanism as a source for
gravitational radiation \cite{Gaertig:2011bm}.

In the last decade most work on secular instabilities has focused on
the inertial r-modes. These modes are interesting as they radiate
mainly through current multipoles, not mass multipoles (as in the case
of virtually all other \gw{} sources). That these modes would also be
unstable due to the emission of \gw{s} came as some surprise, and by
now many aspects of the associated instability have been considered:
see \cite{2001IJMPD..10..381A,2003CQGra..20R.105A} for exhaustive
reviews. The r-mode instability is interesting for many reasons. It
may provide a natural explanation for the absence of \ns{s} spinning
faster that 720~Hz by preventing further spin-up once an accreting
(recycled) \ns{} reaches the instability threshold. This mechanism
would lead to \ns{s} in Low-Mass X-ray Binaries being interesting
targets for \gw{} searches \cite{1999ApJ...516..307A}. However, due to
the varying accretion rate in these systems and the many unknown
parameters, such searches will be very
difficult \cite{2008MNRAS.389..839W}. As in the case of the f-modes,
the unstable r-modes are counteracted by a range of dissipative
mechanisms. It is generally thought that the most important damping
mechanisms are associated with i) a viscous boundary layer at the
crust-core interface, ii) superfluid mutual friction, and iii) hyperon
bulk viscosity in the deep core of the star. The star's magnetic field
may also have a decisive importance; this issue has not been studied
in sufficient detail yet, but see \cite{2000ApJ...531L.139R}. Very
recent work \cite{2011PhRvL.107j1101H}, comparing the predicted r-mode
instability window to observed accreting systems, suggests that our
understanding is far from complete. The generally accepted r-mode
model would lead to a large number of observed systems in fact being
unstable. This is an obvious problem that needs to be addressed by
improving our models. It is, however, not clear what the missing piece
of the puzzle may be.
 
Possibly in contrast with the f-mode, the r-mode is expected to
saturate at low amplitudes due to nonlinear mode-coupling. The upshot
of this is that the associated \gw{s} are unlikely to be observed from
outside our galaxy \cite{2007PhRvD..76f4019B}. The results also imply
that the spin evolution of a \ns{} with an unstable r-mode (at
saturation) may be rather complex, making a \gw{} search even more
challenging.
 
\subsubsection{Asteroseismology}

In order to understand the observed \ns{} phenomenology we need to
account for much extreme physics, many aspects of which are poorly
constrained. In fact, many relevant issues will never be tested in
terrestrial laboratories. Consider, for example, the possibility of
quark deconfinement at high densities. While colliders like the LHC at
CERN and RHIC at Brookhaven probe the properties of quark-gluon
plasma, they can not reproduce the high-density/low temperature
environment of a \ns{} core. By exploring \ns{} physics (making
``sense'' of observations) we can hope to constrain theoretical
physics in many useful ways. In fact, this is an exciting promise
of \gw{} astronomy. The basic idea is simple. If we observe \gw{s}
from an oscillating \ns{}, then we can use the data to infer the state
of matter in the star's core. This prospect is particularly exciting
as it provides a probe of the high-density region, not the surface
(where most electromagnetic phenomena arise). Of course, there is a
downside to this as well. It means that we need to construct models
that faithfully represent the core physics.  This is far from easy.

Due to their complex interior structure, \ns{s} have many (more or
less) distinct families of oscillation modes. Roughly speaking, one
can associate different mode families with different pieces of
physics \cite{1999LRR.....2....2K}. Pressure gradients lead to the
acoustic p-modes, composition (or thermal) stratification leads to
g-modes, rotation leads to inertial modes, the dynamic spacetime leads
to w-modes. There are modes associated with the crust, superfluidity,
the magnetic field and so on. Of course, this means that the spectrum
of a real \ns{} is tremendously complicated and it may be very
difficult to make sense of any data.  However, for \gw{} astronomy the
situation may not be too bad, because most of the possible modes are
unlikely to be efficient emitters of gravitational radiation. There
are, essentially, two questions. Are there viable astrophysical
scenarios where the oscillations of a star are excited to a large
enough amplitude that the associated \gw{s} may be detected? If so,
what can we learn from such observations? So far, most research in
this area has focussed on the second question. It has been
demonstrated that global properties, such as mass and radius, can be
constrained by observing f-modes (perhaps in some combination with
p-modes and
w-modes) \cite{1996PhRvL..77.4134A,1998MNRAS.299.1059A,2004PhRvD..70l4015B}. It
has also been shown that the rotational deformation, which may have a
severe effect on the mode spectrum, can be ``filtered
out'' \cite{2011PhRvD..83f4031G}. Thermal g-modes have been studied
for proto-\ns{s} \cite{2011PhRvD..84d4017B}, and superfluid modes have
also been considered \cite{2011arXiv1105.4787P}, but we do not yet
have sufficiently realistic models that we can consider the
combination of the different effects. However the relevant theory
framework has been developed, so it is just a matter of time until the
models we consider can be considered (at least moderately) realistic.

As far as realistic astrophysical scenarios are concerned, we know
that isolated \ns{s} suffer violent events like pulsar glitches or
magnetar flares.  The energetics of these events is such that they
could plausibly be relevant for \gw{} astronomy. The key question is
whether sufficient energy is released gravitationally. To establish
this, we need to develop models that account for the observed
phenomenology. Following the exciting discovery of quasiperiodic
oscillations (QPOs) in the tails of giant magnetar
flares \cite{2005ApJ...628L..53I,2006ApJ...637L.117W,2007AdSpR..40.1446W,2011A&A...528A..45H},
likely heralding the era of actual \ns{} seismology, there has been
significant activity aimed at understanding the dynamics of these
events.  Magnetars are strongly magnetized, slowly spinning \ns{s}
that exhibit high energy emission, occasionally punctuated by
bursts. The favoured model contends that the energy of the magnetic
field powers the observed activity, and is responsible for the bursts
and occasional giant flares in soft-gamma repeaters and anomalous
X-ray pulsars \cite{2008A&ARv..15..225M}. The observed QPOs have a
complicated oscillation
spectrum \cite{2007MNRAS.374..256S,2007MNRAS.375..261S,2007MNRAS.374..256S,2007MNRAS.377..159L,2008MNRAS.385L...5S,2011MNRAS.412.1730L},
the analysis of which may constrain both crust physics and the
magnetic field
structure \cite{2009MNRAS.396..894A,2011MNRAS.410.1036V,2011MNRAS.410L..37G,2011MNRAS.414.3014C}.
However, the fact that these features are coupled makes the problem
non-trivial.  Perhaps optimistically, one may assume that \gw{s} are
also generated by the large-scale, dynamical rearrangement of the core
magnetic
field~\cite{2001MNRAS.327..639I,2011PhRvD..83h1302K,2011PhRvD..83j4014C,2011arXiv1103.0880L}.
However, recent numerical simulations of magneto-hydrodynamic
instabilities \cite{2011ApJ...735L..20L} suggest that there would not
be any observable gravitational radiation \cite{2012PhRvD..85b4030Z}
unless the magnetic field is unphysically
large \cite{2011ApJ...736L...6C}. In fact, the detection of \gw{s}
from magnetars in the near future seems unlikely when based on
triggering f-modes in the \ns{}
~\cite{2011arXiv1103.0880L,2012PhRvD..85b4030Z}.

The recurrent pulsar glitches are also interesting, especially since
they set a relatively low energy threshold for events that happen
regularly in our galaxy. The general picture is that smaller glitches
may be due to crust cracking, while the largest observed events are
due to a transfer of angular momentum from a superfluid component to
the crust (to which the magnetic field is anchored). These events
could plausibly generate \gw{s} as well, although to make definite
statements about such signals is very difficult. This is not
surprising, since the underlying glitch mechanism is not well
understood. Available estimates range from pessimistic, suggesting
that the radiated \gw{s} will never be
detected \cite{2010MNRAS.405.1061S}, to (overly) optimistic, where the
signal would be borderline detectable with the first generation
detectors \cite{2008CQGra..25v5020V}. It is quite easy to point to the
flaws of each model, but to fix the relevant issues is not so
straightforward. To make progress we need to improve our understanding
of superfluid dynamics.

A closely related problem concerns the tidal interaction in a binary
inspiral. It has been argued that the deviation from point-mass
dynamics may be detectable at the late stages of
inspiral \cite{2008PhRvD..77b1502F,
2008ApJ...677.1216H,2009MNRAS.396..894A}. Tidal stresses may also
crack the crust, possibly leading to an electromagnetic signal that
would precede the
merger \cite{2011arXiv1109.5041P,2011arXiv1110.0467T}. This problem is
interesting, and it requires the same computational technology as the
seismology problem. Again, the challenge is to build truly realistic
\ns{} models, and assess the impact of the many different
pieces of physics involved in the dynamics of the star. Possible
electromagnetic precursors are also of obvious interest for \gw{}
searches.

\section{Detectors and their capabilities}
\label{sec:instrumentation}

The observation of \gw{s}, combined with astronomical observations
from gamma-ray and X-ray satellites, optical/radio telescopes, and
neutrino detectors, will enable a new, comprehensive multi-messenger
astrophysics which will play a transformative role in our
understanding of the Universe, with new constraints to source models
combined with identification of the host galaxy, redshift and
luminosity distance.
In this section, we review present and prospected instrumental
capabilities for multi-messenger observations in the foreseeable
future.

\subsection{Gravitational-Wave Interferometers}

The inception of ground-based interferometers represented significant progress
toward the  detection of \gw{s}. In 2005-2010 the Laser
Interferometer Gravitational-wave Observatory (LIGO)~\cite{LIGO}
and Virgo~\cite{Virgo} operated four detectors,
 sensitive to  the merger of two \ns{s} within $\sim{30}$~Mpc
of Earth~\cite{S5range, S6cbcLowMass}.
A GW interferometer uses lasers to monitor the relative distances between the beam splitter and
mirrors located at the end of its two arms.  A GW signal will
stretch one arm of the interferometer and compress the other, causing
a detectable change in the interference pattern at the output of the
interferometer.  First-generation GW interferometers were capable of
observing a change in the arm length of $10^{-18}$~m, or about 1/1000
the diameter of a proton, around the most sensitive frequency ($\sim $100~Hz).
Despite this impressive sensitivity no
\gw{s} have been detected so far, due to the low expected event rates,
as discussed in Section~\ref{subs:rates} for the merger of compact binary systems.

The second generation of \gw{} interferometers, scheduled to start within this decade,
will have improved seismic isolation, suspension, optics and laser systems,
offering roughly a factor 10 sensitivity improvement
over a wide frequency range ~\cite{advLIGO}.
These detectors could allow us to search for \gw{}s from the rich class of transients discussed in Section \ref{sec:SourceScience}.

In general, transient searches for \gw{} signals fall into two categories: modeled and unmodeled (or weakly modeled).  Modeled searches rely on the availability of precise template waveforms for some signal classes, such as the known inspiral waveforms from compact binaries, against which data can be compared with matched filtering techniques \cite{ihope}.  Unmodeled searches can use the coherence between excess power in multiple detectors to distinguish signals from noise \cite{PhysRevD.83.102001}.  Therefore, it is critical to have several detectors with comparable sensitivies operating with a high-coincidence duty cycle.

A network of detectors brings significant resistance against
nonstationary noise due to environmental and instrumental
disturbances, as well as sensitivity in a greater
volume~\cite{PhysRevD.83.102001, ShutzNET}.  Two advanced LIGO
detectors~\cite{advLIGO} in Livingston, LA and Hanford, WA, and one
advanced Virgo detector \cite{advVirgo,Virgo} are currently under
construction and will start operation in a few years. A large
cryogenic \gw{} telescope, KAGRA,~\cite{KAGRA,2009CQGra..26t4020A}
is being constructed in Japan, and a third advanced LIGO detector is
currently under consideration for being built in India \cite{LIGO:India}.  The LIGO and
Virgo collaboration have released a plan for the commissioning
deployment of a network of second-generation \gw{} detectors, which
will start in 2015 with short (few months) science runs with two
detectors, and will grow to stable operation of a network of 4
detectors by 2022~\cite{ObservingScenarios}.

\begin{figure*}[!hbt]
\begin{center}
\begin{tabular}{cc}
\includegraphics[width=0.45\textwidth]{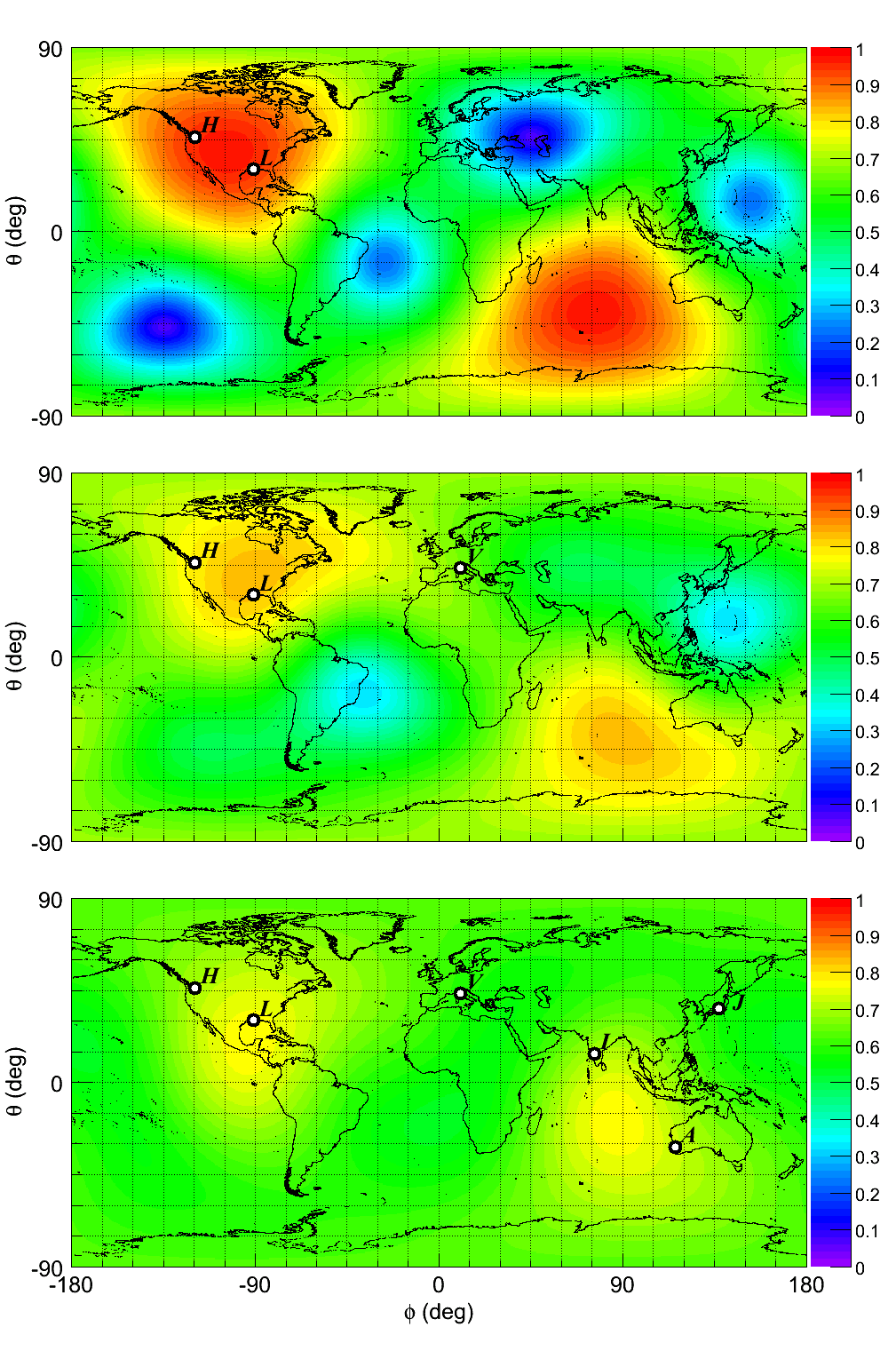} &
\includegraphics[width=0.45\textwidth]{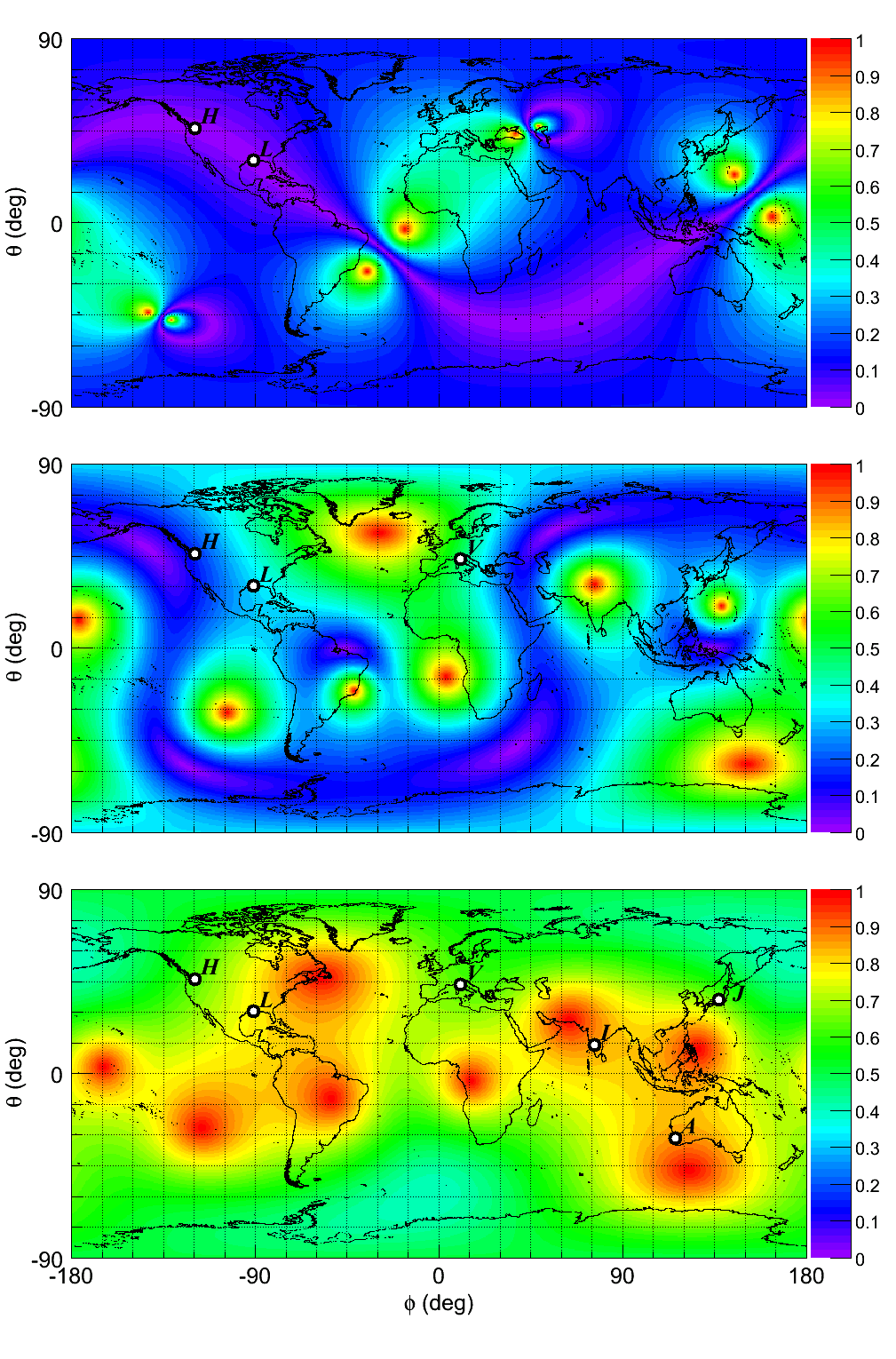}
\end{tabular}
\end{center}
\caption{\small{The distributions of the network antenna factor
(left plots) and the network alignment factor (right plots) as a
function of latitude ($\theta$) and longitude ($\phi$) for the
Hanford-Livingston network, Hanford-Livingston-Virgo network and a
six-detector network (from top to bottom).}}
\label{Fig:factorF}
\end{figure*}

The \gw{} detector network performance greatly depends on the number
of detectors in the network, their geographical location, and the
relative orientation of the detector arms.
For each direction in the sky, the performance of the detectors is characterized by their
antenna patterns, which can be combined into the network antenna
factor~\cite{PhysRevD.83.102001}.  For example,
Figure~\ref{Fig:factorF} (left plots) shows the distribution of the
network antenna factor as a function of the sky coordinates. Note that
the six-detector network provides more uniform coverage of the sky
than the Hanford-Livingston or Hanford-Livingston-Virgo networks.  The right-hand panel of Fig.~\ref{Fig:factorF} shows the alignment factor between detectors in the network; networks of few detectors with similar arm orientations may only be sensitive to one of the two \gw{} polarizations for some sources.

A worldwide GW detector network enables the reconstruction of source
parameters and the localization of \gw{} events on the
sky~\cite{Fa:09, PhysRevD.81.082001,
PhysRevD.83.102001,2011PhRvD..83j2001K,2011ApJ...739...99N,Veitch:2012}.
Accurate source localization is key to enabling multi-messenger
astronomy, e.g., via joint observations with electromagnetic
telescopes \cite{followupS6,SwiftS6,Bloom:2009,Mandel:2011}.
Meanwhile, measurements of other source parameters can enable studies
of astrophysics and tests of general
relativity \cite{BulikBelczynski:2003,MandelOShaughnessy:2010,Yunes:2011,Gerosa2013}.
Reconstruction of source location and polarization requires
geographically-separated detectors to make independent observations of
the same GW event; recovery of intrinsic parameters such as binary
component masses by modeled searches primarily depend only on the
total network SNR \cite{Veitch:2012}.

Proposed future \gw{} instruments will further enhance our astrophysical reach.  These include third-generation ground-based detectors~\cite{0264-9381-27-8-084007,2011GReGr..43..437C,2011GReGr..43..437C,2011CQGra..28i4013H,2010CQGra..27u5006S} that would be an order of magnitude more sensitive than second-generation instruments.  Meanwhile, pulsar timing arrays~\cite{1997asxo.proc..279K} would be sensitive to binaries composed of the most massive black holes in the Universe, while the Laser Interferometer Space Antenna or a similar space-based
detector~\cite{LISA,AmaroSeoane:2012km} could detect massive black hole mergers, extreme mass ratio inspirals, and tens of thousands of Galactic white dwarf binaries.

\subsection{Gamma-Ray and X-Ray Instruments}

The continued operation of high-energy, electromagnetic satellites in the
coming decades is essential for the realization of the full science potential
of  joint GW-GRB observations discussed in Section~\ref{sec:SourceScience}.
A GRB/X-ray trigger, possibly followed by optical observation
to identify the host galaxy and  source redshift, can be used as input for
a \gw{} transient search at a known time and location.
In turn, the worldwide network of \gw{} observatories
will be able to reconstruct in near real time the sky position of
\gw{} candidates and trigger multi-messenger observations for apparent afterglows.
The electromagnetic observations may identify the source's host galaxy,
redshift and, assuming standard cosmology, the luminosity distance, while the \gw{} measurement may offer
enhanced constraints on the source engine; a joint observation will ultimately
enable us to decipher science that would otherwise be inaccessible.

Currently, it is hard to know which GRB and X-ray sensitive instruments
and all-sky transient surveys will be operating in the next two decades.
Present GRB transient observation satellites such as AGILE~\cite{agile},
Fermi~\cite{fermi}, INTEGRAL~\cite{integral} and Swift~\cite{swift}
are not guaranteed to operate in the era of Advanced LIGO and Virgo.
It is possible that the lifetime of some
of these instruments will be extended, unless an operational constraint
or accidental equipment failure necessitates end-of-life procedures
for the satellites.

Ongoing X-ray missions such as Chandra~\cite{chandra} and
XMM-Newton~\cite{xmmnewton} are expected to continue to provide
information on GRB properties by observing their X-ray afterglows, at
least during the early years of the advanced \gw{} detector era, but
do require prior localization of the GRB. The Suzaku~\cite{suzaku}
orbiting X-ray observatory has a capability of detecting GRBs via its
wide-band all-sky monitor. The recently launched NuSTAR~\cite{nustar},
targeting the hard X-ray region, is likely to operate much beyond its
originally planned lifetime of two years.
MAXI~\cite{1997asxo.proc..279K} is an all-sky
X-Ray monitor, installed on the International Space
Station by the Japan Aerospace Exploration Agency~\cite{maxi}.

New projects include the Indian satellite
ASTROSAT~\cite{2006AdSpR..38.2989A,2009AcAau..65....6K,astrosat},
 a multi-wavelength mission that will  monitor  the
X-ray sky for new transients and is currently  scheduled for a 2013 launch,
with an expected lifetime of 5 years.
The next Japanese X-ray Astronomy mission,
ASTRO-H~\cite{Takahashi:2012jn}, is scheduled to be launched in
2014;  it can contribute to follow-up observations of GRB
afterglows at high resolution~\cite{Kawai2010}.
The launch of SVOM~\cite{2011CRPhy..12..298P,2010arXiv1005.5008S,svom}
-- a joint Chinese-French GRB monitor mission with an extended
spectral coverage from the visible to a few MeVs, and with good GRB
localization capability -- is planned towards the end of the decade.

In addition, there are numerous high-energy astrophysics mission
concepts targeting the X-ray and gamma-ray spectrum~\cite{heasarc}.
Some (e.g. AXTAR, EXIST, Xenia, GRIPS, A-STAR, JANUS) would allow the
detection of X-ray or gamma-ray transients with potential \gw{} burst
counterparts.  NASA's X-ray Mission Concept Study
Report~\cite{xraymissionsconceptstudy} states that it is feasible to
start a next X-ray mission toward the end of the decade. In the study
report, simplified missions that capture most of the recently
terminated IXO (International X-ray Observatory) mission science goal
elements were identified.
The Lobster Transient X-ray Detector is a mature concept that
was proposed to be deployed on the International Space Station in
three to four years. Its unique technology would allow to detect
transient X-ray emissions from a large portion of the sky at a wide
field of view and high sensitivity. If approved, the instrument
could work in conjunction with ground-based \gw{}
detectors, following up \gw{} candidate events~\cite{ISSlobster}.

\subsection{Electromagnetic Instruments from Radio to UV}

Low-latency electromagnetic follow-up observations of \gw{} event candidates will enable
the identification of possible optical and other electromagnetic
counterparts~\cite{2003ApJ...591.1152S,2001MNRAS.322..695H,2008CQGra..25r4033S,2009arXiv0902.1527B,2009astro2010S.235P,2009aaxo.conf..312K}.
The infrastructure for this type of analysis was tested during the most recent LIGO-Virgo data run, in 2010, when
observation requests were sent to wide-field
optical telescopes and other instruments, including QUEST, TAROT,
ZADKO, Pi in the sky, ROTSE, SkyMapper, LOFAR, the Palomar Transient
Factory~\cite{followupS6} and the Swift
satellite~\cite{SwiftS6}.

Radio telescope arrays, such as LOFAR~\cite{lofar}, EVLA~\cite{evla},
ASKAP~\cite{askap} and the future Square Kilometre Array~\cite{ska},
have in most cases a wide field of view and are able to provide
sub-arc-second angular resolution, that is superior to the
pointing of the advanced GW detector networks of the future.

In some theoretical models various mechanisms may give rise to a
prompt pulse (see e.g.~\cite{2001MNRAS.322..695H}), strong winds or
bursts \cite{2012PhRvD..86j4035L},
flares \cite{2011arXiv1109.5041P,2011arXiv1110.0467T}, intense
Poynting fluxes and emissions through shocks
\cite{2013arXiv1301.7074P,2013MNRAS.tmp.1023M}
or afterglow radio emission (see e.g.~\cite{2011Natur.478...82N}) from
some expected \gw{} sources, particularly coalescing compact binaries,
thus motivating coordinated observations. In addition, the use of GW
detectors as a trigger for follow-up radio searches could provide a
method of detecting faint radio transients that might otherwise be
missed.

Prospects of electromagnetic counterpart observations of \gw{} events
were recently discussed
in~\cite{MetzgerBerger2012,PiranNakarRosswog2013,2012arXiv1212.2289B,Samaya2012,Kelley:2012,2013arXiv1303.5788K}.
Gamma-ray observations will be critical for confirming a connection
between short GRBs and NS-NS/NS-BH mergers. Optical and radio
afterglows, even off-axis, as well as r-process nucleosynthesis
``kilonovae'' are detectable in principle, provided that an optimized
search strategy is used, taking into account the emission timescale
and instrument parameters.

\subsection{Low-Energy Neutrino Detectors}

Core collapse supernovae were discussed in Section~\ref{sec:core} as
sources of \gw{} radiation and engines for long gamma-ray bursts.
However most of their energy, about 99\%
of $\sim 10^{53}$ergs, is released in the form of  neutrinos, within
a few tens of seconds immediately following the  collapse.
The neutrinos in this burst are of all flavors, and their energy
ranges from a few to tens of MeV.
This prediction was confirmed by the detection of
a burst of 19 neutrinos from SN1987A in the Large Magellanic Cloud
 by two water Cherenkov detectors: IMB in the United States~\cite{Bionta:1987qt}
 and Kamiokande~II in Japan~\cite{Hirata:1987hu}.
Scintillation detectors also reported observations
\cite{Alekseev:1987ej, Aglietta:1987it}, and the main features of the
signal~\cite{bahcall1989neutrino,Pagliaroli:2008ur} confirmed the baseline
model of stellar collapse (see e.g. \cite{bethe:90,kotake:06a}).

Several existing neutrino detectors are sensitive to a neutrino burst
from a galactic supernova~\cite{Scholberg:2012id,fulgione10:status}.
Super-Kamiokande~\cite{Fukuda2003418}, a 50-kiloton water Cherenkov
detector in Japan, would observe $\sim$8000 events from a supernova
$\sim$8.5~kpc away~\cite{ikeda:07}.  The
LVD~\cite{LVD2006MSAIS...9..388F,Agafonova:2007hn} and
Borexino~\cite{borexino,Cadonati:2000kq} scintillation detectors at
Gran Sasso in Italy, KamLAND~\cite{kamlandweb,Piepke:2001tg} in Japan,
and the upcoming SNO+~\cite{snoplusweb} in Canada would also observe
hundreds of neutrino events interacting in 300-1000 tons of liquid
scintillator.  The IceCube detector \cite{IceCubeAhrens2004507} is a
cubic-kilometer detector located at the geographic South Pole.
IceCube is nominally a multi-GeV neutrino detector, but it is also
sensitive to MeV-neutrinos from a Galactic supernova and could observe
an increase in the count rate due to a diffuse burst of Cherenkov
photons in the ice~\cite{Halzen:1995ex}.  Super-Kamiokande, LVD,
IceCube and Borexino operate as part of the SNEWS (SuperNova Early
Warning System) network
\cite{Antonioli:2004zb,Scholberg:2008fa}, for a
prompt alert to astronomers in the case of a supernova neutrino
burst.

The distance reach of the global network of neutrino detectors
covers the Milky Way and a
significant fraction of its Satellite System, up to $\sim$O(100~kpc).
This can be considered as a good match to the \gw{} network
reach for core-collapse supernovae~\cite{Sutton:10,Ott2008,2010JPhCS.203a2077F,2009CQGra..26t4015O,2004APh....21..201A}
of GW detectors.  A 50-kiloton detector like Super-Kamiokande has a very low
chance of detecting a SN in M31 (at $\sim$770~kpc) on its own.
Low-energy neutrino detectors have angular resolution comparable to
advanced \gw{} detector networks~\cite{Fa:09}, and this makes
synergetic observations particularly
desirable~\cite{Leonor:2010,2009PhRvL.103c1102P}.

Long-term plans call
for neutrino detectors with up to $\sim$5 megatons of fiducial
mass~\cite{2008arXiv0804.1500S,autiero-2007-0711}
to enable observation of neutrinos from M31 and
M33~\cite{Kistler:2008us,PhysRevLett.95.171101}.
Proposals include
MEMPHYS~\cite{debellefon-2006}, LENA~\cite{LENA}, and
GLACIER~\cite{1742-6596-171-1-012020} in Europe,
DUSEL LBNE~\cite{2010JPhCS.203a2109M} in the US,
 Hyper-Kamiokande~\cite{nakamura03} and
Deep-TITAND~\cite{Suzuki:2001rb,Kistler:2008us} in Japan.
 It is not
unreasonable to expect that the lifetime of these detectors will
coincide with the operation of third-generation \gw{} detectors.

\subsection{High-Energy Neutrino Telescopes}

High-energy neutrinos (HENs) in the GeV--PeV range could
also  unveil new physics in joint observations with
electromagnetic and \gw{} signatures \cite{2013arXiv1304.2553B,2012PhRvD..86h3007B}.
The detection of HENs is pursued in large Cherenkov detectors
that exploit their charged-current interaction in large volumes of water or ice.
Most of the neutrino energy is
transferred to a single high-energy electron, muon, or tau particle,
which will emit Cherenkov radiation as it travels through the detector
medium.
High-energy muons are most useful for neutrino astronomy, since
they don't lose energy as rapidly as
electrons and have longer lifetime than  taus, so their path can be  several
kilometers long.
The Cherenkov light emitted along this path can be detected
and used to measure the direction and energy of the muon, and thus of
the primary neutrino.

There are three HEN observatories currently in operation.  The IceCube
observatory \cite{IceCubeAhrens2004507} has recently been extended
with an additional component called
DeepCore~\cite{2012APh....35..615A}, designed to be sensitive to
neutrino energies below IceCube's lower limit of $\sim100$\,GeV, down
to 10\,GeV, effectively increasing the detector's astrophysics
reach \cite{2013arXiv1301.4232B}.
\textsc{Antares}~\cite{ANTARES}, located in the Mediterranean sea,
is scheduled for an upgrade to a cubic-kilometer detector called KM3NeT in the coming
years~\cite{deJong2010445}.
A third HEN detector operating at lake
Baikal is also planned to be upgraded to a km$^3$ volume~\cite{Avrorin2011S13}.

The distance reach of high-energy neutrino detectors is virtually infinite,
although at larger distances the probability of detecting a neutrino from an
individual source diminishes. The Waxman-Bahcall model~\cite{waxmanbachall},
the benchmark model of HEN emission from GRBs, predicts about
$n_{\textsc{hen}} \approx 100$ neutrinos detected in a km$^3$ detector for a
typical GRB at 10 Mpc \cite{NeutrinoBATSEGuetta2004429}, although recent
upper limits from the IceCube detector disfavor
GRB fireball models with strong HEN emission associated with cosmic ray acceleration~\citep{2004APh....20..507A}.
However, milder HEN fluxes or alternative acceleration scenarios are not ruled
out~\cite{2012Natur.484..351A}. Moreover, the constraints weaken substantially
when uncertainties in GRB astrophysics and inaccuracies in older calculations
are taken into account, and the standard fireball picture remains
viable~\cite{PhysRevLett.108.231101,2012ApJ...752...29H}.

Models of high-energy neutrino emission from mildly relativistic jets of
core-collapse supernovae, and potentially from choked GRBs, predict
HEN emission of $n_{\textsc{hen}} \approx 10$~\cite{HENAndoPhysRevLett.95.061103}
(note that the result presented in~\cite{HENAndoPhysRevLett.95.061103} is three
times higher, as it does not take into account neutrino flavor mixing).
Horiuchi and Ando~\cite{chokedfromreverseshockPhysRevD.77.063007}
estimate $n_{\textsc{hen}}$ from reverse shocks in mildly relativistic
jets to be $n_{\textsc{hen}} \approx 0.7-7$ for a km$^3$ neutrino detector
(after taking into account neutrino flavor mixing).
Razzaque \emph{et al.}~\cite{HeneventratePhysRevLett.93.181101}
obtain $n_{\textsc{hen}}\approx 0.15$ for supernovae with mildly relativistic
jets with jet energy of $E\sim10^{51.5}$ erg.

The most stringent observational constraints on transient GW+HEN sources so far 
has been obtained using searches with the latest Initial LIGO-Virgo (S6-VSR2/VSR3) 
detectors and the IceCube detector in its 40-string configuration~\cite{PhysRevLett.107.251101}. 
The derived constraints were also used to estimate the science reach of the Advanced 
LIGO-Virgo detectors in combination with the completed IceCube detector, with promising 
results. The first joint search of ANTARES, LIGO, and Virgo data for coincident \gw{} 
and HEN using LIGO, Virgo and ANTARES data derived limits on the rate density of joint 
GW-HEN emitting systems in the local universe, comparing them with densities of merger 
and core-collapse events~\cite{LVCantares} that are compatible with previous results. 
While the available upper limits for initial detectors impose no constraints on joint 
emission models, the results~\cite{PhysRevLett.107.251101} show that advanced detectors 
will be able to constrain some emission and population models, therefore also having the 
potential of detecting joint sources.

\section{Challenges and Open Questions}
\label{sec:challenges}

The initial generation of \gw{} detectors has targeted most of the transient
sources described in Section~\ref{sec:SourceScience}, yielding
observational limits which already bear astrophysical interest. For
instance, for two nearby GRBs, the non-detection of a \gw{} signal
made it possible to exclude a merger as the GRB central
engine~\cite{S5GRB070201,GRB051103}, while constraints on \gw{}
emissions were produced for magnetars~\cite{6magnetars} and in a study
of the 2006 Vela pulsar glitch~\cite{VelaGlitch}.  Population
constraints have been produced for \gw{s} in coincidence with
gamma-ray bursts~\cite{S6GRB}, and all-sky limits have been set on
rates of binary mergers~\cite{S6cbcLowMass,S6cbcHighMass,S6IMBH} and
generic bursts~\cite{S6burstAllSky}.  A first coincidence search with
high-energy neutrinos has been performed~\cite{LVCantares}, and in the
most recent data run (2009-2010) transient candidates have been
broadcast for EM follow-up~\cite{followupS6,SwiftS6}. 
The potential for extracting fundamental physics and astrophysics from \gw{} data
should be dramatically enhanced by the next generation of \gw{}
detectors, with a projected $\sim$1000 times larger sensitive volume and
the newly accessible $\sim 10-40$~Hz frequency band. Predictions for compact
binary detection rates with next generation detectors, based on astrophysical observations, population synthesis and source models, are available in \cite{Abadie:2010cf}.

The engagement between the experimental and theoretical communities in
the coming years will shape the future of \gw{} astrophysics: to
maximize their scientific output, \gw{} transient searches will need
to include more information from theoretical and computational
astrophysics.  In turn, robust source modeling will be required to provide
a theoretical understanding of the mapping between
signal characteristics and physics parameters, including the knowledge
of potential degeneracies that may be broken by complementary
information from multi-messenger observations.
In this section we elaborate on what we identify as the main open
challenges in the theoretical understanding of \gw{} transient sources
and in the ability to identify their signature.

\subsection{Observations of Gravitational-Wave Transients}

Despite steady progress in relativistic astrophysics, there is still
significant uncertainty in predicted waveforms for most \gw{}
transient sources. 
In some special cases, when an
accurate signal model is available, the search relies on matched
filtering with a bank of
templates~\cite{S6cbcLowMass,S6cbcHighMass,S4ringdown,S4CosmicString}.
In the more general case, \gw{} bursts can be identified in the
detector output data as unmodeled excess power localized in the time-frequency
domain~\cite{S6burstAllSky}.

The principal challenge for the instruments is to achieve a 10-fold
improvement in sensitivity compared to the first generation of \gw{}
interferometers, including a reliable calibration and a low rate of noise
transients \cite[e.g.,][]{Calibration:S5,DetChar:S6}. As soon as the second generation
of \gw{} detectors reaches design sensitivity, the main
experimental challenge to \gw{} burst science will be to 
discriminate between real \gw{} signals and noise transients that
happen to coincide in multiple detectors.

A fourth site, in addition to the existing facilities in the USA (LIGO
Hanford and LIGO Livingston) and Italy (Virgo), is an additional
instrumental challenge that will enable source localization and
increase detection confidence.  As discussed
in Section~\ref{sec:instrumentation}, the identification of a consistent
signal in a network of instruments is a key ingredient in \gw{}
searches, dramatically increasing the confidence in a candidate event.
This is especially important for a \gw{} burst, which otherwise may not
be distinguishable from noise fluctuations of instrumental or
environmental origin.
A statistic built from the coherent sum over the detector responses
is used to rank candidate events and discriminate between signal and
noise, yielding better sensitivity (at the same false alarm rate) than
individual detector statistics.  Accurate timing information from
multiple detectors also makes it possible to reconstruct the two \gw{}
polarizations and localize the source on the sky.  The LIGO and
Virgo collaborations outlined a timeline for the localization of \gw{}
transients by advanced detectors \cite{ObservingScenarios}, in order to facilitate
the formulation of joint detection strategies \cite{AMON} with electromagnetic,
neutrino, or other observing facilities.

The level of background, and, hence, the significance of a candidate event,
is estimated empirically with
time slides.  Tests with simulated signals injected into LIGO and Virgo
noise show that when a signal model can be assumed, this technique
makes it possible to achieve high detection confidence.  For instance, in a
detection exercise during the most recent LIGO-Virgo run, the
simulated coalescence of two compact objects with network
signal-to-noise ratio of 12.5 was identified with a false alarm rate
of 1 in 7000 years using matched filtering~\cite{S6cbcLowMass}. A
completely un-modeled search instead yielded ~1/1.1 years,
after accounting for trial factors~\cite{S6burstAllSky}.

An important challenge for \gw{} burst searches in the advanced
detector era will be to incorporate partial information from
theoretical models in order to constrain the false alarm rate and increase
detection confidence in analyses tuned to specific sources, with perfectly modeled and completely general searches as extreme cases.
Understanding what information can be incorporated in searches, and
learning how to make inferences on source parameters from
transients, starting from the frequency, bandwidth and duration of a
coherent event, will require a close synergy with the modeling
community.  One particularly exciting possibility is the use of \gw{} signals, possibly in coincidence with electromagnetic observations, as probes of strong-field dynamics and tests of general relativity itself \cite[e.g.,][]{YunesSiemens:2013}.

\vspace{0.2cm}
\noindent
We highlighted three key areas of cooperation for the relativistic astrophysics
community:

\vspace{0.1cm}
\noindent
{\it 1. What can we understand about the source from the data?}
Theorists and modelers can guide targeted analyses for a specific
source, interpret the astrophysics of general searches, and (together
with instrumentalists) determine what science can be extracted from
the detector response.
In return, the data analysis community should instruct
astrophysicists/modelers on how to combine detector capabilities and
theoretical waveforms into a prediction about the physics that can be
extracted from a detection, and a realistic assessment of the extent
to which these predictions are reliable.

\vspace{0.1cm}
\noindent
{\it 2.  Can we improve the pipeline that moves theoretical advances
in source modeling into the GW data analysis search methodologies?}
Both theoretical and computational understanding of \gw{} sources and
\gw{} search methodologies are rapidly advancing.  
There is, however, a time lag in
implementing some of these advances into actual LIGO searches.  One of the
challenges in anticipation of the advent of \gw{} astronomy is the development of better strategies to rapidly get the best models to data analysts.

\vspace{0.2cm}

\noindent
{\it 3. How to bring together \gw{}, EM and astroparticle physics?}
Searches for \gw{} bursts in the advanced detector era
will be a combination of untriggered, all-sky searches and externally
triggered, localized searches.
All-sky searches have the greatest potential for finding
electromagnetically dark sources, and may discover unexpected
signatures. They also provide triggers for follow-up studies of
candidate events with EM observations.
Externally triggered searches will have EM and/or
neutrino counterpart observations. 
Strategies are needed to combine EM, neutrino, and \gw{}
information and foster collaborations between the \gw{} community and
``traditional'' astronomers, from data sharing to physics extraction.

\subsection{Double Neutron Star and Neutron Star -- Black Hole Binaries and Short Gamma-ray Bursts}

While the most widely accepted model for short \grb{} progenitors is
the merger of a \ns{} with another \ns{} or
a \bh{}~\cite{Berger2009}, Swift and Fermi have observed only a few
short \grb{s} per year, and therefore the prospect of observing a
short \grb{} in association with a \gw{} counterpart appears
challenging. Furthermore, it is not easy to identify the host galaxy of
a short \grb because the sky localization is poor unless an afterglow is detected, and even then,
natal kicks could displace the binary system from the host galaxy~\cite{2006ApJ...648.1110B,Kelley:2010}. The
nearest short \grb{s} may be the hardest to associate with their host,
since angular offsets could be large. This uncertainty in host galaxy
identification leads to an uncertainty in the progenitor distance,
which, in turn, leads to an uncertainty in \gw{} signal association.
However, given the relatively small reach of advanced \gw{} detectors,
it is possible for \gw{} observations to rule out potential host
galaxies that are too distant.

There are many outstanding questions about the nature of
observed \grb{s}.  It is unclear how to best distinguish short \grb{s}
associated with merging compact binary progenitors from soft gamma-ray
repeater hyperflares.  The energy budget for short \grb{s} is also
unclear, although they have been identified (via their host galaxies)
up to high redshifts. In particular, there are several uncertainties
about the beaming of short \grb{s}, with very few measurements up to
now~\cite{Fong:2012}. Beaming is also a crucial ingredient to
understand how many \gw{} events could be associated to the observed
rates of short \grb{s} (for a recent review of the properties of
short \grb{s}, and the implications for their progenitors,
see~\cite{Berger11}).

There are several open problems in the astrophysics of the merging
systems that are very important in the modeling of compact binary
mergers. If magnetic fields are indeed responsible for the energy
budget, where do these large fields come from in a merger scenario?
The magnetic fields of the progenitor \ns{s} are not large enough and
the merger event can amplify them. Is this amplification sufficiently
strong? Does the spin-down of the final \bh{} cause any late emission
signature? Shall we be able via more complete modeling to use \gw{} or
EM observations to distinguish between a \ns{} or a \bh{} as the
companion compact object in the merger?

Observing \gw{s} from compact binaries also challenges our
understanding of binary evolution. If we can better understand merger
events, we may be able to use \gw{} and EM observations to understand
physical mechanisms that are responsible for compact binary formation
and the corresponding distribution of observed events. Another
challenge is posed by the poorly known nuclear physics at the
densities typical of \ns{} interiors. Do we understand the physics of
nuclear processes at \ns{} densities well enough to accurately
predict \gw{s} from these systems?  Even with an improved
understanding of the properties of matter within \ns{s}, adding an
accurate treatment of magnetic fields, together with realistic \ns{}
EOS and neutrino radiation to get the right post-merger dynamics will
be a significant technical and computational challenge to the
numerical relativity community. It will be especially important to be
able to efficiently generate a large number of data sets to be
compared with \gw{} observations. While the \gw{} inspiral signal
could probably be modeled analytically (e.g. using variants of the
effective one-body model~\cite{Baiotti2010,Bernuzzi2012}) facilitating
the production of a large number of templates, the merger and
post-merger dynamics will require a large and accurate set of
numerical simulations.

\subsection{Binary Black-Hole Mergers as Transient Sources}
Numerical relativity provides direct solutions of the Einstein equations for
binary \bh{} spacetimes.  Numerical relativity plays a major role in
two large collaborations using binary \bh{} waveforms.  The NRAR (Numerical
Relativity/Analytical Relativity) collaboration is working to provide
templates that cover the inspiral, merger and ringdown phases of
binary \bh{} coalescence.  The NINJA (Numerical INJection Analysis)
project~\cite{Aylott:2009ya,2009CQGra..26k4008C} is using hybrid
numerical relativity waveforms to test data analysis algorithms for
the detection and characterization of \gw{s}. The hybrid waveforms are
made by stitching post-Newtonian and numerical relativity waveforms
together.  While the waveforms and templates created in these projects
are useful for burst analysis, and in the case of NINJA unmodeled burst search
algorithms are being tested, a burst-only focused study offers an
opportunity and challenge to theory and data analysis.

For stellar-mass binary \bh{} systems the challenge is
to provide template coverage for modeled searches over a large parameter space using a
combination of post-Newtonian techniques and numerical relativity.
This includes the challenge of stitching the post-Newtonian and
numerical waveforms together. The choice of where to make the
stitching (i.e., how many cycles are supplied by post-Newtonian theory
and how many cycles myst be simulated numerically) depends on the
physical parameters of the binary. The NRAR and NINJA collaborations
are working systematically to address these problems.

More massive systems 
offer a
different challenge to the numerical relativity and data analysis
communities.  Unmodeled or weakly modeled burst searches are most appropriate when we know little
about the signal model.
Rather than the exquisite accuracy in phase necessary for template
building, a burst-only focus in numerical relativity offers the
opportunity to survey the physical parameter space of binary \bh{s} in
a timely manner by reducing the resolution required compared with
inspiral-focused simulations.  The challenge, however, is learning to
identify key features of the merger that will be useful to the
data-analysis algorithms for transient sources, potentially improving
glitch identification and reducing false-alarm rates. Only one of
these studies, aimed at investigating the sensitivity of the Omega
burst algorithm~\cite{omega} to changes in the \bh{} spin
vector~\cite{2011PhRvD..83d4019F}, has been completed, and more
studies of this kind are needed.

Prior to the first science runs of the advanced detectors, we need
information about the merging systems, including the energy, angular
and linear momentum radiated (in total and in each harmonic mode), the
polarization (such as ratios of circular to linear) and mode dependent
merger time-scales.  This provides great opportunities: can we
increase the likelihood of detecting \gw{s} from binary \bh{} mergers
by employing weakly modeled burst searches?  Can information gleaned from NR
waveforms of merging BHs reduce the false-alarm rate for burst
searches?  Can a burst algorithm find sources of merging \bh{s} for
physical parameters that have not yet been modeled, or have not been
modeled well enough?  In order to answer these questions, we need to
determine the smallest number of cycles we can simulate in numerical
relativity while still representing actual signals reasonably well.
We need to figure out how the inevitable lack of phase accuracy or
length of numerical simulations propagates in the search for a
transient source.
This leads us directly to perhaps the most interesting and important
challenge for a transient search of the binary \bh{} merger sky: how
well can such a search determine the physical parameters of the
merging sources?  Perhaps a burst-first approach (targeted at first
finding a burst signal, and then digging through the data with 
matched-filtering parameter estimation techniques) can help in this
regard.

\subsection{Core-Collapse Supernovae and Long Gamma-ray Bursts}

Uncertainties abound in current \gw{} estimates from core-collapse
supernovae.  These uncertainties include poorly known initial
conditions, the relevance of 3D effects in the explosion mechanism, or
the impact of detailed 3D neutrino transport on the \gw{} signal from
anisotropic neutrino emission.  Here we focus on two major issues in
determining accurate \gw{} signals: rotation of the progenitor and
convective instabilities.

\subsubsection{The Role of Rotation.}

Rapidly rotating systems have the potential to produce strong \gw{}
signals, especially if nonaxisymmetric instabilities develop.  Whether
or not such signals occur in nature depends upon the rotation profile
of the precollapse star.  Classic models of single stars produce a
range of core rotation speeds, depending upon the initial angular
momentum of the star and the magnetic field generation
~\cite{heger00,heger05,hirschi04,hirschi05,Yoon05b,woosley:06}.  In
the absence of magnetic fields, the core can decouple from the rest of
the star, retaining a high spin rate even as the stellar envelope
expands and decreases its spin rate.  But if magnetic fields are
produced at boundary layers, they tie the core to the stellar
envelope, causing its spin to decelerate with the envelope.

The high-spin requirements of most long GRB progenitors have pushed
stellar modelers to seek new ways to produce fast-spinning cores.
Fast-rotating single star models often invoke extended mixing and
generally require low metallicity, below $\sim 0.1
Z_\odot$~\cite{Yoon05b,Yoon06,woosley:06}.  Others have devised a
number of binary progenitors of long GRBs that may also produce fast
rotating cores invoking tidal interactions, mass transfer, and common
envelope
mergers~\cite{fryer99,podsiadlowski04,fryer05,vdH07,Cantiello07}.  The
fastest-rotating cores collapsing to form \ns{s} could be produced in
the merger of two white
dwarfs~\cite{Yoon04,Yoon05,Yoon07,fryer09,dessart:06aic,abdikamalov:10},
but we currently do not know whether such systems produce
thermonuclear or core-collapse supernovae.

Uncertainties in the evolution of massive stars (especially rotating
stars) and in binary evolution make it very difficult to predict the
distribution of spin rates of collapsing cores.  If fast-spinning
pulsars also develop strong magnetic fields, selection biases make it
difficult to use the observed pulsar distribution to constrain the
birth spin rate of \ns{s}.  Similarly, high spin rates are just
one requirement for long GRB engines, and extrapolations to the spin
rates of newly formed \ns{s} from long GRB rates can only be
done with specific model assumptions.

\subsubsection{Convective Instabilities.}

The detailed nature of the convective instabilities (both within and
above the PNS) also remains an open question in estimates of \gw{}
emission.  If low-mode large-scale convection develops and is
modulated by the SASI, as is borne out by current simulations, we have
a fairly strong picture of the \gw{} signal. But the nature of this
convection (and of the resulting \gw{} signal) is different between 2D
and 3D models.  The current resolution in core-collapse supernova
simulations, in particular in the first 3D simulations, remains
roughly an order of magnitude lower (per dimension) than what
computational fluid dynamicists argue is required to accurately model
convective modes and turbulence~\cite{dimonte04,porter06}.  In
addition, with few total turnover timescales in the model, initial
perturbations (caused by convection in the Silicon burning layer) may
have a strong impact on the convective motions.  To more reliably
predict the \gw{} signal from convection, it will be necessary to gain
a better understanding of convection, the SASI, and of the impact of
perturbations from late convective burning stages.

\subsection{Isolated Neutron Stars}

It should be clear from our discussion that transients in isolated \ns{s} provide a range of
interesting prospects for \gw{} astronomy. At the same time, it must
be appreciated that the expected signals are likely to be
weak. Because of the complexity of the physics involved in most
realistic scenarios, these signals are also extremely difficult to
model in detail. This means that it may be unrealistic to expect
modeling to produce accurate search templates for matched
filtering. Dealing with this difficulty is one of the main challenges
for the next few years. Realistically, the sources that we have
discussed here may only be borderline detectable with
second-generation detectors. They may require future developments,
like the Einstein Telescope, with enhanced sensitivity in the
high-frequency regime. However, such detectors could provide us with
an unprecedented view into \ns{} interiors
\cite{1996PhRvL..77.4134A,1998MNRAS.299.1059A,2001PhRvL..87x1101A,2011PhRvD..83f4031G},
especially if the signals are strong enough that the asteroseismology
strategy is viable. The hope is that \gw{} observations will
complement EM data, helping us to place constraints on
physics under extreme conditions.

On the theory side, it is easy to identify directions in which the
current models need to be improved. Basically, we need to continue
developing both the computational technology and our
understanding/implementation of the relevant microphysics. In order to
model the relevant astrophysical scenarios, we obviously need to build
more realistic models. This is associated with a number of challenging
issues.  For the purposes of this summary, let us consider two
specific problems. 

First of all, it is clear that magnetars may be interesting target
sources given their, sometimes rather violent, activity. However, even
though there has been clear progress in recent years, we are still
quite far from modeling ``realistic'' magnetized stars. The difficulties
range from fundamental to technical. We do not yet have reliable
models of compact magnetic equilibria with the expected composition,
and we must account for superconductivity (and the associated magnetic
flux tubes) in the star's core.  

Secondly, recent work on instabilities, such as those of the r-mode,
has shown that we do not, in fact, understand the relevant
dissipation mechanisms well enough to explain the observed population
of fast spinning, accreting \ns{s}. Given the amount of effort that
has gone into modeling the r-mode instability, this is somewhat
embarrassing. Another major problem concerns the nonlinear evolution
of this kind of instability. The associated timescale is too long for
the problem to be within reach of nonlinear simulations. At the same
time, we need to account for nonlinear aspects (like the mode-coupling
that is thought to saturate the r-mode). This is truly problematic,
especially if we want to make the stellar models more realistic
by including the appropriate microphysics. Similar issues relate to
pulsar glitches, with a relaxation time of weeks to months but an
initial rise time of tens of seconds, and \ns{} mountains, which may
evolve due to plastic flow.

These problems are clearly far from trivial, but they provide a good
illustration of the kind of issues that we need to deal with if we
want to extract as much physics as possible from future \gw{} signals
from \ns{s}.

\subsection{Multi-Messenger Astrophysics of Transient Sources}
The success of the \gw{} burst endeavor relies on a continued dialogue
between source theory and detector expertise, or between predicted
sources and what can be achieved in a practical instrument. On one
side this drives detector design, on the other it determines which
parts of parameter space theorists should focus on.

While \gw{} bursts generally do not have well-defined waveforms, they
have robust signal features when characterized by their energy.  The
duration, central frequency or even time-frequency characteristics of
the signal energy can be used to tune source-specific searches.
A \gw{} burst search can be informed by source population or emission
models to identify relevant regions of the parameter space to focus
over.  Moreover, any waveform estimates from potential sources can be
use to characterize the efficiency of the developed searches.  When a
population of sources or repeating \gw{} emitters are considered, it
is possible to design an analysis to target the energy features to
build up SNR
and reject noise events.

While all-sky, untriggered \gw{} searches are the principal mode of operation for \gw{} detectors, triggered searches based on observations in other bands or theoretical guidance on the most important potential sources can yield to improved sensitivity.  We will want to revisit the issue of optimal sources for triggered \gw{} searches as new astrophysical discoveries are made and \gw{} detector performance and data analysis capabilities improve \cite[e.g.,][]{Kelley:2012}.  Meanwhile, once potential \gw{} sources have been identified, to facilitate efficient observations, it is highly desirable to exchange triggers between the \gw{} and astronomy communities within 1 hour of the trigger, or even less.

Research on \grb{s} is now a mature subject. There are redshifts for
hundreds of \grb{s} (all extragalactic).  Despite this, many
uncertainties remain.  What is the nature of their progenitors? At
least some long \grb{s} are confidently associated with CCSNe, but
what about the other observations?  \grb{} emission processes (both
non-thermal and thermal) are quite uncertain, and the total energy
budget is unclear.

A lot of infrastructure has been put in place and communication
channels between the \gw{} and astronomy communities opened in the
preparation for the Advanced LIGO era.  It is worth asking, in the
multi-messenger era, how should this momentum be maintained?  A
post-Swift mission with a comparable source localization error box
will complement \gw{} observations.  Facilities with wide-field search
capability (snapshot vs. tiling) and a field of view covering Advanced
LIGO error regions may allow a rapid discovery of the EM transient
counterpart, but it is equally important to have high-resolution
instruments that can follow up the transient discovery with other
observations, such as obtaining a source spectrum.  There is also
tension on the \grb{} requirements between the \gw{} community and the
broader \grb{} community: missions that suit multi-messenger astronomy
would focus on rapid observations of closer \grb{s}, which may not be
what other \grb{} astronomers (who may want to explore the
oldest \grb{s} and cosmology) are after.

\section{Conclusions and Peroration}
\label{sec:concl}

The era of \gw{} astronomy is upon us.  Advanced ground-based
detectors are expected to reach their design sensitivity near 2019.
The space-based \gw{} detector eLISA may be selected in the EU
as one of the next L2/L3 missions, with a launch possibly scheduled
before the end of the next decade.  IPTA may also detect \gw{s} in the
next few years~\cite{mcwilliams2012,sesana2012}.  This paper has
focused on transient \gw{} sources for second-generation
Earth-based detectors, particularly those amenable 
to unmodeled or weakly-modeled burst searches.

We have reviewed the state of the art of theoretical and observational
studies of several \gw{} transient sources. Compact binaries composed
of \ns{s} and/or \bh{s} are among the most promising sources, because
they are detectable by advanced ground-based interferometers at
distances of hundred of Mpc. In recent years, numerical simulations of
these objects have increased our knowledge of their \gw{} signals and,
more recently, of their possible connection to EM observations.  In
particular, \ns{}-\ns{} and \ns{}-\bh{} binaries are thought to be
linked to SGRB engines. A coincident detection of \gw{s} and SGRBs
would be the smoking gun that SGRBs are indeed caused by compact
object mergers. Another \gw{} source that has a very important
connection with EM observations are CCSNe, which are currently the
main candidate LGRB engines. Unfortunately, \gw{s} generated by such
systems will be detectable only for events localated in our Galaxy,
hence the expected event rates are quite low.

Due to the possibility that the first \gw{} detection may happen in
the next few years, it is important to increase the current efforts in
modeling \gw{} sources.  Significant progress has been made in
modeling binary \bh{s}, but covering the template space of generic
precessing binaries is still a big challenge. New theoretical and
computation insights are needed to tackle \bh{s} spinning close to the
Kerr limit, even for nonprecessing binaries with aligned spins.  The
study of non-vacuum sources, due to the higher level of complexity,
still requires significant effort in order to better understand the
effects of several different physical mechanisms on their evolution
(e.g. EOS, neutrino emission, magnetic fields, convection,
turbulence). While part of this work will require the development of
better and more efficient numerical codes, several studies will also
require the use of larger computational resources. For example, the
study of convective instabilities in CCSNe and of magnetic
instabilities in \ns{} mergers will require much higher resolutions,
and hence more powerful computational resources than those used so far.
Moreover, since the properties of matter in both \ns{} mergers and
CCSNe are not yet well known, it will be imperative to build a large
and publicly available database containing \gw{} signals generated by
different models. This will be crucial in order to be able to extract
physical information from the \gw{s} that will be detected from these
systems.

Detecting EM counterparts to \gw{} events will lend greater confidence
to \gw{{} searches and provide useful complementary information.  This
will considerably help in source localization and, in the case for
example of SGRBs and LGRBs, allow us to better understand the engine
of some of the most fascinating astrophysical phenomena. It will be
particularly important that both X-ray and $\gamma$-ray satellites
will be available and ready when the second generation of \gw{}
detectors will start to take data. For these studies to be successful,
it will be necessary to increase the current level of collaboration
between the \gw{} and EM communities, and ideally better coordination
between funding bodies.

After almost 100 years since the prediction of their existence, the
direct detection of \gw{s} is on the horizon.  In the next
decades \gw{} astronomy may become one of the most important sources
of information on high-energy astrophysics, providing us with an
unprecedented level of information on \ns{s} and \bh{s}.

\begin{acknowledgments}

The ``Gravitational Wave Bursts'' workshops in Chichen-Itza, Mexico
(December 9-11, 2009) and Tobermory, Scotland (May 29-31, 2012) were
supported by National Science Foundation Grants Number 0946361 and
1231548.
The authors gratefully acknowledge the support of the United States
National Science Foundation for the construction and operation of the
LIGO Laboratory and the Science and Technology Facilities Council of the
United Kingdom, the Max-Planck-Society, and the State of
Niedersachsen/Germany for support of the construction and operation of
the GEO600 detector. The authors also gratefully acknowledge the support
of the research by these agencies and by the Australian Research Council,
the International Science Linkages program of the Commonwealth of Australia,
the Council of Scientific and Industrial Research of India, the Istituto
Nazionale di Fisica Nucleare of Italy, the Spanish Ministerio de
Educaci\'on y Ciencia, the Conselleria d'Economia, Hisenda i Innovaci\'o of
the Govern de les Illes Balears, the Royal Society, the Scottish Funding
Council, the Scottish Universities Physics Alliance, The National Aeronautics
and Space Administration, the Carnegie Trust, the Leverhulme Trust, the David
and Lucile Packard Foundation, the Research Corporation, and the Alfred
P. Sloan Foundation.
We thank the Kavli Institute for Theoretical Physics at UC-Santa
Barbara, supported in part by NSF grant PHY11-25915, for hosting the
workshop ``Chirps, Mergers and Explosions: The Final Moments of
Coalescing Compact Binaries.''
The Columbia Experimental Gravity group is grateful for the generous
support from Columbia University in the City of New York and from the
National Science Foundation under cooperative agreement PHY-0847182.
E.B. is supported by National Science Foundation through CAREER Award
Number PHY-1055103.
K.B. acknowledges NASA Grant Number NNX09AV06A and NSF Grant Number
HRD 1242090 awarded to the Center for Gravitational Wave Astronomy,
UTB.
S.B. and K.K. acknowledge support from the German Science Foundation
SFB/TR7 ``Gravitational Wave Astronomy.''
L.C. acknowledges National Science Foundation Grants Number PHY-0653550
and PHY-0955773.
LL was supported in part by an NSERC through Discovery Grant.
Research at Perimeter Institute is supported through Industry Canada
and by the Province of Ontario through the Ministry of Research \&
Innovation.
P.C.D. acknowledges Spanish Ministerio de Educaci\'on y Ciencia grant
number AYA 2010-21097-C03-01, Generalitat Valenciana grant number
PROMETEO-2009-103 and ERC Starting Grant number CAMAP-259276.
The work of C.F. is under the auspices of the US Department of Energy,
and supported by its contract W-7405-ENG-36 to Los Alamos National
Laboratory.
L.S.F. acknowledges National Science Foundation Grants Number
PHY-0653462, CBET-0940924 and PHY-0969857.
B.G. and R.P. acknowledge support from NSF Grant Number AST-1009396 and
NASA Grant Number NNX12AO67G.
P.L. acknowledges National Science Foundation Grants Number 0903973
and 1205864.
D.M.S. acknowledges National Science Foundation Grants Number PHY-0925345
and PHY-0955825.
We wish to thank Christian Ott for useful contributions to this
review.

\end{acknowledgments}

\bibliographystyle{unsrt}


\begin{thebibliography}{100}

\bibitem{Weisberg:2004hi}
J.~M. Weisberg and J.~H. Taylor.
\newblock {Relativistic Binary Pulsar B1913+16: Thirty Years of Observations
  and Analysis}.
\newblock {\em ASP Conf. Ser.}, 328:25, 2005.

\bibitem{advLIGO}
G.~M. {Harry} and {the LIGO Scientific Collaboration}.
\newblock {Advanced LIGO: the next generation of gravitational wave detectors}.
\newblock {\em Classical and Quantum Gravity}, 27(8):084006--+, April 2010.

\bibitem{advVirgo}
{Virgo Collaboration}.
\newblock Advanced virgo baseline design.
\newblock Virgo Technical Report VIR-0027A-09, 2009.
\newblock https://tds.ego-gw.it/itf/tds/file.php?callFile=VIR-0027A-09.pdf.

\bibitem{GEOHF}
B.~{Willke}, P.~{Ajith}, B.~{Allen}, P.~{Aufmuth}, C.~{Aulbert}, S.~{Babak},
  R.~{Balasubramanian}, et~al.
\newblock {The GEO-HF project}.
\newblock {\em Classical and Quantum Gravity}, 23:207, April 2006.

\bibitem{janssen:633}
G.~H. Janssen et~al.
\newblock {European Pulsar Timing Array}.
\newblock {\em AIP Conference Proceedings}, 983(1):633, 2008.

\bibitem{Jenet:2009hk}
F.~Jenet et~al.
\newblock {The North American Nanohertz Observatory for Gravitational Waves},
  2009.
\newblock arXiv:0909.1058.

\bibitem{Hobbs:2008yn}
G.~B. Hobbs et~al.
\newblock {Gravitational wave detection using pulsars: status of the Parkes
  Pulsar Timing Array project}, 2008.
\newblock arXiv:0812.2721.

\bibitem{LIGO:India}
B.~{Iyer} et~al.
\newblock {LIGO India}.
\newblock Technical report, 2011.
\newblock https://dcc.ligo.org/cgi-bin/DocDB/ShowDocument?docid=75988.

\bibitem{KAGRA}
K.~{Somiya}.
\newblock {Detector configuration of KAGRA-the Japanese cryogenic
  gravitational-wave detector}.
\newblock {\em Classical and Quantum Gravity}, 29(12):124007, June 2012.

\bibitem{ET}
M.~{Punturo}, M.~{Abernathy}, F.~{Acernese}, B.~{Allen}, N.~{Andersson},
  K.~{Arun}, F.~{Barone}, B.~{Barr}, et~al.
\newblock {The third generation of gravitational wave observatories and their
  science reach}, April 2010.

\bibitem{ET:2011}
S.~{Hild}, M.~{Abernathy}, F.~{Acernese}, P.~{Amaro-Seoane}, N.~{Andersson},
  K.~{Arun}, F.~{Barone}, B.~{Barr}, et~al.
\newblock {Sensitivity studies for third-generation gravitational wave
  observatories}.
\newblock {\em Classical and Quantum Gravity}, 28(9):094013, May 2011.

\bibitem{LISA}
K.~Danzmann et~al.
\newblock {LISA: laser interferometer space antenna for gravitational wave
  measurements}.
\newblock {\em Classical and Quantum Gravity}, 13(11A):A247, 1996.

\bibitem{AmaroSeoane:2012km}
P.~Amaro-Seoane et~al.
\newblock {eLISA: Astrophysics and cosmology in the millihertz regime}, 2012.
\newblock arXiv:1201.3621.

\bibitem{LIGO}
B.~Abbott et~al.
\newblock {LIGO: The Laser Interferometer Gravitational-Wave Observatory}.
\newblock {\em Rept.~Prog.~Phys.}, 72:076901, 2009.

\bibitem{Virgo}
T.~{Accadia}, F.~{Acernese}, M.~{Alshourbagy}, P.~{Amico}, F.~{Antonucci},
  S.~{Aoudia}, N.~{Arnaud}, C.~{Arnault}, K.~G. {Arun}, P.~{Astone}, et~al.
\newblock {Virgo: a laser interferometer to detect gravitational waves}.
\newblock {\em Journal of Instrumentation}, 7:3012, March 2012.

\bibitem{GEO600:2010}
H.~{Grote} and {LIGO Scientific Collaboration}.
\newblock {The GEO 600 status}.
\newblock {\em Classical and Quantum Gravity}, 27(8):084003, April 2010.

\bibitem{national2010New}
{Committee for a Decadal Survey of Astronomy and Astrophysics; National
  Research Council}.
\newblock {\em {New Worlds, New Horizons in Astronomy and Astrophysics}}.
\newblock The National Academies Press, 2010.

\bibitem{2008LRR....11....8L}
D.~R. {Lorimer}.
\newblock {Binary and Millisecond Pulsars}.
\newblock {\em Living Reviews in Relativity}, 11:8, 2008.

\bibitem{2010ApJ...715L.138B}
K.~{Belczynski}, M.~{Dominik}, T.~{Bulik}, R.~{O'Shaughnessy}, C.~{Fryer}, and
  D.~E. {Holz}.
\newblock {The Effect of Metallicity on the Detection Prospects for
  Gravitational Waves}.
\newblock {\em \apjl}, 715:L138, 2010.

\bibitem{Dominik:2012}
M.~{Dominik}, K.~{Belczynski}, C.~{Fryer}, D.~{Holz}, E.~{Berti}, T.~{Bulik},
  I.~{Mandel}, and R.~{O'Shaughnessy}.
\newblock {Double Compact Objects I: The Significance Of The Common Envelope On
  Merger Rates}, February 2012.
\newblock {arXiv:1202.4901}.

\bibitem{2002ApJ...570..252P}
R.~{Perna} and K.~{Belczynski}.
\newblock {Short Gamma-Ray Bursts and Mergers of Compact Objects: Observational
  Constraints}.
\newblock {\em \apj}, 570:252, 2002.

\bibitem{2006ApJ...648.1110B}
K.~{Belczynski}, R.~{Perna}, T.~{Bulik}, V.~{Kalogera}, N.~{Ivanova}, and D.~Q.
  {Lamb}.
\newblock {A Study of Compact Object Mergers as Short Gamma-Ray Burst
  Progenitors}.
\newblock {\em \apj}, 648:1110, 2006.

\bibitem{2008ApJ...675..566O}
R.~{O'Shaughnessy}, K.~{Belczynski}, and V.~{Kalogera}.
\newblock {Short Gamma-Ray Bursts and Binary Mergers in Spiral and Elliptical
  Galaxies: Redshift Distribution and Hosts}.
\newblock {\em \apj}, 675:566, 2008.

\bibitem{2012arXiv1202.2179C}
D.~{Coward}, E.~{Howell}, T.~{Piran}, G.~{Stratta}, M.~{Branchesi},
  O.~{Bromberg}, B.~{Gendre}, R.~{Burman}, and D.~{Guetta}.
\newblock {The Swift short gamma-ray burst rate density: implications for
  binary neutron star merger rates}, February 2012.
\newblock arXiv:1202.2179.

\bibitem{2012arXiv1205.4621K}
I.~{Kowalska}, T.~{Regimbau}, T.~{Bulik}, M.~{Dominik}, and K.~{Belczynski}.
\newblock {Effect of metallicity on the gravitational-wave signal from the
  cosmological population of compact binary coalescences}, May 2012.
\newblock ArXiv:1205.4621.

\bibitem{Fong:2012}
W.~{Fong}, E.~{Berger}, R.~{Margutti}, B.~A. {Zauderer}, E.~{Troja},
  I.~{Czekala}, R.~{Chornock}, N.~{Gehrels}, T.~{Sakamoto}, D.~B. {Fox}, and
  P.~{Podsiadlowski}.
\newblock {A Jet Break in the X-Ray Light Curve of Short GRB 111020A:
  Implications for Energetics and Rates}.
\newblock {\em \apj}, 756:189, 2012.

\bibitem{Berger2009}
E.~{Berger}.
\newblock {The Host Galaxies of Short-Duration Gamma-Ray Bursts: Luminosities,
  Metallicities, and Star-Formation Rates}.
\newblock {\em \apj}, 690:231, 2009.

\bibitem{Fong:2009bd}
W.~Fong, E.~Berger, and D.~B. Fox.
\newblock {Hubble Space Telescope Observations of Short GRB Host Galaxies:
  Morphologies, Offsets, and Local Environments}.
\newblock {\em \apj}, 708:9, 2010.

\bibitem{Troja2010}
E.~{Troja}, S.~{Rosswog}, and N.~{Gehrels}.
\newblock {Precursors of Short Gamma-ray Bursts}.
\newblock {\em \apj}, 723:1711, 2010.

\bibitem{Margutti2011}
R.~{Margutti}, G.~{Chincarini}, J.~{Granot}, C.~{Guidorzi}, E.~{Berger}, M.~G.
  {Bernardini}, N.~{Gehrels}, A.~M. {Soderberg}, M.~{Stamatikos}, and
  E.~{Zaninoni}.
\newblock {X-ray flare candidates in short gamma-ray bursts}, 2011.
\newblock arXiv:1107.1740.

\bibitem{Faber2012}
J.~A. {Faber} and F.~A. {Rasio}.
\newblock {Binary Neutron Star Mergers}.
\newblock {\em Living Reviews in Relativity}, 15:8, 2012.

\bibitem{Shibata2000}
M.~{Shibata} and K.~{\= o}. {Ury{\= u}}.
\newblock {Simulation of merging binary neutron stars in full general
  relativity: {$\Gamma$}=2 case}.
\newblock {\em \prd}, 61(6):064001, 2000.

\bibitem{Rezzolla2010}
L.~{Rezzolla}, L.~{Baiotti}, B.~{Giacomazzo}, D.~{Link}, and J.~A. {Font}.
\newblock {Accurate evolutions of unequal-mass neutron-star binaries:
  properties of the torus and short GRB engines}.
\newblock {\em Class.Quant.Grav.}, 27:114105, 2010.

\bibitem{Anderson2008}
M.~Anderson, E.~W. Hirschmann, L.~Lehner, S.~L. Liebling, P.~M. Motl,
  D.~Neilsen, C.~Palenzuela, and J.~E. Tohline.
\newblock {Simulating binary neutron stars: dynamics and gravitational waves}.
\newblock {\em Phys. Rev.}, D77:024006, 2008.

\bibitem{Baiotti2008}
L.~{Baiotti}, B.~{Giacomazzo}, and L.~{Rezzolla}.
\newblock {Accurate evolutions of inspiralling neutron-star binaries: prompt
  and delayed collapse to black hole}.
\newblock {\em Phys. Rev. D}, 78:084033, 2008.

\bibitem{Yamamoto2008}
T.~{Yamamoto}, M.~{Shibata}, and K.~{Taniguchi}.
\newblock {Simulating coalescing compact binaries by a new code SACRA}.
\newblock {\em Phys. Rev. D}, 78:064054--1--38, 2008.

\bibitem{Baiotti2009}
L.~Baiotti, B.~Giacomazzo, and L.~Rezzolla.
\newblock {Accurate evolutions of inspiralling neutron-star binaries:
  assessment of the truncation error}.
\newblock {\em Class. Quantum Grav.}, 26:114005, 2009.

\bibitem{Liu2008MHD}
Y.~T. {Liu}, S.~L. {Shapiro}, Z.~B. {Etienne}, and K.~{Taniguchi}.
\newblock {General relativistic simulations of magnetized binary neutron star
  mergers}.
\newblock {\em Phys. Rev. D}, 78:024012, 2008.

\bibitem{Giacomazzo2009}
B.~{Giacomazzo}, L.~{Rezzolla}, and L.~{Baiotti}.
\newblock {Can magnetic fields be detected during the inspiral of binary
  neutron stars?}
\newblock {\em MNRAS}, 399:L164, 2009.

\bibitem{Shibata2005}
M.~{Shibata}, K.~{Taniguchi}, and K.~{Ury{\= u}}.
\newblock {Merger of binary neutron stars with realistic equations of state in
  full general relativity}.
\newblock {\em \prd}, 71(8):084021, 2005.

\bibitem{Shibata2006}
M.~{Shibata} and K.~{Taniguchi}.
\newblock {Merger of binary neutron stars to a black hole: Disk mass, short
  gamma-ray bursts, and quasinormal mode ringing}.
\newblock {\em \prd}, 73(6):064027, 2006.

\bibitem{2009CQGra..26p3001B}
E.~{Berti}, V.~{Cardoso}, and A.~O. {Starinets}.
\newblock {TOPICAL REVIEW: Quasinormal modes of black holes and black branes}.
\newblock {\em Classical and Quantum Gravity}, 26(16):163001, 2009.

\bibitem{Sekiguchi2011b}
Y.~{Sekiguchi}, K.~{Kiuchi}, K.~{Kyutoku}, and M.~{Shibata}.
\newblock {Effects of Hyperons in Binary Neutron Star Mergers}.
\newblock {\em Physical Review Letters}, 107(21):211101, 2011.

\bibitem{Kiuchi2012}
K.~{Kiuchi}, Y.~{Sekiguchi}, K.~{Kyutoku}, and M.~{Shibata}.
\newblock {Gravitational waves, neutrino emissions and effects of hyperons in
  binary neutron star mergers}.
\newblock {\em Classical and Quantum Gravity}, 29(12):124003, 2012.

\bibitem{Kiuchi2009}
K.~{Kiuchi}, Y.~{Sekiguchi}, M.~{Shibata}, and K.~{Taniguchi}.
\newblock {Long-term general relativistic simulation of binary neutron stars
  collapsing to a black hole}.
\newblock {\em \prd}, 80(6):064037, 2009.

\bibitem{Kiuchi2010}
K.~{Kiuchi}, Y.~{Sekiguchi}, M.~{Shibata}, and K.~{Taniguchi}.
\newblock {Exploring Binary-Neutron-Star-Merger Scenario of Short-Gamma-Ray
  Bursts by Gravitational-Wave Observation}.
\newblock {\em Physical Review Letters}, 104(14):141101, 2010.

\bibitem{enrico}
{Enrico Ramirez-Ruiz}.
\newblock \url{http://online.kitp.ucsb.edu/online/chirps_c12/ramirezruiz/}.

\bibitem{Read2009}
J.~S. {Read}, C.~{Markakis}, M.~{Shibata}, K.~{Ury{\= u}}, J.~D.~E.
  {Creighton}, and J.~L. {Friedman}.
\newblock {Measuring the neutron star equation of state with gravitational wave
  observations}.
\newblock {\em \prd}, 79(12):124033, 2009.

\bibitem{Hinderer2009}
T.~{Hinderer}, B.~D. {Lackey}, R.~N. {Lang}, and J.~S. {Read}.
\newblock {Tidal deformability of neutron stars with realistic equations of
  state and their gravitational wave signatures in binary inspiral}, 2009.
\newblock {arXiv:0911.3535}.

\bibitem{Markakis:2011}
C.~Markakis, J.~S. Read, M.~Shibata, K.~Uryu, J.~Creighton, et~al.
\newblock {Neutron star equation of state via gravitational wave observations}.
\newblock {\em J.Phys.Conf.Ser.}, 189:012024, 2009.

\bibitem{Ozel2010}
F.~{{\"O}zel}, G.~{Baym}, and T.~{G{\"u}ver}.
\newblock {Astrophysical measurement of the equation of state of neutron star
  matter}.
\newblock {\em \prd}, 82(10):101301, 2010.

\bibitem{2012ApJ...748....5O}
F.~{{\"O}zel}, A.~{Gould}, and T.~{G{\"u}ver}.
\newblock {The Mass and Radius of the Neutron Star in the Bulge Low-mass X-Ray
  Binary KS 1731-260}.
\newblock {\em \apj}, 748:5, 2012.

\bibitem{Baiotti2010}
L.~{Baiotti}, T.~{Damour}, B.~{Giacomazzo}, A.~{Nagar}, and L.~{Rezzolla}.
\newblock {Analytic Modeling of Tidal Effects in the Relativistic Inspiral of
  Binary Neutron Stars}.
\newblock {\em Physical Review Letters}, 105(26):261101, 2010.

\bibitem{Baiotti2011}
L.~{Baiotti}, T.~{Damour}, B.~{Giacomazzo}, A.~{Nagar}, and L.~{Rezzolla}.
\newblock {Accurate numerical simulations of inspiralling binary neutron stars
  and their comparison with effective-one-body analytical models}.
\newblock {\em \prd}, 84(2):024017, 2011.

\bibitem{Bernuzzi2012}
S.~{Bernuzzi}, A.~{Nagar}, M.~{Thierfelder}, and B.~{Br{\"u}gmann}.
\newblock {Tidal effects in binary neutron star coalescence}.
\newblock {\em \prd}, 86(4):044030, 2012.

\bibitem{Bauswein:2011tp}
A.~Bauswein and H.-Th. Janka.
\newblock {Measuring neutron-star properties via gravitational waves from
  binary mergers}.
\newblock {\em Phys.Rev.Lett.}, 108:011101, 2012.

\bibitem{Bauswein:2012ya}
A.~Bauswein, H.T. Janka, K.~Hebeler, and A.~Schwenk.
\newblock {Equation-of-state dependence of the gravitational-wave signal from
  the ring-down phase of neutron-star mergers}.
\newblock {\em Phys.Rev.}, D86:063001, 2012.

\bibitem{Giacomazzo2011}
B.~{Giacomazzo}, L.~{Rezzolla}, and L.~{Baiotti}.
\newblock {Accurate evolutions of inspiralling and magnetized neutron stars:
  Equal-mass binaries}.
\newblock {\em \prd}, 83(4):044014, 2011.

\bibitem{Anderson2008MHD}
M.~{Anderson}, E.~W. {Hirschmann}, L.~{Lehner}, S.~L. {Liebling}, P.~M. {Motl},
  D.~{Neilsen}, C.~{Palenzuela}, and J.~E. {Tohline}.
\newblock {Magnetized Neutron Star Mergers and Gravitational Wave Signals}.
\newblock {\em Phys. Rev. Lett.}, 100:191101, 2008.

\bibitem{Rezzolla2011}
L.~{Rezzolla}, B.~{Giacomazzo}, L.~{Baiotti}, J.~{Granot}, C.~{Kouveliotou},
  and M.~A. {Aloy}.
\newblock {The Missing Link: Merging Neutron Stars Naturally Produce Jet-like
  Structures and Can Power Short Gamma-ray Bursts}.
\newblock {\em \apjl}, 732:L6, 2011.

\bibitem{2012PhRvD..86j4035L}
L.~{Lehner}, C.~{Palenzuela}, S.~L. {Liebling}, C.~{Thompson}, and C.~{Hanna}.
\newblock {Intense electromagnetic outbursts from collapsing hypermassive
  neutron stars}.
\newblock {\em \prd}, 86(10):104035, November 2012.

\bibitem{2013arXiv1301.7074P}
C.~{Palenzuela}, L.~{Lehner}, M.~{Ponce}, S.~L. {Liebling}, M.~{Anderson},
  D.~{Neilsen}, and P.~{Motl}.
\newblock {Gravitational and electromagnetic outputs from binary neutron star
  mergers}.
\newblock {\em ArXiv e-prints}, January 2013.

\bibitem{Sekiguchi2011}
Y.~{Sekiguchi}, K.~{Kiuchi}, K.~{Kyutoku}, and M.~{Shibata}.
\newblock {Gravitational Waves and Neutrino Emission from the Merger of Binary
  Neutron Stars}.
\newblock {\em Physical Review Letters}, 107(5):051102, 2011.

\bibitem{Gold2011}
R.~{Gold}, S.~{Bernuzzi}, M.~{Thierfelder}, B.~{Bruegmann}, and F.~{Pretorius}.
\newblock {Eccentric binary neutron star mergers}, 2011.
\newblock {arXiv:1109.5128}.

\bibitem{2012arXiv1212.0837E}
W.~E. {East}, S.~T. {McWilliams}, J.~{Levin}, and F.~{Pretorius}.
\newblock {Observing complete gravitational wave signals from dynamical capture
  binaries}, December 2012.
\newblock {arXiv:1212.0837}.

\bibitem{Lee2010}
W.~H. {Lee}, E.~{Ramirez-Ruiz}, and G.~{van de Ven}.
\newblock {Short Gamma-ray Bursts from Dynamically Assembled Compact Binaries
  in Globular Clusters: Pathways, Rates, Hydrodynamics, and Cosmological
  Setting}.
\newblock {\em \apj}, 720:953, 2010.

\bibitem{ShibataLRR}
M.~{Shibata} and K.~{Taniguchi}.
\newblock {Coalescence of Black Hole-Neutron Star Binaries}.
\newblock {\em Living Rev. Relativity}, 14:6, 2011.

\bibitem{Shibata06d}
M.~{Shibata} and K.~{Uryu}.
\newblock {Merger of black hole-neutron star binaries: nonspinning black hole
  case}.
\newblock {\em Phys. Rev. D}, 74:{121503}, 2006.

\bibitem{ShibataUryu2007}
M.~{Shibata} and K.~{Uryu}.
\newblock {Merger of black hole neutron star binaries in full general
  relativity}.
\newblock {\em Classical and Quantum Gravity}, 24:125, 2007.

\bibitem{ShibataTaniguchi2008}
M.~{Shibata} and K.~{Taniguchi}.
\newblock {Merger of black hole and neutron star in general relativity: Tidal
  disruption, torus mass, and gravitational waves}.
\newblock {\em Phys. Rev. D}, 77(8):084015, 2008.

\bibitem{Etienne2008}
Z.~B. {Etienne}, J.~A. {Faber}, Y.~T. {Liu}, S.~L. {Shapiro}, K.~{Taniguchi},
  and T.~W. {Baumgarte}.
\newblock {Fully general relativistic simulations of black hole-neutron star
  mergers}.
\newblock {\em \prd}, 77(8):084002, 2008.

\bibitem{Duez2008}
M.~D. {Duez}, F.~{Foucart}, L.~E. {Kidder}, H.~P. {Pfeiffer}, M.~A. {Scheel},
  and S.~A. {Teukolsky}.
\newblock {Evolving black hole-neutron star binaries in general relativity
  using pseudospectral and finite difference methods}.
\newblock {\em \prd}, 78(10):104015, 2008.

\bibitem{Giacomazzo2013}
B.~{Giacomazzo}, R.~{Perna}, L.~{Rezzolla}, E.~{Troja}, and D.~{Lazzati}.
\newblock {Compact Binary Progenitors of Short Gamma-Ray Bursts}.
\newblock {\em \apjl}, 762:L18, 2013.

\bibitem{Etienne2009}
Z.~B. {Etienne}, Y.~T. {Liu}, S.~L. {Shapiro}, and T.~W. {Baumgarte}.
\newblock {General relativistic simulations of black-hole-neutron-star mergers:
  Effects of black-hole spin}.
\newblock {\em \prd}, 79(4):044024, 2009.

\bibitem{Foucart2012a}
F.~{Foucart}, M.~D. {Duez}, L.~E. {Kidder}, M.~A. {Scheel}, B.~{Szilagyi}, and
  S.~A. {Teukolsky}.
\newblock {Black hole-neutron star mergers for 10M$M_{\odot}$ black holes}.
\newblock {\em \prd}, 85(4):044015, 2012.

\bibitem{2010PhRvL.105k1101C}
S.~{Chawla}, M.~{Anderson}, M.~{Besselman}, L.~{Lehner}, S.~L. {Liebling},
  P.~M. {Motl}, and D.~{Neilsen}.
\newblock {Mergers of Magnetized Neutron Stars with Spinning Black Holes:
  Disruption, Accretion, and Fallback}.
\newblock {\em Physical Review Letters}, 105(11):111101, September 2010.

\bibitem{Pannarale2011}
F.~{Pannarale}, A.~{Tonita}, and L.~{Rezzolla}.
\newblock {Black Hole-Neutron Star Mergers and Short Gamma-ray Bursts: A
  Relativistic Toy Model to Estimate the Mass of the Torus}.
\newblock {\em \apj}, 727:95, 2011.

\bibitem{Foucart2012}
F.~{Foucart}.
\newblock {Black Hole-Neutron Star Mergers: Disk Mass Predictions}, 2012.
\newblock {arXiv:1207.6304}.

\bibitem{Pannarale:2012ux}
Francesco Pannarale.
\newblock {The Black Hole Remnant of Black Hole-Neutron Star Coalescing
  Binaries}, 2012.
\newblock arxiv:1208.5869.

\bibitem{Shibata2009}
M.~{Shibata}, K.~{Kyutoku}, T.~{Yamamoto}, and K.~{Taniguchi}.
\newblock {Gravitational waves from black hole-neutron star binaries:
  Classification of waveforms}.
\newblock {\em \prd}, 79(4):044030, 2009.

\bibitem{Pannarale2011b}
F.~{Pannarale}, L.~{Rezzolla}, F.~{Ohme}, and J.~S. {Read}.
\newblock {Will black hole-neutron star binary inspirals tell us about the
  neutron star equation of state?}
\newblock {\em \prd}, 84(10):104017, 2011.

\bibitem{Lackey2012}
B.~D. {Lackey}, K.~{Kyutoku}, M.~{Shibata}, P.~R. {Brady}, and J.~L.
  {Friedman}.
\newblock {Extracting equation of state parameters from black hole-neutron star
  mergers: Nonspinning black holes}.
\newblock {\em \prd}, 85(4):044061, 2012.

\bibitem{Lackey2013}
B.~D. {Lackey}, K.~{Kyutoku}, M.~{Shibata}, P.~R. {Brady}, and J.~L.
  {Friedman}.
\newblock {Extracting equation of state parameters from black hole-neutron star
  mergers: aligned-spin black holes and a preliminary waveform model}.
\newblock {\em ArXiv e-prints}, March 2013.

\bibitem{Duez2010}
M.~D. {Duez}, F.~{Foucart}, L.~E. {Kidder}, C.~D. {Ott}, and S.~A. {Teukolsky}.
\newblock {Equation of state effects in black hole-neutron star mergers}.
\newblock {\em Classical and Quantum Gravity}, 27(11):114106, 2010.

\bibitem{Kyutoku2010}
K.~{Kyutoku}, M.~{Shibata}, and K.~{Taniguchi}.
\newblock {Gravitational waves from nonspinning black hole-neutron star
  binaries: Dependence on equations of state}.
\newblock {\em \prd}, 82(4):044049, 2010.

\bibitem{Kyutoku2011}
K.~{Kyutoku}, H.~{Okawa}, M.~{Shibata}, and K.~{Taniguchi}.
\newblock {Gravitational waves from spinning black hole-neutron star binaries:
  dependence on black hole spins and on neutron star equations of state}.
\newblock {\em \prd}, 84(6):064018, 2011.

\bibitem{Foucart2011}
F.~{Foucart}, M.~D. {Duez}, L.~E. {Kidder}, and S.~A. {Teukolsky}.
\newblock {Black hole-neutron star mergers: Effects of the orientation of the
  black hole spin}.
\newblock {\em \prd}, 83(2):024005, 2011.

\bibitem{Chawla2010}
S.~{Chawla}, M.~{Anderson}, M.~{Besselman}, L.~{Lehner}, S.~L. {Liebling},
  P.~M. {Motl}, and D.~{Neilsen}.
\newblock {Mergers of Magnetized Neutron Stars with Spinning Black Holes:
  Disruption, Accretion, and Fallback}.
\newblock {\em Physical Review Letters}, 105(11):111101, 2010.

\bibitem{Etienne2012}
Z.~B. {Etienne}, Y.~T. {Liu}, V.~{Paschalidis}, and S.~L. {Shapiro}.
\newblock {General relativistic simulations of black-hole-neutron-star mergers:
  Effects of magnetic fields}.
\newblock {\em \prd}, 85(6):064029, 2012.

\bibitem{2001MNRAS.322..695H}
B.~M.~S. {Hansen} and M.~{Lyutikov}.
\newblock {Radio and X-ray signatures of merging neutron stars}.
\newblock {\em \mnras}, 322:695--701, April 2001.

\bibitem{2010PhRvD..82d4045P}
C.~{Palenzuela}, T.~{Garrett}, L.~{Lehner}, and S.~L. {Liebling}.
\newblock {Magnetospheres of black hole systems in force-free plasma}.
\newblock {\em \prd}, 82(4):044045, August 2010.

\bibitem{2011PNAS..10812641N}
D.~{Neilsen}, L.~{Lehner}, C.~{Palenzuela}, E.~W. {Hirschmann}, S.~L.
  {Liebling}, P.~M. {Motl}, and T.~{Garrett}.
\newblock {Boosting jet power in black hole spacetimes}.
\newblock {\em Proceedings of the National Academy of Science},
  108:12641--12646, August 2011.

\bibitem{2011ApJ...742...90M}
S.~T. {McWilliams} and J.~{Levin}.
\newblock {Electromagnetic Extraction of Energy from Black-hole-Neutron-star
  Binaries}.
\newblock {\em \apj}, 742:90, December 2011.

\bibitem{Stephens2011}
B.~C. {Stephens}, W.~E. {East}, and F.~{Pretorius}.
\newblock {Eccentric Black-hole-Neutron-star Mergers}.
\newblock {\em \apjl}, 737:L5, 2011.

\bibitem{East2012}
W.~E. {East}, F.~{Pretorius}, and B.~C. {Stephens}.
\newblock {Eccentric black hole-neutron star mergers: Effects of black hole
  spin and equation of state}.
\newblock {\em \prd}, 85(12):124009, 2012.

\bibitem{MandelOShaughnessy:2010}
I.~{Mandel} and R.~{O'Shaughnessy}.
\newblock {Compact binary coalescences in the band of ground-based
  gravitational-wave detectors}.
\newblock {\em Classical and Quantum Gravity}, 27(11):114007, 2010.

\bibitem{2003ApJ...584..985K}
C.~{Kim}, V.~{Kalogera}, and D.~R. {Lorimer}.
\newblock {The Probability Distribution of Binary Pulsar Coalescence Rates. I.
  Double Neutron Star Systems in the Galactic Field}.
\newblock {\em \apj}, 584:985, 2003.

\bibitem{OShaughnessy:2007}
R.~{O'Shaughnessy}, V.~{Kalogera}, and K.~{Belczynski}.
\newblock {Mapping Population Synthesis Event Rates on Model Parameters. II.
  Convergence and Accuracy of Multidimensional Fits}.
\newblock {\em \apj}, 667:1048, 2007.

\bibitem{CygnusX3:2012}
K.~{Belczynski}, T.~{Bulik}, I.~{Mandel}, B.~S. {Sathyaprakash},
  A.~{Zdziarski}, and J.~{Mikolajewska}.
\newblock {Cyg X-3: a Galactic double black hole or black hole-neutron star
  progenitor}.
\newblock {\em ArXiv e-prints}, September 2012.

\bibitem{2012ApJ...759...52D}
M.~{Dominik}, K.~{Belczynski}, C.~{Fryer}, D.~E. {Holz}, E.~{Berti},
  T.~{Bulik}, I.~{Mandel}, and R.~{O'Shaughnessy}.
\newblock {Double Compact Objects. I. The Significance of the Common Envelope
  on Merger Rates}.
\newblock {\em \apj}, 759:52, 2012.

\bibitem{Abadie:2010cf}
J.~Abadie et~al.
\newblock {Predictions for the Rates of Compact Binary Coalescences Observable
  by Ground-based Gravitational-wave Detectors}.
\newblock {\em Class.Quant.Grav.}, 27:173001, 2010.

\bibitem{OShaughnessy:2009}
R.~{O'Shaughnessy}, V.~{Kalogera}, and K.~{Belczynski}.
\newblock {Binary Compact Object Coalescence Rates: The Role of Elliptical
  Galaxies}.
\newblock {\em \apj}, 716:615, 2010.

\bibitem{lrr-2008-9}
Dimitrios Psaltis.
\newblock Probes and tests of strong-field gravity with observations in the
  electromagnetic spectrum.
\newblock {\em Living Reviews in Relativity}, 11(9), 2008.

\bibitem{Abramowicz:2002vt}
M.~A. Abramowicz, W.~Kluzniak, and J.~P. Lasota.
\newblock {No observational proof of the black hole event-horizon}.
\newblock {\em Astron.Astrophys.}, 396:L31, 2002.

\bibitem{Sperhake:2011xk}
U.~Sperhake, E.~Berti, and V.~Cardoso.
\newblock {Numerical simulations of black-hole binaries and gravitational wave
  emission}, 2011.
\newblock {arXiv:1107.2819}.

\bibitem{Bulik:2008}
T.~{Bulik}, K.~{Belczynski}, and A.~{Prestwich}.
\newblock {IC10 X-1/NGC300 X-1: The Very Immediate Progenitors of BH-BH
  Binaries}.
\newblock {\em \apj}, 730:140, 2011.

\bibitem{Sadowski}
A.~{Sadowski}, K.~{Belczynski}, T.~{Bulik}, N.~{Ivanova}, F.~A. {Rasio}, and
  R.~{O'Shaughnessy}.
\newblock {The Total Merger Rate of Compact Object Binaries in the Local
  Universe}.
\newblock {\em \apj}, 676:1162, 2008.

\bibitem{MillerLauburg:2008}
M.~C. {Miller} and V.~M. {Lauburg}.
\newblock {Mergers of Stellar-Mass Black Holes in Nuclear Star Clusters}.
\newblock {\em \apj}, 692:917, 2009.

\bibitem{OLeary:2008}
R.~M. {O'Leary}, B.~{Kocsis}, and A.~{Loeb}.
\newblock {Gravitational waves from scattering of stellar-mass black holes in
  galactic nuclei}.
\newblock {\em \mnras}, 395:2127, 2009.

\bibitem{Banerjee:2010}
S.~{Banerjee}, H.~{Baumgardt}, and P.~{Kroupa}.
\newblock {Stellar-mass black holes in star clusters: implications for
  gravitational wave radiation}.
\newblock {\em \mnras}, 402:371, 2010.

\bibitem{Pretorius:2005gq}
F.~Pretorius.
\newblock {Evolution of binary black hole spacetimes}.
\newblock {\em Phys. Rev. Lett.}, 95:121101, 2005.

\bibitem{Campanelli:2005dd}
M.~Campanelli, C.~O. Lousto, P.~Marronetti, and Y.~Zlochower.
\newblock {Accurate evolutions of orbiting black-hole binaries without
  excision}.
\newblock {\em Phys. Rev. Lett.}, 96:111101, 2006.

\bibitem{2006PhRvL..96k1102B}
J.~G. {Baker}, J.~{Centrella}, {D.-I.} {Choi}, M.~{Koppitz}, and J.~{van
  Meter}.
\newblock {Gravitational-Wave Extraction from an Inspiraling Configuration of
  Merging Black Holes}.
\newblock {\em Physical Review Letters}, 96(11):111102, March 2006.

\bibitem{Pretorius:2007nq}
Frans Pretorius.
\newblock {Binary Black Hole Coalescence}.
\newblock 2007.

\bibitem{Pfeiffer:2012pc}
Harald~P. Pfeiffer.
\newblock {Numerical simulations of compact object binaries}.
\newblock {\em Class.Quant.Grav.}, 29:124004, 2012.

\bibitem{Buonanno:2007pf}
A.~Buonanno et~al.
\newblock {Toward faithful templates for non-spinning binary black holes using
  the effective-one-body approach}.
\newblock {\em Phys. Rev. D}, 76:104049, 2007.

\bibitem{Ajith:2009bn}
P.~Ajith et~al.
\newblock {Inspiral-merger-ringdown waveforms for black-hole binaries with
  non-precessing spins}.
\newblock {\em Phys.Rev.Lett.}, 106:241101, 2011.

\bibitem{Santamaria:2010yb}
L.~Santamaria, F.~Ohme, P.~Ajith, B.~Bruegmann, N.~Dorband, et~al.
\newblock {Matching post-Newtonian and numerical relativity waveforms:
  systematic errors and a new phenomenological model for non-precessing black
  hole binaries}.
\newblock {\em Phys.Rev.}, D82:064016, 2010.

\bibitem{Sturani:2010yv}
R.~Sturani et~al.
\newblock {Complete phenomenological gravitational waveforms from spinning
  coalescing binaries}.
\newblock {\em J.Phys.Conf.Ser.}, 243:012007, 2010.

\bibitem{Taracchini:2012ig}
A.~Taracchini et~al.
\newblock {Prototype effective-one-body model for nonprecessing spinning
  inspiral-merger-ringdown waveforms}.
\newblock {\em Phys.Rev.}, D86:024011, 2012.

\bibitem{2012arXiv1212.4357D}
T.~{Damour}, A.~{Nagar}, and S.~{Bernuzzi}.
\newblock {Improved effective-one-body description of coalescing nonspinning
  black-hole binaries and its numerical-relativity completion}, December 2012.
\newblock {arXiv:1212.4357}.

\bibitem{MillerColbert:2004}
M.~C. {Miller} and E.~J.~M. {Colbert}.
\newblock {Intermediate-Mass Black Holes}.
\newblock {\em International Journal of Modern Physics D}, 13:1, 2004.

\bibitem{Miller:2009}
M.~C. {Miller}.
\newblock {Intermediate-mass black holes as LISA sources}.
\newblock {\em Classical and Quantum Gravity}, 26(9):094031, 2009.

\bibitem{Brown:2007}
D.~A. {Brown}, J.~{Brink}, H.~{Fang}, J.~R. {Gair}, C.~{Li}, G.~{Lovelace},
  I.~{Mandel}, and K.~S. {Thorne}.
\newblock {Prospects for Detection of Gravitational Waves from
  Intermediate-Mass-Ratio Inspirals}.
\newblock {\em Physical Review Letters}, 99(20):201102, 2007.

\bibitem{Mandel:2007rates}
I.~{Mandel}, D.~A. {Brown}, J.~R. {Gair}, and M.~C. {Miller}.
\newblock {Rates and Characteristics of Intermediate Mass Ratio Inspirals
  Detectable by Advanced LIGO}.
\newblock {\em \apj}, 681:1431, 2008.

\bibitem{imbhlisa-2006}
J.~M. {Fregeau}, S.~L. {Larson}, M.~C. {Miller}, R.~{O'Shaughnessy}, and F.~A.
  {Rasio}.
\newblock {Observing IMBH-IMBH Binary Coalescences via Gravitational
  Radiation}.
\newblock {\em Astrophysical Journal Letters}, 646:L135, 2006.

\bibitem{Amaro:2006imbh}
P.~{Amaro-Seoane} and M.~{Freitag}.
\newblock {Intermediate-Mass Black Holes in Colliding Clusters: Implications
  for Lower Frequency Gravitational-Wave Astronomy}.
\newblock {\em \apjl}, 653:L53, 2006.

\bibitem{2002PhRvD..66d4002G}
K.~{Glampedakis} and D.~{Kennefick}.
\newblock {Zoom and whirl: Eccentric equatorial orbits around spinning black
  holes and their evolution under gravitational radiation reaction}.
\newblock {\em \prd}, 66(4):044002, 2002.

\bibitem{2008PhRvD..77j3005L}
J.~{Levin} and G.~{Perez-Giz}.
\newblock {A periodic table for black hole orbits}.
\newblock {\em \prd}, 77(10):103005, 2008.

\bibitem{2009PhRvD..79d3016L}
J.~{Levin} and R.~{Grossman}.
\newblock {Dynamics of black hole pairs. I. Periodic tables}.
\newblock {\em \prd}, 79(4):043016, 2009.

\bibitem{2009CQGra..26w5010L}
J.~{Levin}.
\newblock {Energy level diagrams for black hole orbits}.
\newblock {\em Classical and Quantum Gravity}, 26(23):235010, 2009.

\bibitem{2009PhRvL.103m1101H}
J.~{Healy}, J.~{Levin}, and D.~{Shoemaker}.
\newblock {Zoom-Whirl Orbits in Black Hole Binaries}.
\newblock {\em Physical Review Letters}, 103(13):131101, 2009.

\bibitem{2007CQGra..24...83P}
F.~{Pretorius} and D.~{Khurana}.
\newblock {Black hole mergers and unstable circular orbits}.
\newblock {\em Classical and Quantum Gravity}, 24:83, 2007.

\bibitem{Gold:2009hr}
R.~Gold and B.~Bruegmann.
\newblock {Radiation from low-momentum zoom-whirl orbits}.
\newblock {\em Class.Quant.Grav.}, 27:084035, 2010.

\bibitem{2006ApJ...648..411K}
B.~{Kocsis}, M.~E. {G{\'a}sp{\'a}r}, and S.~{M{\'a}rka}.
\newblock {Detection Rate Estimates of Gravity Waves Emitted during Parabolic
  Encounters of Stellar Black Holes in Globular Clusters}.
\newblock {\em \apj}, 648:411, 2006.

\bibitem{2009MNRAS.395.2127O}
R.~M. {O'Leary}, B.~{Kocsis}, and A.~{Loeb}.
\newblock {Gravitational waves from scattering of stellar-mass black holes in
  galactic nuclei}.
\newblock {\em \mnras}, 395:2127, 2009.

\bibitem{wb06}
S.~E. {Woosley} and J.~S. {Bloom}.
\newblock {The Supernova Gamma-Ray Burst Connection}.
\newblock {\em \araa}, 44:507, 2006.

\bibitem{2012grbu.book..169H}
J.~{Hjorth} and J.~S. {Bloom}.
\newblock {\em {The Gamma-Ray Burst - Supernova Connection}}, pages 169--190.
\newblock November 2012.

\bibitem{fil97}
A.~V. {Filippenko}.
\newblock {Optical Spectra of Supernovae}.
\newblock {\em \araa}, 35:309, 1997.

\bibitem{kfw+98}
S.~R. {Kulkarni}, D.~A. {Frail}, M.~H. {Wieringa}, R.~D. {Ekers}, E.~M.
  {Sadler}, R.~M. {Wark}, J.~L. {Higdon}, E.~S. {Phinney}, and J.~S. {Bloom}.
\newblock {Radio emission from the unusual supernova 1998bw and its association
  with the {$\gamma$}-ray burst of 25 April 1998}.
\newblock {\em \nat}, 395:663, 1998.

\bibitem{piran04}
T.~{Piran}.
\newblock {The physics of gamma-ray bursts}.
\newblock {\em Reviews of Modern Physics}, 76:1143, October 2004.

\bibitem{skn+06}
A.~M. {Soderberg}, , et~al.
\newblock {Relativistic ejecta from X-ray flash XRF 060218 and the rate of
  cosmic explosions}.
\newblock {\em \nat}, 442:1014, 2006.

\bibitem{scp+10}
A.~M. {Soderberg} et~al.
\newblock {A relativistic type Ibc supernova without a detected {$\gamma$}-ray
  burst}.
\newblock {\em \nat}, 463:513, 2010.

\bibitem{smartt09}
S.~J. {Smartt}.
\newblock {Progenitors of Core-Collapse Supernovae}.
\newblock {\em \araa}, 47:63, 2009.

\bibitem{wlw95}
S.~E. {Woosley}, N.~{Langer}, and T.~A. {Weaver}.
\newblock {The Presupernova Evolution and Explosion of Helium Stars That
  Experience Mass Loss}.
\newblock {\em \apj}, 448:315, 1995.

\bibitem{pjh92}
P.~{Podsiadlowski}, P.~C. {Joss}, and J.~J.~L. {Hsu}.
\newblock {Presupernova evolution in massive interacting binaries}.
\newblock {\em \apj}, 391:246, 1992.

\bibitem{wh06}
S.~E. {Woosley} and A.~{Heger}.
\newblock {The Progenitor Stars of Gamma-Ray Bursts}.
\newblock {\em \apj}, 637:914, 2006.

\bibitem{tvb+09}
S.~{Taubenberger}, S.~{Valenti}, S.~{Benetti}, E.~{Cappellaro}, M.~{Della
  Valle}, N.~{Elias-Rosa}, S.~{Hachinger}, W.~{Hillebrandt}, K.~{Maeda}, P.~A.
  {Mazzali}, A.~{Pastorello}, F.~{Patat}, S.~A. {Sim}, and M.~{Turatto}.
\newblock {Nebular emission-line profiles of Type Ib/c supernovae - probing the
  ejecta asphericity}.
\newblock {\em \mnras}, 397:677, 2009.

\bibitem{cfl+10}
R.~{Chornock}, A.~V. {Filippenko}, W.~{Li}, and J.~M. {Silverman}.
\newblock {Large Late-Time Asphericities in Three Type IIP Supernovae}.
\newblock {\em \apj}, 713:1363, 2010.

\bibitem{fhm+06}
R.~A. {Fesen}, M.~C. {Hammell}, J.~{Morse}, R.~A. {Chevalier}, K.~J.
  {Borkowski}, M.~A. {Dopita}, C.~L. {Gerardy}, S.~S. {Lawrence}, J.~C.
  {Raymond}, and S.~{van den Bergh}.
\newblock {The Expansion Asymmetry and Age of the Cassiopeia A Supernova
  Remnant}.
\newblock {\em \apj}, 645:283, 2006.

\bibitem{bethe:90}
H.~A. {Bethe}.
\newblock {Supernova mechanisms}.
\newblock {\em Rev. Mod. Phys.}, 62:801, 1990.

\bibitem{janka:07}
H.-T. {Janka}, K.~{Langanke}, A.~{Marek}, G.~{Mart{\'{\i}}nez-Pinedo}, and
  B.~{M{\"u}ller}.
\newblock {Theory of core-collapse supernovae}.
\newblock {\em \physrep}, 442:38, 2007.

\bibitem{fryerwarren:04}
C.~L. {Fryer} and M.~S. {Warren}.
\newblock {The Collapse of Rotating Massive Stars in Three Dimensions}.
\newblock {\em \apj}, 601:391, 2004.

\bibitem{marek:09}
A.~{Marek} and H.-T. {Janka}.
\newblock {Delayed Neutrino-Driven Supernova Explosions Aided by the Standing
  Accretion-Shock Instability}.
\newblock {\em \apj}, 694:664, 2009.

\bibitem{yakunin:10}
K.~N. {Yakunin}, P.~{Marronetti}, A.~{Mezzacappa}, S.~W. {Bruenn}, C.-T. {Lee},
  M.~A. {Chertkow}, W.~R. {Hix}, J.~M. {Blondin}, E.~J. {Lentz}, O.~E. {Bronson
  Messer}, and S.~{Yoshida}.
\newblock {Gravitational waves from core collapse supernovae}.
\newblock {\em Class. Quantum Grav.}, 27(19):194005, 2010.

\bibitem{nordhaus:10}
J.~{Nordhaus}, A.~{Burrows}, A.~{Almgren}, and J.~{Bell}.
\newblock {Dimension as a Key to the Neutrino Mechanism of Core-Collapse
  Supernova Explosions}.
\newblock {\em \apj}, 720:694, 2010.

\bibitem{murphy:08}
J.~W. {Murphy} and A.~{Burrows}.
\newblock {Criteria for Core-Collapse Supernova Explosions by the Neutrino
  Mechanism}.
\newblock {\em \apj}, 688:1159, 2008.

\bibitem{hanke:11}
F.~{Hanke}, A.~{Marek}, B.~{Mueller}, and H.-T. {Janka}.
\newblock {Is Strong SASI Activity the Key to Successful Neutrino-Driven
  Supernova Explosions?}
\newblock {\em Submitted to the Astrophys. J., arXiv:1108.4355}, August 2011.

\bibitem{takiwaki:11b}
T.~{Takiwaki}, K.~{Kotake}, and Y.~{Suwa}.
\newblock {Three-dimensional Hydrodynamic Core-Collapse Supernova Simulations
  for an \$11.2 M\_$\{$$\backslash$odot$\}$\$ Star with Spectral Neutrino
  Transport}.
\newblock {\em Submitted to the Astrophys. J., arXiv:1108.3989}, August 2011.

\bibitem{burrows:07b}
A.~{Burrows}, L.~{Dessart}, E.~{Livne}, C.~D. {Ott}, and J.~{Murphy}.
\newblock {Simulations of Magnetically Driven Supernova and Hypernova
  Explosions in the Context of Rapid Rotation}.
\newblock {\em \apj}, 664:416, 2007.

\bibitem{bh:91}
S.~A. {Balbus} and J.~F. {Hawley}.
\newblock {A powerful local shear instability in weakly magnetized disks. I -
  Linear analysis. II - Nonlinear evolution}.
\newblock {\em \apj}, 376:214, 1991.

\bibitem{obergaulinger:09}
M.~{Obergaulinger}, P.~{Cerd{\'a}-Dur{\'a}n}, E.~{M{\"u}ller}, and M.~A.
  {Aloy}.
\newblock {Semi-global simulations of the magneto-rotational instability in
  core collapse supernovae}.
\newblock {\em \aap}, 498:241, 2009.

\bibitem{cerda:08}
P.~{Cerd{\'a}-Dur{\'a}n}, J.~A. {Font}, L.~{Ant{\'o}n}, and E.~{M{\"u}ller}.
\newblock {A new general relativistic magnetohydrodynamics code for dynamical
  spacetimes}.
\newblock {\em \aap}, 492:937, 2008.

\bibitem{takiwaki:11a}
T.~{Takiwaki} and K.~{Kotake}.
\newblock {Gravitational-Wave Signatures in Magnetically-driven Supernova
  Explosions}.
\newblock {\em to appear in ApJ; arXiv:1004.2896}, April 2010.

\bibitem{couch:09}
S.~M. {Couch}, J.~C. {Wheeler}, and M.~{Milosavljevi{\'c}}.
\newblock {Aspherical Core-Collapse Supernovae in Red Supergiants Powered by
  Nonrelativistic Jets}.
\newblock {\em \apj}, 696:953, 2009.

\bibitem{heger05}
A.~Heger, S.E. Woosley, and H.C. Spruit.
\newblock Presupernova evolution of differentially rotating massive stars
  including magnetic fields.
\newblock {\em Astrophys. J.}, 626:350, 2005.

\bibitem{ott:06spin}
C.~D. {Ott}, A.~{Burrows}, T.~A. {Thompson}, E.~{Livne}, and R.~{Walder}.
\newblock {The Spin Periods and Rotational Profiles of Neutron Stars at Birth}.
\newblock {\em Astrophys. J. Suppl. Ser.}, 164:130, 2006.

\bibitem{burrows:06}
A.~{Burrows}, E.~{Livne}, L.~{Dessart}, C.~D. {Ott}, and J.~{Murphy}.
\newblock {A New Mechanism for Core-Collapse Supernova Explosions}.
\newblock {\em Astrophys. J.}, 640:878, 2006.

\bibitem{burrows:07a}
A.~{Burrows}, E.~{Livne}, L.~{Dessart}, C.~D. {Ott}, and J.~{Murphy}.
\newblock {Features of the Acoustic Mechanism of Core-Collapse Supernova
  Explosions}.
\newblock {\em Astrophys. J.}, 655:416, 2007.

\bibitem{weinberg:08}
N.~N. {Weinberg} and E.~{Quataert}.
\newblock {Non-linear saturation of g-modes in proto-neutron stars: quieting
  the acoustic engine}.
\newblock {\em \mnras}, 387:L64, 2008.

\bibitem{woosley:93}
S.~E. {Woosley}.
\newblock {Gamma-ray bursts from stellar mass accretion disks around black
  holes}.
\newblock {\em \apj}, 405:273, 1993.

\bibitem{oconnor:11}
E.~{O'Connor} and C.~D. {Ott}.
\newblock {Black Hole Formation in Failing Core-Collapse Supernovae}.
\newblock {\em \apj}, 730:70, 2011.

\bibitem{nagataki:07}
S.~{Nagataki}, R.~{Takahashi}, A.~{Mizuta}, and T.~{Takiwaki}.
\newblock {Numerical Study of Gamma-Ray Burst Jet Formation in Collapsars}.
\newblock {\em \apj}, 659:512, 2007.

\bibitem{komissarov:09}
S.~S. {Komissarov}, N.~{Vlahakis}, A.~{K{\"o}nigl}, and M.~V. {Barkov}.
\newblock {Magnetic acceleration of ultrarelativistic jets in gamma-ray burst
  sources}.
\newblock {\em \mnras}, 394:1182, 2009.

\bibitem{birkl:07}
R.~{Birkl}, M.~A. {Aloy}, {H.-T.} {Janka}, and E.~{M{\"u}ller}.
\newblock {Neutrino pair annihilation near accreting, stellar-mass black
  holes}.
\newblock {\em \aap}, 463:51, 2007.

\bibitem{harikae:10}
S.~{Harikae}, K.~{Kotake}, T.~{Takiwaki}, and {Y.-i.} {Sekiguchi}.
\newblock {A General Relativistic Ray-tracing Method for Estimating the Energy
  and Momentum Deposition by Neutrino Pair Annihilation in Collapsars}.
\newblock {\em \apj}, 720:614, 2010.

\bibitem{macfadyen:01}
A.~I. {MacFadyen}, S.~E. {Woosley}, and A.~{Heger}.
\newblock {Supernovae, Jets, and Collapsars}.
\newblock {\em \apj}, 550:410, 2001.

\bibitem{lindner:11}
C.~C. {Lindner}, M.~{Milosavljevic}, R.~{Shen}, and P.~{Kumar}.
\newblock {Simulations of Accretion Powered Supernovae in the Progenitors of
  Gamma Ray Bursts}.
\newblock {\em Submitted to ApJ. ArXiv:1108.1415}, August 2011.

\bibitem{dessart:08a}
L.~{Dessart}, A.~{Burrows}, E.~{Livne}, and C.~D. {Ott}.
\newblock {The Proto-Neutron Star Phase of the Collapsar Model and the Route to
  Long-Soft Gamma-Ray Bursts and Hypernovae}.
\newblock {\em \apjl}, 673:L43, 2008.

\bibitem{thompson:04}
T.~A. {Thompson}, P.~{Chang}, and E.~{Quataert}.
\newblock {Magnetar Spin-Down, Hyperenergetic Supernovae, and Gamma-Ray
  Bursts}.
\newblock {\em \apj}, 611:380, 2004.

\bibitem{bucciantini:08}
N.~{Bucciantini}, E.~{Quataert}, J.~{Arons}, B.~D. {Metzger}, and T.~A.
  {Thompson}.
\newblock {Relativistic jets and long-duration gamma-ray bursts from the birth
  of magnetars}.
\newblock {\em \mnras}, 383:L25, 2008.

\bibitem{metzger:11}
B.~D. {Metzger}, D.~{Giannios}, T.~A. {Thompson}, N.~{Bucciantini}, and
  E.~{Quataert}.
\newblock {The protomagnetar model for gamma-ray bursts}.
\newblock {\em \mnras}, 413:2031, 2011.

\bibitem{weber:66}
J.~{Weber}.
\newblock {Observation of the Thermal Fluctuations of a Gravitational-Wave
  Detector}.
\newblock {\em \prl}, 17:1228, 1966.

\bibitem{ruffini:71}
R.~{Ruffini} and J.~A. {Wheeler}.
\newblock {Relativistic Cosmology and Space Platforms}.
\newblock In V.~{Hardy} and H.~{Moore}, editors, {\em Proceedings of the
  Conference on Space Physics, ESRO, Paris, France}, page~45, 1971.

\bibitem{fryernew:11}
C.~Fryer and K.~C.~B. New.
\newblock Gravitational waves from gravitational collapse.
\newblock {\em Liv. Rev. Rel.}, 14:1, 2011.

\bibitem{ott:09}
C.~D. {Ott}.
\newblock {TOPICAL REVIEW: The gravitational-wave signature of core-collapse
  supernovae}.
\newblock {\em Class. Quant. Grav.}, 26(6):063001, 2009.

\bibitem{kotake:06a}
K.~{Kotake}, K.~{Sato}, and K.~{Takahashi}.
\newblock {Explosion mechanism, neutrino burst and gravitational wave in
  core-collapse supernovae.}
\newblock {\em Rep. Prog. Phys.}, 69:971, 2006.

\bibitem{marek:09b}
A.~{Marek}, H.-T. {Janka}, and E.~{M{\"u}ller}.
\newblock {Equation-of-state dependent features in shock-oscillation modulated
  neutrino and gravitational-wave signals from supernovae}.
\newblock {\em \aap}, 496:475, 2009.

\bibitem{murphy:09}
J.~Murphy, C.~D. Ott, and A.~Burrows.
\newblock Gravitational waves from convection and sasi in core-collapse
  supernovae.
\newblock {\em To be submitted to the Astrophys. J.}, 2009.

\bibitem{mueller:11}
E.~{M{\"u}ller}, H.-T. {Janka}, and A.~{Wongwathanarat}.
\newblock {Parametrized 3D models of neutrino-driven supernova explosions:
  Neutrino emission asymmetries and gravitational-wave signals}.
\newblock {\em Submitted to Astron. Astrophys. arXiv:1106.6301 [astro-ph]},
  June 2011.

\bibitem{kotake:11}
K.~{Kotake}, W.~{Iwakami Nakano}, and N.~{Ohnishi}.
\newblock {Effects of Rotation on Stochasticity of Gravitational Waves in
  Nonlinear Phase of Core-Collapse Supernovae}.
\newblock {\em Submitted to ApJ. arXiv:1106.0544 [astro-ph]}, June 2011.

\bibitem{epstein:78}
R.~{Epstein}.
\newblock {The generation of gravitational radiation by escaping supernova
  neutrinos}.
\newblock {\em \apj}, 223:1037, 1978.

\bibitem{ott:06prl}
C.~D. {Ott}, A.~{Burrows}, L.~{Dessart}, and E.~{Livne}.
\newblock {A New Mechanism for Gravitational-Wave Emission in Core-Collapse
  Supernovae}.
\newblock {\em \prl}, 96(20):201102, 2006.

\bibitem{dimmelmeier:08}
H.~{Dimmelmeier}, C.~D. {Ott}, A.~{Marek}, and H.-T. {Janka}.
\newblock {Gravitational wave burst signal from core collapse of rotating
  stars}.
\newblock {\em \prd}, 78(6):064056, 2008.

\bibitem{dimmelmeier:07}
H.~{Dimmelmeier}, C.~D. {Ott}, H.-T. {Janka}, A.~{Marek}, and E.~{M{\"u}ller}.
\newblock {Generic Gravitational-Wave Signals from the Collapse of Rotating
  Stellar Cores}.
\newblock {\em \prl}, 98(25):251101, 2007.

\bibitem{kotake:03}
K.~{Kotake}, S.~{Yamada}, and K.~{Sato}.
\newblock {Gravitational radiation from axisymmetric rotational core collapse}.
\newblock {\em \prd}, 68(4):044023, 2003.

\bibitem{scheidegger:10b}
S.~{Scheidegger}, R.~{K{\"a}ppeli}, S.~C. {Whitehouse}, T.~{Fischer}, and
  M.~{Liebend{\"o}rfer}.
\newblock {The influence of model parameters on the prediction of gravitational
  wave signals from stellar core collapse}.
\newblock {\em Astron. Astrophys.}, 514:A51, 2010.

\bibitem{ott:07prl}
C.~D. {Ott}, H.~{Dimmelmeier}, A.~{Marek}, H.-T. {Janka}, I.~{Hawke},
  B.~{Zink}, and E.~{Schnetter}.
\newblock {3D Collapse of Rotating Stellar Iron Cores in General Relativity
  Including Deleptonization and a Nuclear Equation of State}.
\newblock {\em \prl}, 98:261101, 2007.

\bibitem{kotake:04}
K.~{Kotake}, S.~{Yamada}, K.~{Sato}, K.~{Sumiyoshi}, H.~{Ono}, and H.~{Suzuki}.
\newblock {Gravitational radiation from rotational core collapse: Effects of
  magnetic fields and realistic equations of state}.
\newblock {\em \prd}, 69(12):124004, 2004.

\bibitem{obergaulinger:06a}
M.~{Obergaulinger}, M.~A. {Aloy}, and E.~{M{\"u}ller}.
\newblock {Axisymmetric simulations of magneto-rotational core collapse:
  dynamics and gravitational wave signal}.
\newblock {\em \aap}, 450:1107, 2006.

\bibitem{obergaulinger:06b}
M.~{Obergaulinger}, M.~A. {Aloy}, H.~{Dimmelmeier}, and E.~{M{\"u}ller}.
\newblock {Axisymmetric simulations of magnetorotational core collapse:
  approximate inclusion of general relativistic effects}.
\newblock {\em \aap}, 457:209, 2006.

\bibitem{shibata:06}
M.~{Shibata}, Y.~T. {Liu}, S.~L. {Shapiro}, and B.~C. {Stephens}.
\newblock {Magnetorotational collapse of massive stellar cores to neutron
  stars: Simulations in full general relativity}.
\newblock {\em \prd}, 74(10):104026, 2006.

\bibitem{stergioulas:03}
N.~Stergioulas.
\newblock Rotating stars in relativity.
\newblock {\em Liv. Rev. Rel.}, 6:3, 2003.

\bibitem{centrella:01}
J.~M. {Centrella}, K.~C.~B. {New}, L.~L. {Lowe}, and J.~D. {Brown}.
\newblock {Dynamical Rotational Instability at Low T/W}.
\newblock {\em \apjl}, 550:L193, 2001.

\bibitem{saijo:03}
M.~{Saijo}, T.~W. {Baumgarte}, and S.~L. {Shapiro}.
\newblock {One-armed Spiral Instability in Differentially Rotating Stars}.
\newblock {\em \apj}, 595:352, 2003.

\bibitem{watts:05}
A.~L. {Watts}, N.~{Andersson}, and D.~I. {Jones}.
\newblock {The Nature of Low $T/|W|$ Dynamical Instabilities in Differentially
  Rotating Stars}.
\newblock {\em \apjl}, 618:L37, 2005.

\bibitem{rotinst:05}
C.~D. {Ott}, S.~{Ou}, J.~E. {Tohline}, and A.~{Burrows}.
\newblock {One-armed Spiral Instability in a Low-T/$|$W$|$ Postbounce Supernova
  Core}.
\newblock {\em Astrophys. J.}, 625:L119, 2005.

\bibitem{saijo:06}
M.~{Saijo} and S.~{Yoshida}.
\newblock {Low $T/|W|$ dynamical instability in differentially rotating stars:
  diagnosis with canonical angular momentum}.
\newblock {\em \mnras}, 368:1429, 2006.

\bibitem{scheidegger:08}
S.~{Scheidegger}, T.~{Fischer}, S.~C. {Whitehouse}, and M.~{Liebend{\"o}rfer}.
\newblock {Gravitational waves from 3D MHD core collapse simulations}.
\newblock {\em \aap}, 490:231, 2008.

\bibitem{corvino:10}
G.~{Corvino}, L.~{Rezzolla}, S.~{Bernuzzi}, R.~{De Pietri}, and
  B.~{Giacomazzo}.
\newblock {On the shear instability in relativistic neutron stars}.
\newblock {\em Class. Quantum Grav.}, 27(11):114104, 2010.

\bibitem{ott:11a}
C.~D. {Ott}, C.~{Reisswig}, E.~{Schnetter}, E.~{O'Connor}, U.~{Sperhake},
  F.~{L{\"o}ffler}, P.~{Diener}, E.~{Abdikamalov}, I.~{Hawke}, and
  A.~{Burrows}.
\newblock {Dynamics and Gravitational Wave Signature of Collapsar Formation}.
\newblock {\em Phys. Rev. Lett.}, 106(16):161103, 2011.

\bibitem{kiuchi:11}
K.~{Kiuchi}, M.~{Shibata}, P.~J. {Montero}, and J.~A. {Font}.
\newblock {Gravitational Waves from the Papaloizou-Pringle Instability in
  Black-Hole-Torus Systems}.
\newblock {\em Phys. Rev. Lett.}, 106:251102, 2011.

\bibitem{piro:07}
A.~L. {Piro} and E.~{Pfahl}.
\newblock {Fragmentation of Collapsar Disks and the Production of Gravitational
  Waves}.
\newblock {\em \apj}, 658:1173, 2007.

\bibitem{korobkin:11}
O.~{Korobkin}, E.~B. {Abdikamalov}, E.~{Schnetter}, N.~{Stergioulas}, and
  B.~{Zink}.
\newblock {Stability of general-relativistic accretion disks}.
\newblock {\em \prd}, 83(4):043007, 2011.

\bibitem{corsi:09}
A.~{Corsi} and P.~{M{\'e}sz{\'a}ros}.
\newblock {Gamma-ray Burst Afterglow Plateaus and Gravitational Waves:
  Multi-messenger Signature of a Millisecond Magnetar?}
\newblock {\em \apj}, 702:1171, 2009.

\bibitem{piro:11}
A.~L. {Piro} and C.~D. {Ott}.
\newblock {Supernova Fallback onto Magnetars and Propeller-powered Supernovae}.
\newblock {\em \apj}, 736:108, 2011.

\bibitem{1986ApJ...307..178B}
A.~{Burrows} and J.~M. {Lattimer}.
\newblock {The birth of neutron stars}.
\newblock {\em \apj}, 307:178, 1986.

\bibitem{1988PhR...163...51B}
A.~{Burrows} and J.~M. {Lattimer}.
\newblock {Convection, Type II supernovae, and the early evolution of neutron
  stars.}
\newblock {\em \physrep}, 163:51, 1988.

\bibitem{1999ApJ...513..780P}
J.~A. {Pons}, S.~{Reddy}, M.~{Prakash}, J.~M. {Lattimer}, and J.~A. {Miralles}.
\newblock {Evolution of Proto-Neutron Stars}.
\newblock {\em \apj}, 513:780, 1999.

\bibitem{2011PhRvD..84d4017B}
G.~F. {Burgio}, V.~{Ferrari}, L.~{Gualtieri}, and H.-J. {Schulze}.
\newblock {Oscillations of hot, young neutron stars: Gravitational wave
  frequencies and damping times}.
\newblock {\em \prd}, 84(4):044017, 2011.

\bibitem{2002PhRvD..66d1303G}
P.~{Gressman}, L.-M. {Lin}, W.-M. {Suen}, N.~{Stergioulas}, and J.~L.
  {Friedman}.
\newblock {Nonlinear r-modes in neutron stars: Instability of an unstable
  mode}.
\newblock {\em \prd}, 66(4):041303, 2002.

\bibitem{2007PhRvD..76f4019B}
R.~{Bondarescu}, S.~A. {Teukolsky}, and I.~{Wasserman}.
\newblock {Spin evolution of accreting neutron stars: Nonlinear development of
  the r-mode instability}.
\newblock {\em \prd}, 76(6):064019, 2007.

\bibitem{2009PhRvD..79j4003B}
R.~{Bondarescu}, S.~A. {Teukolsky}, and I.~{Wasserman}.
\newblock {Spinning down newborn neutron stars: Nonlinear development of the
  r-mode instability}.
\newblock {\em \prd}, 79(10):104003, 2009.

\bibitem{2010PhRvD..82j4036K}
W.~{Kastaun}, B.~{Willburger}, and K.~D. {Kokkotas}.
\newblock {Saturation amplitude of the f-mode instability}.
\newblock {\em \prd}, 82(10):104036, 2010.

\bibitem{2001IJMPD..10..381A}
N.~{Andersson} and K.~D. {Kokkotas}.
\newblock {The R-Mode Instability in Rotating Neutron Stars}.
\newblock {\em International Journal of Modern Physics D}, 10:381, 2001.

\bibitem{2011MNRAS.414.1679E}
C.~M. {Espinoza}, A.~G. {Lyne}, B.~W. {Stappers}, and M.~{Kramer}.
\newblock {A study of 315 glitches in the rotation of 102 pulsars}.
\newblock {\em \mnras}, 414:1679, 2011.

\bibitem{2010ApJ...713..671A}
B.~{Abbott} et~al.
\newblock {Searches for Gravitational Waves from Known Pulsars with Science Run
  5 LIGO Data}.
\newblock {\em \apj}, 713:671, 2010.

\bibitem{2008ApJ...683L..45A}
B.~{Abbott} et~al.
\newblock {Beating the Spin-Down Limit on Gravitational Wave Emission from the
  Crab Pulsar}.
\newblock {\em \apjl}, 683:L45, 2008.

\bibitem{2011ApJ...737...93A}
J.~{Abadie} et~al.
\newblock {Beating the Spin-down Limit on Gravitational Wave Emission from the
  Vela Pulsar}.
\newblock {\em \apj}, 737:93, 2011.

\bibitem{2010MNRAS.407L..54C}
A.~I. {Chugunov} and C.~J. {Horowitz}.
\newblock {Breaking stress of neutron star crust}.
\newblock {\em \mnras}, 407:L54, 2010.

\bibitem{1998ApJ...501L..89B}
L.~{Bildsten}.
\newblock {Gravitational Radiation and Rotation of Accreting Neutron Stars}.
\newblock {\em \apjl}, 501:L89, 1998.

\bibitem{2002MNRAS.337.1224A}
N.~{Andersson}, D.~I. {Jones}, and K.~D. {Kokkotas}.
\newblock {Strange stars as persistent sources of gravitational waves}.
\newblock {\em \mnras}, 337:1224, 2002.

\bibitem{2007PhRvD..76f2003A}
B.~{Abbott} et~al.
\newblock {Search for gravitational wave radiation associated with the
  pulsating tail of the SGR 1806-20 hyperflare of 27 December 2004 using LIGO}.
\newblock {\em \prd}, 76(6):062003, 2007.

\bibitem{2011ApJ...734L..35A}
J.~{Abadie} et~al.
\newblock {Search for Gravitational Wave Bursts from Six Magnetars}.
\newblock {\em \apjl}, 734:L35, 2011.

\bibitem{2011PhRvD..83d2001A}
J.~{Abadie} et~al.
\newblock {Search for gravitational waves associated with the August 2006
  timing glitch of the Vela pulsar}.
\newblock {\em \prd}, 83(4):042001, 2011.

\bibitem{2010ApJ...722.1504A}
J.~{Abadie} et~al.
\newblock {First Search for Gravitational Waves from the Youngest Known Neutron
  Star}.
\newblock {\em \apj}, 722:1504, 2010.

\bibitem{2008PhRvL.101u1102A}
B.~{Abbott} et~al.
\newblock {Search for Gravitational-Wave Bursts from Soft Gamma Repeaters}.
\newblock {\em Physical Review Letters}, 101(21):211102, 2008.

\bibitem{2010CQGra..27s4002P}
M.~{Punturo} et~al.
\newblock {The Einstein Telescope: a third-generation gravitational wave
  observatory}.
\newblock {\em Classical and Quantum Gravity}, 27(19):194002, 2010.

\bibitem{2008RvMA...20..140K}
K.~D. {Kokkotas}.
\newblock {Gravitational Wave Astronomy (With 2 Figures)}.
\newblock In {S.~R{\"o}ser}, editor, {\em Reviews in Modern Astronomy},
  volume~20 of {\em Reviews in Modern Astronomy}, page 140, 2008.

\bibitem{2011GReGr..43..409A}
N.~{Andersson}, V.~{Ferrari}, D.~I. {Jones}, K.~D. {Kokkotas}, B.~{Krishnan},
  J.~S. {Read}, L.~{Rezzolla}, and B.~{Zink}.
\newblock {Gravitational waves from neutron stars: promises and challenges}.
\newblock {\em General Relativity and Gravitation}, 43:409, 2011.

\bibitem{Prix:2011}
R.~{Prix}, S.~{Giampanis}, and C.~{Messenger}.
\newblock {Search method for long-duration gravitational-wave transients from
  neutron stars}.
\newblock {\em \prd}, 84(2):023007, July 2011.

\bibitem{Thrane:2011}
E.~{Thrane}, S.~{Kandhasamy}, C.~D. {Ott}, W.~G. {Anderson}, N.~L.
  {Christensen}, M.~W. {Coughlin}, S.~{Dorsher}, S.~{Giampanis}, V.~{Mandic},
  A.~{Mytidis}, T.~{Prestegard}, P.~{Raffai}, and B.~{Whiting}.
\newblock {Long gravitational-wave transients and associated detection
  strategies for a network of terrestrial interferometers}.
\newblock {\em \prd}, 83(8):083004, April 2011.

\bibitem{1970ApJ...161..561C}
S.~{Chandrasekhar}.
\newblock {The Effect of Gravitational Radiation on the Secular Stability of
  the Maclaurin Spheroid}.
\newblock {\em \apj}, 161:561, 1970.

\bibitem{1978ApJ...222..281F}
J.~L. {Friedman} and B.~F. {Schutz}.
\newblock {Secular instability of rotating Newtonian stars}.
\newblock {\em \apj}, 222:281, 1978.

\bibitem{2001ApJ...548..919S}
M.~{Saijo}, M.~{Shibata}, T.~W. {Baumgarte}, and S.~L. {Shapiro}.
\newblock {Dynamical Bar Instability in Rotating Stars: Effect of General
  Relativity}.
\newblock {\em \apj}, 548:919, 2001.

\bibitem{2007PhRvD..75d4023B}
L.~{Baiotti}, R.~{de Pietri}, G.~M. {Manca}, and L.~{Rezzolla}.
\newblock {Accurate simulations of the dynamical bar-mode instability in full
  general relativity}.
\newblock {\em \prd}, 75(4):044023, 2007.

\bibitem{2009ApJ...707.1610C}
K.~D. {Camarda}, P.~{Anninos}, P.~C. {Fragile}, and J.~A. {Font}.
\newblock {Dynamical Bar-Mode Instability in Differentially Rotating Magnetized
  Neutron Stars}.
\newblock {\em \apj}, 707:1610, 2009.

\bibitem{2001ApJ...550L.193C}
J.~M. {Centrella}, K.~C.~B. {New}, L.~L. {Lowe}, and J.~D. {Brown}.
\newblock {Dynamical Rotational Instability at Low T/W}.
\newblock {\em \apjl}, 550:L193, 2001.

\bibitem{2002MNRAS.334L..27S}
M.~{Shibata}, S.~{Karino}, and Y.~{Eriguchi}.
\newblock {Dynamical instability of differentially rotating stars}.
\newblock {\em \mnras}, 334:L27, 2002.

\bibitem{2005ApJ...618L..37W}
A.~L. {Watts}, N.~{Andersson}, and D.~I. {Jones}.
\newblock {The Nature of Low T/|W| Dynamical Instabilities in Differentially
  Rotating Stars}.
\newblock {\em \apjl}, 618:L37, 2005.

\bibitem{2007CoPhC.177..288C}
P.~{Cerd{\'a}-Dur{\'a}n}, V.~{Quilis}, and J.~A. {Font}.
\newblock {AMR simulations of the low T/|W| bar-mode instability of neutron
  stars}.
\newblock {\em Computer Physics Communications}, 177:288, 2007.

\bibitem{2010CQGra..27k4104C}
G.~{Corvino}, L.~{Rezzolla}, S.~{Bernuzzi}, R.~{De Pietri}, and
  B.~{Giacomazzo}.
\newblock {On the shear instability in relativistic neutron stars}.
\newblock {\em Classical and Quantum Gravity}, 27(11):114104, 2010.

\bibitem{1983PhRvL..51..718F}
J.~L. {Friedman}.
\newblock {Upper Limit on the Frequency of Fast Pulsars}.
\newblock {\em Physical Review Letters}, 51:718, 1983.

\bibitem{1991ApJ...373..213I}
J.~R. {Ipser} and L.~{Lindblom}.
\newblock {The oscillations of rapidly rotating Newtonian stellar models. II -
  Dissipative effects}.
\newblock {\em \apj}, 373:213, 1991.

\bibitem{1995ApJ...444..804L}
L.~{Lindblom} and G.~{Mendell}.
\newblock {Does gravitational radiation limit the angular velocities of
  superfluid neutron stars}.
\newblock {\em \apj}, 444:804, 1995.

\bibitem{1995ApJ...442..259L}
D.~{Lai} and S.~L. {Shapiro}.
\newblock {Gravitational radiation from rapidly rotating nascent neutron
  stars}.
\newblock {\em \apj}, 442:259, 1995.

\bibitem{Noutsos:2013}
A.~{Noutsos}, D.~H.~F.~M. {Schnitzeler}, E.~F. {Keane}, M.~{Kramer}, and
  S.~{Johnston}.
\newblock {Pulsar spin-velocity alignment: kinematic ages, birth periods and
  braking indices}.
\newblock {\em \mnras}, 430:2281--2301, April 2013.

\bibitem{2009PhRvD..79j3009A}
N.~{Andersson}, K.~{Glampedakis}, and B.~{Haskell}.
\newblock {Oscillations of dissipative superfluid neutron stars}.
\newblock {\em \prd}, 79(10):103009, 2009.

\bibitem{2008PhRvD..78f4063G}
E.~{Gaertig} and K.~D. {Kokkotas}.
\newblock {Oscillations of rapidly rotating relativistic stars}.
\newblock {\em \prd}, 78(6):064063, 2008.

\bibitem{PhysRevD.81.084055}
Burkhard Zink, Oleg Korobkin, Erik Schnetter, and Nikolaos Stergioulas.
\newblock Frequency band of the $f$-mode chandrasekhar-friedman-schutz
  instability.
\newblock {\em Phys. Rev. D}, 81(8):084055, 2010.

\bibitem{Gaertig:2011bm}
E.~Gaertig, K.~Glampedakis, K.~D. Kokkotas, and B.~Zink.
\newblock {The f-mode instability in relativistic neutron stars}.
\newblock {\em Phys.Rev.Lett.}, 107:101102, 2011.

\bibitem{2003CQGra..20R.105A}
N.~{Andersson}.
\newblock {TOPICAL REVIEW: Gravitational waves from instabilities in
  relativistic stars}.
\newblock {\em Classical and Quantum Gravity}, 20:105, 2003.

\bibitem{1999ApJ...516..307A}
N.~{Andersson}, K.~D. {Kokkotas}, and N.~{Stergioulas}.
\newblock {On the Relevance of the R-Mode Instability for Accreting Neutron
  Stars and White Dwarfs}.
\newblock {\em \apj}, 516:307, 1999.

\bibitem{2008MNRAS.389..839W}
A.~L. {Watts}, B.~{Krishnan}, L.~{Bildsten}, and B.~F. {Schutz}.
\newblock {Detecting gravitational wave emission from the known accreting
  neutron stars}.
\newblock {\em \mnras}, 389:839, 2008.

\bibitem{2000ApJ...531L.139R}
L.~{Rezzolla}, F.~K. {Lamb}, and S.~L. {Shapiro}.
\newblock {R-Mode Oscillations in Rotating Magnetic Neutron Stars}.
\newblock {\em \apjl}, 531:L139, 2000.

\bibitem{2011PhRvL.107j1101H}
W.~C.~G. {Ho}, N.~{Andersson}, and B.~{Haskell}.
\newblock {Revealing the Physics of r Modes in Low-Mass X-Ray Binaries}.
\newblock {\em Physical Review Letters}, 107(10):101101, 2011.

\bibitem{1999LRR.....2....2K}
K.~{Kokkotas} and B.~{Schmidt}.
\newblock {Quasi-Normal Modes of Stars and Black Holes}.
\newblock {\em Living Reviews in Relativity}, 2:2, 1999.

\bibitem{1996PhRvL..77.4134A}
N.~{Andersson} and K.~D. {Kokkotas}.
\newblock {Gravitational Waves and Pulsating Stars: What Can We Learn from
  Future Observations?}
\newblock {\em Physical Review Letters}, 77:4134, 1996.

\bibitem{1998MNRAS.299.1059A}
N.~{Andersson} and K.~D. {Kokkotas}.
\newblock {Towards gravitational wave asteroseismology}.
\newblock {\em \mnras}, 299:1059, 1998.

\bibitem{2004PhRvD..70l4015B}
O.~{Benhar}, V.~{Ferrari}, and L.~{Gualtieri}.
\newblock {Gravitational wave asteroseismology reexamined}.
\newblock {\em \prd}, 70(12):124015, 2004.

\bibitem{2011PhRvD..83f4031G}
E.~{Gaertig} and K.~D. {Kokkotas}.
\newblock {Gravitational wave asteroseismology with fast rotating neutron
  stars}.
\newblock {\em \prd}, 83(6):064031, 2011.

\bibitem{2011arXiv1105.4787P}
A.~{Passamonti} and N.~{Andersson}.
\newblock {Towards real neutron star seismology: Accounting for elasticity and
  superfluidity}.
\newblock {\em ArXiv e-prints}, May 2011.

\bibitem{2005ApJ...628L..53I}
G.~L. {Israel}, T.~{Belloni}, L.~{Stella}, Y.~{Rephaeli}, D.~E. {Gruber},
  P.~{Casella}, S.~{Dall'Osso}, N.~{Rea}, M.~{Persic}, and R.~E. {Rothschild}.
\newblock {The Discovery of Rapid X-Ray Oscillations in the Tail of the SGR
  1806-20 Hyperflare}.
\newblock {\em \apjl}, 628:L53, 2005.

\bibitem{2006ApJ...637L.117W}
A.~L. {Watts} and T.~E. {Strohmayer}.
\newblock {Detection with RHESSI of High-Frequency X-Ray Oscillations in the
  Tailof the 2004 Hyperflare from SGR 1806-20}.
\newblock {\em \apjl}, 637:L117, 2006.

\bibitem{2007AdSpR..40.1446W}
A.~L. {Watts} and T.~E. {Strohmayer}.
\newblock {Neutron star oscillations and QPOs during magnetar flares}.
\newblock {\em Advances in Space Research}, 40:1446, 2007.

\bibitem{2011A&A...528A..45H}
V.~{Hambaryan}, R.~{Neuh{\"a}user}, and K.~D. {Kokkotas}.
\newblock {Bayesian timing analysis of giant flare of SGR 180620 by RXTE PCA}.
\newblock {\em \aap}, 528:A45, 2011.

\bibitem{2008A&ARv..15..225M}
S.~{Mereghetti}.
\newblock {The strongest cosmic magnets: soft gamma-ray repeaters and anomalous
  X-ray pulsars}.
\newblock {\em Astron. Astrophys. Rev.}, 15:225, 2008.

\bibitem{2007MNRAS.374..256S}
L.~{Samuelsson} and N.~{Andersson}.
\newblock {Neutron star asteroseismology. Axial crust oscillations in the
  Cowling approximation}.
\newblock {\em \mnras}, 374:256, 2007.

\bibitem{2007MNRAS.375..261S}
H.~{Sotani}, K.~D. {Kokkotas}, and N.~{Stergioulas}.
\newblock {Torsional oscillations of relativistic stars with dipole magnetic
  fields}.
\newblock {\em \mnras}, 375:261, 2007.

\bibitem{2007MNRAS.377..159L}
Y.~{Levin}.
\newblock {On the theory of magnetar QPOs}.
\newblock {\em \mnras}, 377:159, 2007.

\bibitem{2008MNRAS.385L...5S}
H.~{Sotani}, K.~D. {Kokkotas}, and N.~{Stergioulas}.
\newblock {Alfv{\'e}n quasi-periodic oscillations in magnetars}.
\newblock {\em \mnras}, 385:L5, 2008.

\bibitem{2011MNRAS.412.1730L}
S.~K. {Lander} and D.~I. {Jones}.
\newblock {Oscillations and instabilities in neutron stars with poloidal
  magnetic fields}.
\newblock {\em \mnras}, 412:1730, 2011.

\bibitem{2009MNRAS.396..894A}
N.~{Andersson}, K.~{Glampedakis}, and L.~{Samuelsson}.
\newblock {Superfluid signatures in magnetar seismology}.
\newblock {\em \mnras}, 396:894, 2009.

\bibitem{2011MNRAS.410.1036V}
M.~{van Hoven} and Y.~{Levin}.
\newblock {Magnetar oscillations - I. Strongly coupled dynamics of the crust
  and the core}.
\newblock {\em \mnras}, 410:1036, 2011.

\bibitem{2011MNRAS.410L..37G}
M.~{Gabler}, P.~{Cerd{\'a} Dur{\'a}n}, J.~A. {Font}, E.~{M{\"u}ller}, and
  N.~{Stergioulas}.
\newblock {Magneto-elastic oscillations and the damping of crustal shear modes
  in magnetars}.
\newblock {\em \mnras}, 410:L37, 2011.

\bibitem{2011MNRAS.414.3014C}
A.~{Colaiuda} and K.~D. {Kokkotas}.
\newblock {Magnetar oscillations in the presence of a crust}.
\newblock {\em \mnras}, 414:3014, 2011.

\bibitem{2001MNRAS.327..639I}
K.~{Ioka}.
\newblock {Magnetic deformation of magnetars for the giant flares of the soft
  gamma-ray repeaters}.
\newblock {\em \mnras}, 327:639, 2001.

\bibitem{2011PhRvD..83h1302K}
K.~{Kashiyama} and K.~{Ioka}.
\newblock {Magnetar asteroseismology with long-term gravitational waves}.
\newblock {\em \prd}, 83(8):081302, 2011.

\bibitem{2011PhRvD..83j4014C}
A.~{Corsi} and B.~J. {Owen}.
\newblock {Maximum gravitational-wave energy emissible in magnetar flares}.
\newblock {\em \prd}, 83(10):104014, 2011.

\bibitem{2011arXiv1103.0880L}
Y.~{Levin} and M.~{van Hoven}.
\newblock {On the excitation of f-modes and torsional modes by magnetar giant
  flares}.
\newblock {\em ArXiv e-prints}, March 2011.

\bibitem{2011ApJ...735L..20L}
P.~D. {Lasky}, B.~{Zink}, K.~D. {Kokkotas}, and K.~{Glampedakis}.
\newblock {Hydromagnetic Instabilities in Relativistic Neutron Stars}.
\newblock {\em \apjl}, 735:L20, 2011.

\bibitem{2012PhRvD..85b4030Z}
B.~{Zink}, P.~D. {Lasky}, and K.~D. {Kokkotas}.
\newblock {Are gravitational waves from giant magnetar flares observable?}
\newblock {\em \prd}, 85(2):024030, January 2012.

\bibitem{2011ApJ...736L...6C}
R.~{Ciolfi}, S.~K. {Lander}, G.~M. {Manca}, and L.~{Rezzolla}.
\newblock {Instability-driven Evolution of Poloidal Magnetic Fields in
  Relativistic Stars}.
\newblock {\em \apjl}, 736:L6, 2011.

\bibitem{2010MNRAS.405.1061S}
T.~{Sidery}, A.~{Passamonti}, and N.~{Andersson}.
\newblock {The dynamics of pulsar glitches: contrasting phenomenology with
  numerical evolutions}.
\newblock {\em \mnras}, 405:1061, 2010.

\bibitem{2008CQGra..25v5020V}
C.~A. {van Eysden} and A.~{Melatos}.
\newblock {Gravitational radiation from pulsar glitches}.
\newblock {\em Classical and Quantum Gravity}, 25(22):225020, 2008.

\bibitem{2008PhRvD..77b1502F}
{\'E}.~{\'E}. {Flanagan} and T.~{Hinderer}.
\newblock {Constraining neutron-star tidal Love numbers with gravitational-wave
  detectors}.
\newblock {\em \prd}, 77(2):021502, 2008.

\bibitem{2008ApJ...677.1216H}
T.~{Hinderer}.
\newblock {Tidal Love Numbers of Neutron Stars}.
\newblock {\em \apj}, 677:1216, 2008.

\bibitem{2011arXiv1109.5041P}
A.~J. {Penner}, N.~{Andersson}, D.~I. {Jones}, L~{Samuelsson}, and I.~{Hawke}.
\newblock {Crustal failure during binary inspiral}.
\newblock {\em ArXiv e-prints}, September 2011.

\bibitem{2011arXiv1110.0467T}
D.~{Tsang}, J.~S. {Read}, T.~{Hinderer}, A.~L. {Piro}, and R.~{Bondarescu}.
\newblock {Resonant Shattering of Neutron Star Crusts}.
\newblock {\em ArXiv e-prints}, October 2011.

\bibitem{S5range}
J.~Abadie et~al.
\newblock {Sensitivity to Gravitational Waves from Compact Binary Coalescences
  Achieved during LIGO's Fifth and Virgo's First Science Run}, 2010.
\newblock arXiv:1003.2481.

\bibitem{S6cbcLowMass}
J.~Abadie et~al.
\newblock {Search for Gravitational Waves from Low Mass Compact Binary
  Coalescence in LIGO's Sixth Science Run and Virgo's Science Runs 2 and 3}.
\newblock {\em Phys.Rev.}, D85:082002, 2012.

\bibitem{ihope}
S.~{Babak} et~al.
\newblock {Searching for gravitational waves from binary coalescence}.
\newblock {\em ArXiv e-prints}, August 2012.

\bibitem{PhysRevD.83.102001}
S.~Klimenko et~al.
\newblock Localization of gravitational wave sources with networks of advanced
  detectors.
\newblock {\em \prd}, 83:102001, 2011.

\bibitem{ShutzNET}
B.~F. Schutz.
\newblock Networks of gravitational wave detectors and three figures of merit.
\newblock {\em Classical and Quantum Gravity}, 28(12):125023, 2011.

\bibitem{2009CQGra..26t4020A}
K.~{Arai} et~al.
\newblock {Status of Japanese gravitational wave detectors}.
\newblock {\em Classical and Quantum Gravity}, 26(20):204020, 2009.

\bibitem{ObservingScenarios}
J.~Aasi et~al.
\newblock {Prospects for Localization of Gravitational Wave Transients by the
  Advanced LIGO and Advanced Virgo Observatories}, 2013.
\newblock LIGO-DCC-P1200087, arXiv:1304.0670.

\bibitem{Fa:09}
S.~Fairhurst.
\newblock Triangulation of gravitational wave sources with a network of
  detectors.
\newblock {\em NJP}, 11:123006, 2009.

\bibitem{PhysRevD.81.082001}
L.~Wen and Y.~Chen.
\newblock Geometrical expression for the angular resolution of a network of
  gravitational-wave detectors.
\newblock {\em Phys. Rev. D}, 81:082001, 2010.

\bibitem{2011PhRvD..83j2001K}
S.~{Klimenko} et~al.
\newblock {Localization of gravitational wave sources with networks of advanced
  detectors}.
\newblock {\em \prd}, 83(10):102001, 2011.

\bibitem{2011ApJ...739...99N}
S.~{Nissanke} et~al.
\newblock {Localizing Compact Binary Inspirals on the Sky Using Ground-based
  Gravitational Wave Interferometers}.
\newblock {\em \apj}, 739:99, 2011.

\bibitem{Veitch:2012}
J.~{Veitch} et~al.
\newblock {Estimating parameters of coalescing compact binaries with proposed
  advanced detector networks}.
\newblock {\em \prd}, 85:104045, 2012.

\bibitem{followupS6}
{LIGO Scientific Collaboration}, {Virgo Collaboration}, J.~{Abadie}, B.~P.
  {Abbott}, R.~{Abbott}, T.~D. {Abbott}, M.~{Abernathy}, T.~{Accadia},
  F.~{Acernese}, C.~{Adams}, and et~al.
\newblock {Implementation and testing of the first prompt search for
  gravitational wave transients with electromagnetic counterparts}.
\newblock {\em \aap}, 539:A124, 2012.

\bibitem{SwiftS6}
P.~A. {Evans}, J.~K. {Fridriksson}, N.~{Gehrels}, J.~{Homan}, J.~P. {Osborne},
  M.~{Siegel}, A.~{Beardmore}, P.~{Handbauer}, J.~{Gelbord}, J.~A. {Kennea},
  and et~al.
\newblock {Swift Follow-up Observations of Candidate Gravitational-wave
  Transient Events}.
\newblock {\em \apjs}, 203:28, 2012.

\bibitem{Bloom:2009}
J.~S. {Bloom} et~al.
\newblock {Astro2010 Decadal Survey Whitepaper: Coordinated Science in the
  Gravitational and Electromagnetic Skies}.
\newblock {\em ArXiv e-prints}, February 2009.
\newblock 0902.1527.

\bibitem{Mandel:2011}
I.~{Mandel}, L.~Z. {Kelley}, and E.~{Ramirez-Ruiz}.
\newblock {Towards improving the prospects for coordinated gravitational-wave
  and electromagnetic observations}.
\newblock {\em ArXiv e-prints}, October 2011.

\bibitem{BulikBelczynski:2003}
T.~{Bulik} and K.~{Belczy{\'n}ski}.
\newblock {Constraints on the Binary Evolution from Chirp Mass Measurements}.
\newblock {\em \apjl}, 589:L37, 2003.

\bibitem{Yunes:2011}
N.~{Yunes}.
\newblock {Gravitational Waves from Compact Binaries as Probes of the
  Universe}.
\newblock {\em ArXiv e-prints}, December 2011.

\bibitem{Gerosa2013}
D.~{Gerosa}, M.~{Kesden}, E.~{Berti}, R.~{O'Shaughnessy}, and U.~{Sperhake}.
\newblock {Resonant-plane locking and spin alignment in stellar-mass black-hole
  binaries: a diagnostic of compact-binary formation}.
\newblock {\em ArXiv e-prints}, February 2013.

\bibitem{0264-9381-27-8-084007}
M.~Punturo et~al.
\newblock {The third generation of gravitational wave observatories and their
  science reach}.
\newblock {\em Classical and Quantum Gravity}, 27(8):084007, 2010.

\bibitem{2011GReGr..43..437C}
E.~{Chassande-Mottin} et~al.
\newblock {Multimessenger astronomy with the Einstein Telescope}.
\newblock {\em General Relativity and Gravitation}, 43:437, 2011.

\bibitem{2011CQGra..28i4013H}
S.~{Hild} et~al.
\newblock {Sensitivity studies for third-generation gravitational wave
  observatories}.
\newblock {\em Classical and Quantum Gravity}, 28(9):094013, 2011.

\bibitem{2010CQGra..27u5006S}
B.~S. {Sathyaprakash}, B.~F. {Schutz}, and C.~{Van Den Broeck}.
\newblock {Cosmography with the Einstein Telescope}.
\newblock {\em Classical and Quantum Gravity}, 27(21):215006, 2010.

\bibitem{1997asxo.proc..279K}
N.~{Kawai} et~al.
\newblock {X-ray All-Sky Monitor on JEM of the Space Station - MAXI (Monitor of
  All-sky X-ray Image)}.
\newblock In {M.~Matsuoka \& N.~Kawai}, editor, {\em All-Sky X-Ray Observations
  in the Next Decade}, page 279, 1997.

\bibitem{agile}
{AGILE}.
\newblock \url{http://agile.asdc.asi.it/}.

\bibitem{fermi}
{FERMI}.
\newblock \url{http://fermi.gsfc.nasa.gov/}.

\bibitem{integral}
{INTEGRAL}.
\newblock \url{http://www.rssd.esa.int/index.php?project=INTEGRAL&page=index}.

\bibitem{swift}
{SWIFT}.
\newblock \url{http://swift.gsfc.nasa.gov/}.

\bibitem{chandra}
{CHANDRA}.
\newblock \url{http://chandra.harvard.edu/}.

\bibitem{xmmnewton}
{XMM-Newton}.
\newblock \url{http://heasarc.gsfc.nasa.gov/docs/xmm/xmmgof.html}.

\bibitem{suzaku}
{SUZAKU}.
\newblock \url{http://www.astro.isas.jaxa.jp/suzaku/index.html.en/}.

\bibitem{nustar}
{NUSTAR}.
\newblock \url{http://www.nustar.caltech.edu/}.

\bibitem{maxi}
{MAXI}.
\newblock \url{http://maxi.riken.jp}.

\bibitem{2006AdSpR..38.2989A}
P.~C. {Agrawal}.
\newblock {A broad spectral band Indian Astronomy satellite {\em Astrosat}}.
\newblock {\em Advances in Space Research}, 38:2989, 2006.

\bibitem{2009AcAau..65....6K}
V.~{Koteswara Rao} et~al.
\newblock {The scientific objectives of the ASTROSAT mission of ISRO}.
\newblock {\em Acta Astronautica}, 65:6, 2009.

\bibitem{astrosat}
{ASTROSAT}.
\newblock \url{http://astrosat.iucaa.in}.

\bibitem{Takahashi:2012jn}
T.~Takahashi et~al.
\newblock {The ASTRO-H X-ray Observatory}, 2012.
\newblock arXiv:1210.4378.

\bibitem{Kawai2010}
N.~{Kawai} and D.~{Yonetoku}.
\newblock {Observing Gamma-Ray Bursts with ASTRO-H}.
\newblock In N.~{Kawai} and S.~{Nagataki}, editors, {\em American Institute of
  Physics Conference Series}, volume 1279 of {\em American Institute of Physics
  Conference Series}, page 204, October 2010.

\bibitem{2011CRPhy..12..298P}
J.~{Paul} et~al.
\newblock {The Chinese--French SVOM mission for gamma-ray burst studies}.
\newblock {\em Comptes Rendus Physique}, 12:298, 2011.

\bibitem{2010arXiv1005.5008S}
S.~{Schanne} et~al.
\newblock {The future Gamma-Ray Burst Mission SVOM}.
\newblock {\em ArXiv e-prints}, May 2010.

\bibitem{svom}
{SVOM}.
\newblock \url{http://smsc.cnes.fr/SVOM/index.htm}.

\bibitem{heasarc}
{Concepts for Future High-Energy Astrophysics Missions}.
\newblock
  \url{http://heasarc.gsfc.nasa.gov/docs/heasarc/missions/concepts.html}.

\bibitem{xraymissionsconceptstudy}
R.~{Petre} et~al.
\newblock {The NASA X-ray Mission concepts study}.
\newblock In {\em Society of Photo-Optical Instrumentation Engineers (SPIE)
  Conference Series}, volume 8443 of {\em Society of Photo-Optical
  Instrumentation Engineers (SPIE) Conference Series}, September 2012.

\bibitem{ISSlobster}
J. Camp, personal communication.

\bibitem{2003ApJ...591.1152S}
J.~{Sylvestre}.
\newblock {Prospects for the Detection of Electromagnetic Counterparts to
  Gravitational Wave Events}.
\newblock {\em \apj}, 591:1152, 2003.

\bibitem{2008CQGra..25r4033S}
C.~W. {Stubbs}.
\newblock {Linking optical and infrared observations with gravitational wave
  sources through transient variability}.
\newblock {\em Classical and Quantum Gravity}, 25(18):184033, 2008.

\bibitem{2009arXiv0902.1527B}
J.~S. {Bloom} et~al.
\newblock {Astro2010 Decadal Survey Whitepaper: Coordinated Science in the
  Gravitational and Electromagnetic Skies}, February 2009.
\newblock arXiv:0902.1527.

\bibitem{2009astro2010S.235P}
E.~S. {Phinney}.
\newblock {Finding and Using Electromagnetic Counterparts of Gravitational Wave
  Sources}.
\newblock In {\em astro2010: The Astronomy and Astrophysics Decadal Survey},
  volume 2010 of {\em Astronomy}, page 235, 2009.

\bibitem{2009aaxo.conf..312K}
S.~{Kulkarni} and M.~M. {Kasliwal}.
\newblock {Transients in the Local Universe}.
\newblock In {N.~Kawai, T.~Mihara, M.~Kohama, \& M.~Suzuki}, editor, {\em
  Astrophysics with All-Sky X-Ray Observations}, page 312, March 2009.

\bibitem{lofar}
{LOFAR}.
\newblock \url{http://www.lofar.org/}.

\bibitem{evla}
{EVLA}.
\newblock \url{http://www.nrao.edu/index.php/about/facilities/vlaevla}.

\bibitem{askap}
{ASKAP}.
\newblock \url{http://www.atnf.csiro.au/projects/askap/index.html}.

\bibitem{ska}
{SKA}.
\newblock \url{http://www.skatelescope.org/}.

\bibitem{2013MNRAS.tmp.1023M}
M.~V. {Medvedev} and A.~{Loeb}.
\newblock {On Poynting-flux-driven bubbles and shocks around merging neutron
  star binaries}.
\newblock {\em \mnras}, March 2013.

\bibitem{2011Natur.478...82N}
E.~{Nakar} and T.~{Piran}.
\newblock {Detectable radio flares following gravitational waves from mergers
  of binary neutron stars}.
\newblock {\em \nat}, 478:82, 2011.

\bibitem{MetzgerBerger2012}
B.~D. {Metzger} and E.~{Berger}.
\newblock {What is the Most Promising Electromagnetic Counterpart of a Neutron
  Star Binary Merger?}
\newblock {\em \apj}, 746:48, 2012.

\bibitem{PiranNakarRosswog2013}
T.~{Piran}, E.~{Nakar}, and S.~{Rosswog}.
\newblock {The electromagnetic signals of compact binary mergers}.
\newblock {\em \mnras}, page 771, 2013.

\bibitem{2012arXiv1212.2289B}
I.~{Bartos}, P.~{Brady}, and S.~{Marka}.
\newblock {How Gravitational-wave Observations Can Shape the Gamma-ray Burst
  Paradigm}.
\newblock {\em ArXiv e-prints}, December 2012.

\bibitem{Samaya2012}
S.~{Nissanke}, M.~{Kasliwal}, and A.~{Georgieva}.
\newblock {Identifying Elusive Electromagnetic Counterparts to Gravitational
  Wave Mergers: an end-to-end simulation}.
\newblock {\em ArXiv e-prints}, October 2012.

\bibitem{Kelley:2012}
L.~Z. {Kelley}, I.~{Mandel}, and E.~{Ramirez-Ruiz}.
\newblock {Electromagnetic transients as triggers in searches for gravitational
  waves from compact binary mergers}.
\newblock {\em ArXiv e-prints}, September 2012.

\bibitem{2013arXiv1303.5788K}
D.~{Kasen}, N.~R. {Badnell}, and J.~{Barnes}.
\newblock {Opacities and Spectra of the r-process Ejecta from Neutron Star
  Mergers}.
\newblock {\em ArXiv e-prints}, March 2013.

\bibitem{Bionta:1987qt}
R.~M. Bionta et~al.
\newblock {Observation of a neutrino burst in coincidence with supernova 1987A
  in the Large Magellanic Cloud}.
\newblock {\em \prl}, 58:1494, 1987.

\bibitem{Hirata:1987hu}
K.~Hirata et~al.
\newblock {Observation of a neutrino burst from the supernova SN1987A}.
\newblock {\em \prl}, 58:1490, 1987.

\bibitem{Alekseev:1987ej}
E.N. Alexeyev, L.N. Alexeyeva, V.I. Volchenko, and I.V. Krivosheina.
\newblock {Possible detection of a neutrino signal on 23 February 1987 at the
  Baksan underground scintillation telescope of the Institute of Nuclear
  Research}.
\newblock {\em JETP Lett.}, 45:589, 1987.

\bibitem{Aglietta:1987it}
M.~Aglietta, G.~Badino, G.~Bologna, C.~Castagnoli, A.~Castellina, et~al.
\newblock {On the event observed in the Mont Blanc Underground Neutrino
  observatory during the occurrence of Supernova 1987a}.
\newblock {\em Europhys.Lett.}, 3:1315, 1987.

\bibitem{bahcall1989neutrino}
J.N. Bahcall.
\newblock {\em {Neutrino Astrophysics}}.
\newblock Cambridge University Press, 1989.

\bibitem{Pagliaroli:2008ur}
G.~Pagliaroli, F.~Vissani, E.~Coccia, and W.~Fulgione.
\newblock {Neutrinos from Supernovae as a Trigger for Gravitational Wave
  Search}.
\newblock {\em \prl}, 103:031102, 2009.

\bibitem{Scholberg:2012id}
Kate Scholberg.
\newblock {Supernova Neutrino Detection}.
\newblock {\em Ann.Rev.Nucl.Part.Sci.}, 62:81, 2012.

\bibitem{fulgione10:status}
W.~Fulgione.
\newblock {Status of supernova neutrino detectors}.
\newblock {\em J. Phys.: Conf. Ser}, 203:012077, 2010.

\bibitem{Fukuda2003418}
S.~Fukuda et~al.
\newblock The super-kamiokande detector.
\newblock {\em Nuclear Instruments and Methods in Physics Research Section A:
  Accelerators, Spectrometers, Detectors and Associated Equipment},
  501(2-3):418, 2003.

\bibitem{ikeda:07}
M.~Ikeda et~al.
\newblock {Search for Supernova Neutrino Bursts at Super-Kamiokande}.
\newblock {\em The Astrophysical Journal}, 669:519, 2007.

\bibitem{LVD2006MSAIS...9..388F}
W.~{for the LVD Collaboration} {Fulgione}.
\newblock {The LVD Neutrino Observatory .}
\newblock {\em Memorie della Societa Astronomica Italiana Supplement}, 9:388,
  2006.

\bibitem{Agafonova:2007hn}
N.~Yu. Agafonova et~al.
\newblock {On-line recognition of supernova neutrino bursts in the LVD
  detector}.
\newblock {\em Astropart. Phys.}, 28:516, 2008.

\bibitem{borexino}
G.~{Alimonti} et~al.
\newblock {The Borexino detector at the Laboratori Nazionali del Gran Sasso}.
\newblock {\em Nuclear Instruments and Methods in Physics Research A}, 600:568,
  2009.

\bibitem{Cadonati:2000kq}
L.~Cadonati, F.~P. Calaprice, and M.~C. Chen.
\newblock {Supernova neutrino detection in Borexino}.
\newblock {\em Astropart. Phys.}, 16:361, 2002.

\bibitem{kamlandweb}
Kamland.
\newblock \url{http://www.awa.tohoku.ac.jp/KamLAND/}.

\bibitem{Piepke:2001tg}
A.~Piepke.
\newblock {KamLAND: A reactor neutrino experiment testing the solar neutrino
  anomaly}.
\newblock {\em Nucl.Phys.Proc.Suppl.}, 91:99, 2001.

\bibitem{snoplusweb}
Sno+.
\newblock \url{http://snoplus.phy.queensu.ca/Home.html}.

\bibitem{IceCubeAhrens2004507}
J~Ahrens et~al.
\newblock Sensitivity of the icecube detector to astrophysical sources of high
  energy muon neutrinos.
\newblock {\em Astroparticle Physics}, 20(5):507, 2004.

\bibitem{Halzen:1995ex}
F.~Halzen, J.~E. Jacobsen, and E.~Zas.
\newblock {Ultra-Transparent Antarctic Ice as a Supernova Detector}.
\newblock {\em Phys. Rev.}, D53:7359, 1996.

\bibitem{Antonioli:2004zb}
P.~Antonioli et~al.
\newblock {SNEWS: The SuperNova Early Warning System}.
\newblock {\em New J. Phys.}, 6:114, 2004.

\bibitem{Scholberg:2008fa}
K.~Scholberg.
\newblock {The SuperNova Early Warning System}.
\newblock {\em Astron. Nachr.}, 329:337, 2008.

\bibitem{Sutton:10}
P.~J. Sutton.
\newblock A rule of thumb for the detectability of gravitational-wave bursts.
\newblock Technical Report LIGO-P1000041-v1, LIGO, 2010.
\newblock
  \url{https://dcc.ligo.org/cgi-bin/private/DocDB/ShowDocument?docid=10614}.

\bibitem{Ott2008}
C.~D. Ott.
\newblock {The Gravitational Wave Signature of Core-Collapse Supernovae}.
\newblock {\em Class. Quantum Grav.}, 26:063001, 2009.
\newblock arXiv:0809.0695.

\bibitem{2010JPhCS.203a2077F}
W.~{Fulgione}.
\newblock {Status of supernova neutrino detectors}.
\newblock {\em Journal of Physics Conference Series}, 203(1):012077, 2010.

\bibitem{2009CQGra..26t4015O}
C.~D. {Ott}.
\newblock {Probing the core-collapse supernova mechanism with gravitational
  waves}.
\newblock {\em Classical and Quantum Gravity}, 26(20):204015, 2009.

\bibitem{2004APh....21..201A}
N.~{Arnaud} et~al.
\newblock {Detection of a close supernova gravitational wave burst in a network
  of interferometers, neutrino and optical detectors}.
\newblock {\em Astroparticle Physics}, 21:201, 2004.

\bibitem{Leonor:2010}
I.~Leonor et~al.
\newblock {Searching for prompt signatures of nearby core-collapse supernovae
  by a joint analysis of neutrino and gravitational-wave data}.
\newblock {\em Class. Quant. Grav.}, 27:084019, 2010.

\bibitem{2009PhRvL.103c1102P}
G.~{Pagliaroli} et~al.
\newblock {Neutrinos from Supernovae as a Trigger for Gravitational Wave
  Search}.
\newblock {\em Physical Review Letters}, 103(3):031102, 2009.

\bibitem{2008arXiv0804.1500S}
C.~{Spiering}.
\newblock {Status and Perspectives of Astroparticle Physics in Europe}, April
  2008.
\newblock arXiv: 0804.1500.

\bibitem{autiero-2007-0711}
D.~Autiero et~al.
\newblock Large underground, liquid based detectors for astro-particle physics
  in europe: scientific case and prospects.
\newblock {\em JCAP}, 0711:011, 2007.

\bibitem{Kistler:2008us}
M.~D. Kistler et~al.
\newblock {Core-Collapse Astrophysics with a Five-Megaton Neutrino Detector},
  2008.
\newblock arXiv:0810.1959.

\bibitem{PhysRevLett.95.171101}
S.~Ando, J.~F. Beacom, and H.~Y\"uksel.
\newblock Detection of neutrinos from supernovae in nearby galaxies.
\newblock {\em Phys. Rev. Lett.}, 95(17):171101, 2005.

\bibitem{debellefon-2006}
A.~{de Bellefon} et~al.
\newblock {MEMPHYS}: A large scale water {C}erenkov detector at {F}r\'ejus,
  2006.
\newblock arxiv: hep-ex/0607026.

\bibitem{LENA}
M.~{Wurm} et~al.
\newblock {The next-generation liquid-scintillator neutrino observatory LENA}.
\newblock {\em ArXiv e-prints}, April 2011.

\bibitem{1742-6596-171-1-012020}
A.~Rubbia.
\newblock Underground neutrino detectors for particle and astroparticle
  science: The giant liquid argon charge imaging experiment (glacier).
\newblock {\em Journal of Physics: Conference Series}, 171(1):012020, 2009.

\bibitem{2010JPhCS.203a2109M}
J.~{Maricic} and {the Lbne Dusel collaboration}.
\newblock {The Long Baseline Neutrino Oscillation Experiment at DUSEL}.
\newblock {\em Journal of Physics Conference Series}, 203(1):012109, 2010.

\bibitem{nakamura03}
K.~{Nakamura}.
\newblock {Hyper-Kamiokande - a Next Generation Water Cherenkov Detector}.
\newblock In R.~{Shrock}, editor, {\em Neutrinos and Implications for Physics
  Beyond the Standard Model}, page 307. World Scientific, 2003.

\bibitem{Suzuki:2001rb}
Y.~{Suzuki}.
\newblock {Multi-Megaton Water Cherenkov Detector for a Proton Decay Search --
  TITAND}, October 2001.
\newblock arXiv:hep-ex/0110005.

\bibitem{2013arXiv1304.2553B}
S.~{B{\"o}ser}, M.~{Kowalski}, L.~{Schulte}, N.~{Linn Strotjohann}, and
  M.~{Voge}.
\newblock {Detecting extra-galactic supernova neutrinos in the Antarctic ice}.
\newblock {\em ArXiv e-prints}, April 2013.

\bibitem{2012PhRvD..86h3007B}
I.~{Bartos}, B.~{Dasgupta}, and S.~{M{\'a}rka}.
\newblock {Probing the structure of jet-driven core-collapse supernova and long
  gamma-ray burst progenitors with high-energy neutrinos}.
\newblock {\em \prd}, 86(8):083007, 2012.

\bibitem{2012APh....35..615A}
R.~{Abbasi} et~al.
\newblock {The design and performance of IceCube DeepCore}.
\newblock {\em Astroparticle Physics}, 35:615, 2012.

\bibitem{2013arXiv1301.4232B}
I.~{Bartos} et~al.
\newblock {Detection Prospects for GeV Neutrinos from Collisionally Heated
  Gamma-ray Bursts with IceCube/DeepCore}.
\newblock {\em ArXiv e-prints}, January 2013.

\bibitem{ANTARES}
M.~{Ageron} et~al.
\newblock {ANTARES: The first undersea neutrino telescope}.
\newblock {\em Nuclear Instruments and Methods in Physics Research A}, 656:11,
  2011.

\bibitem{deJong2010445}
M.~de~Jong.
\newblock {The KM3NeT neutrino telescope}.
\newblock {\em Nucl. Instrum. Methods Phys. Res., Sect. A}, 623(1):445, 2010.

\bibitem{Avrorin2011S13}
A.~Avrorin et~al.
\newblock The baikal neutrino experiment.
\newblock {\em {Nucl. Instrum. Methods Phys. Res., Sect. A}},
  626-627(Supplement 1):S13, 2011.

\bibitem{waxmanbachall}
E.~Waxman and J.~Bahcall.
\newblock High energy neutrinos from cosmological gamma-ray burst fireballs.
\newblock {\em Physical Review Letters}, 78:2292, 1997.

\bibitem{NeutrinoBATSEGuetta2004429}
D.~{Guetta} et~al.
\newblock {Neutrinos from individual gamma-ray bursts in the BATSE catalog}.
\newblock {\em Astropart. Phys.}, 20(4):429, 2004.

\bibitem{2004APh....20..507A}
J.~{Ahrens} et~al.
\newblock {Sensitivity of the IceCube detector to astrophysical sources of high
  energy muon neutrinos}.
\newblock {\em Astroparticle Physics}, 20:507, 2004.

\bibitem{2012Natur.484..351A}
R.~{Abbasi} et~al.
\newblock {An absence of neutrinos associated with cosmic-ray acceleration in
  {$\gamma$}-ray bursts}.
\newblock {\em \nat}, 484:351, 2012.

\bibitem{PhysRevLett.108.231101}
S.~H\"ummer, P.~Baerwald, and W.~Winter.
\newblock Neutrino emission from gamma-ray burst fireballs, revised.
\newblock {\em Phys. Rev. Lett.}, 108:231101, 2012.

\bibitem{2012ApJ...752...29H}
H.-N. {He} et~al.
\newblock {Icecube Nondetection of Gamma-Ray Bursts: Constraints on the
  Fireball Properties}.
\newblock {\em \apj}, 752:29, 2012.

\bibitem{HENAndoPhysRevLett.95.061103}
S.~Ando and J.~F. Beacom.
\newblock {Revealing the Supernova--Gamma-Ray Burst Connection with TeV
  Neutrinos}.
\newblock {\em Phys. Rev. Lett.}, 95(6):061103, 2005.

\bibitem{chokedfromreverseshockPhysRevD.77.063007}
S.~Horiuchi and S.~Ando.
\newblock {High-energy neutrinos from reverse shocks in choked and successful
  relativistic jets}.
\newblock {\em \prd}, 77(6):063007, 2008.

\bibitem{HeneventratePhysRevLett.93.181101}
Soebur Razzaque, Peter M\'esz\'aros, and Eli Waxman.
\newblock {TeV Neutrinos from Core Collapse Supernovae and Hypernovae}.
\newblock {\em Phys. Rev. Lett.}, 93(18):181101, 2004.
\newblock Erratum: Phys. Rev. Lett. 94, 109903 (2005).

\bibitem{PhysRevLett.107.251101}
I.~Bartos et~al.
\newblock Observational constraints on multimessenger sources of gravitational
  waves and high-energy neutrinos.
\newblock {\em \prl}, 107:251101, 2011.

\bibitem{LVCantares}
S.~Adrian-Martinez et~al.
\newblock {A First Search for coincident Gravitational Waves and High Energy
  Neutrinos using LIGO, Virgo and ANTARES data from 2007}, 2012.
\newblock arXiv:1205.3018.

\bibitem{S5GRB070201}
B.~Abbott and for the LIGO~Scientific Collaboration.
\newblock {Implications for the Origin of GRB 070201 from LIGO Observations}.
\newblock {\em Astrophys. J.}, 681:1419, 2008.

\bibitem{GRB051103}
J.~Abadie et~al.
\newblock {Implications For The Origin Of GRB 051103 From LIGO Observations}.
\newblock {\em Astrophys. J.}, 755:2, 2012.

\bibitem{6magnetars}
J.~Abadie et~al.
\newblock {Search for Gravitational Wave Bursts from Six Magnetars}.
\newblock {\em Astrophys. J.}, 734:L35, 2011.

\bibitem{VelaGlitch}
J.~Abadie et~al.
\newblock {A search for gravitational waves associated with the August 2006
  timing glitch of the Vela pulsar}.
\newblock {\em Phys.Rev.}, D83:042001, 2011.

\bibitem{S6GRB}
M.S. Briggs et~al.
\newblock {Search for gravitational waves associated with gamma-ray bursts
  during LIGO science run 6 and Virgo science runs 2 and 3}.
\newblock {\em Astrophys.J.}, 760:12, 2012.

\bibitem{S6cbcHighMass}
J.~Aasi et~al.
\newblock {Search for Gravitational Waves from Binary Black Hole Inspiral,
  Merger and Ringdown in LIGO-Virgo Data from 2009-2010}.
\newblock {\em Phys. Rev. D}, 87:022002, 2013.

\bibitem{S6IMBH}
J.~Abadie et~al.
\newblock {Search for Gravitational Waves from Intermediate Mass Binary Black
  Holes}.
\newblock {\em Phys.Rev.}, D85:102004, 2012.

\bibitem{S6burstAllSky}
J.~Abadie et~al.
\newblock {All-sky search for gravitational-wave bursts in the second joint
  LIGO-Virgo run}.
\newblock {\em Phys.Rev.}, D85:122007, 2012.

\bibitem{S4ringdown}
B.P. Abbott et~al.
\newblock {Search for gravitational wave ringdowns from perturbed black holes
  in LIGO S4 data}.
\newblock {\em Phys.Rev.}, D80:062001, 2009.

\bibitem{S4CosmicString}
B.P. Abbott et~al.
\newblock {First LIGO search for gravitational wave bursts from cosmic
  (super)strings}.
\newblock {\em Phys.Rev.}, D80:062002, 2009.

\bibitem{Calibration:S5}
J.~{Abadie}, B.~P. {Abbott}, R.~{Abbott}, M.~{Abernathy}, C.~{Adams},
  R.~{Adhikari}, P.~{Ajith}, B.~{Allen}, G.~{Allen}, E.~{Amador Ceron}, and
  et~al.
\newblock {Calibration of the LIGO gravitational wave detectors in the fifth
  science run}.
\newblock {\em Nuclear Instruments and Methods in Physics Research A},
  624:223--240, December 2010.

\bibitem{DetChar:S6}
N.~{Christensen}, {LIGO Scientific Collaboration}, and {Virgo Collaboration}.
\newblock {LIGO S6 detector characterization studies}.
\newblock {\em Classical and Quantum Gravity}, 27(19):194010, October 2010.

\bibitem{AMON}
M.~W.~E. {Smith}, D.~B. {Fox}, D.~F. {Cowen}, P.~{M{\'e}sz{\'a}ros}, G.~{Te{\v
  s}i{\'c}}, J.~{Fixelle}, I.~{Bartos}, P.~{Sommers}, A.~{Ashtekar}, G.~{Jogesh
  Babu}, S.~D. {Barthelmy}, S.~{Coutu}, T.~{DeYoung}, A.~D. {Falcone}, L.~S.
  {Finn}, S.~{Gao}, B.~{Hashemi}, A.~{Homeier}, S.~{M{\'a}rka}, B.~J. {Owen},
  and I.~{Taboada}.
\newblock {The Astrophysical Multimessenger Observatory Network (AMON)}.
\newblock {\em Astropart. Phys.}, 45:56, May 2013.

\bibitem{YunesSiemens:2013}
N.~{Yunes} and X.~{Siemens}.
\newblock {Gravitational Wave Tests of General Relativity with Ground-Based
  Detectors and Pulsar Timing Arrays}.
\newblock {\em ArXiv e-prints}, April 2013.

\bibitem{Kelley:2010}
L.~Z. {Kelley}, E.~{Ramirez-Ruiz}, M.~{Zemp}, J.~{Diemand}, and I.~{Mandel}.
\newblock {The Distribution of Coalescing Compact Binaries in the Local
  Universe: Prospects for Gravitational-wave Observations}.
\newblock {\em \apjl}, 725:L91--L96, December 2010.

\bibitem{Berger11}
E.~{Berger}.
\newblock {The environments of short-duration gamma-ray bursts and implications
  for their progenitors}.
\newblock {\em \nar}, page~1, 2011.

\bibitem{Aylott:2009ya}
B.~Aylott et~al.
\newblock {Testing gravitational-wave searches with numerical relativity
  waveforms: Results from the first Numerical INJection Analysis (NINJA)
  project}.
\newblock {\em Class.Quant.Grav.}, 26:165008, 2009.

\bibitem{2009CQGra..26k4008C}
L.~{Cadonati} et~al.
\newblock {Status of NINJA: the Numerical INJection Analysis project}.
\newblock {\em Classical and Quantum Gravity}, 26(11):114008, 2009.

\bibitem{omega}
{The Omega Pipeline}.
\newblock https://geco.phys.columbia.edu/omega.

\bibitem{2011PhRvD..83d4019F}
S.~{Fischetti}, J.~{Healy}, L.~{Cadonati}, L.~{London}, S.~R.~P. {Mohapatra},
  and D.~{Shoemaker}.
\newblock {Exploring the use of numerical relativity waveforms in burst
  analysis of precessing black hole mergers}.
\newblock {\em \prd}, 83(4):044019, 2011.

\bibitem{heger00}
A.~Heger, N.~Langer, and S.E. Woosley.
\newblock {Presupernova Evolution of Rotating Massive Stars. I. Numerical
  Method and Evolution of the Internal Stellar Structure}.
\newblock {\em Astrophys. J.}, 528:368, 2000.

\bibitem{hirschi04}
R.~Hirschi, G.~Meynet, and A.~Maeder.
\newblock Stellar evolution with rotation. xii. pre-supernova models.
\newblock {\em Astron. Astroph.}, 443:649, 2005.

\bibitem{hirschi05}
R.~Hirschi, G.~Meynet, and A.~Maeder.
\newblock Stellar evolution with rotation. xiii. predicted grb rates at various
  z.
\newblock {\em Astron. Astroph.}, 443:581, 2005.

\bibitem{Yoon05b}
S.-C. Yoon and N.~Langer.
\newblock Evolution of rapidly rotating metal-poor massive stars towards
  gamma-ray bursts.
\newblock {\em Astron. Astroph.}, 443:643, 2005.

\bibitem{woosley:06}
S.~E. {Woosley} and A.~{Heger}.
\newblock {The Progenitor Stars of Gamma-Ray Bursts}.
\newblock {\em Astrophys. J.}, 637:914, 2006.

\bibitem{Yoon06}
S.-C. Yoon, N.~Langer, and C.~Norman.
\newblock On the evolution of rapidly rotating massive white dwarfs towards
  supernovae or collapses.
\newblock {\em Astron. Astroph.}, 460:199, 2006.

\bibitem{fryer99}
C.L. Fryer, S.E. Woosley, and D.H. Hartmann.
\newblock Formation rates of black hole accretion disk gamma-ray bursts.
\newblock {\em Astrophys. J.}, 526:152, 1999.

\bibitem{podsiadlowski04}
Ph. Podsiadlowski, P.A. Mazzali, K.~Nomoto, D.~Lazzati, and E.~Cappellaro.
\newblock The rates of hypernovae and gamma-ray brusts: Implications for their
  progenitors.
\newblock {\em Astrophys. J.}, 607:L17, 2004.

\bibitem{fryer05}
C.L. Fryer and A.~Heger.
\newblock {Binary Merger Progenitors for Gamma-Ray Bursts and Hypernovae}.
\newblock {\em Astrophys. J.}, 623:302, 2005.

\bibitem{vdH07}
E.P.J. van~den Heuvel and S.-C. Yoon.
\newblock Long gamma-ray burst progenitors: boundary conditions and binary
  models.
\newblock {\em Astrophysics and Space Science}, 311:177, 2007.

\bibitem{Cantiello07}
M.~Cantiello, S.-C. Yoon, N.~Langer, and M.~Livio.
\newblock Binary star progenitors of long gamma-ray bursts.
\newblock {\em Astron. Astroph.}, 465:L29, 2007.

\bibitem{Yoon04}
S.-C. Yoon, N.~Langer, and S.~Scheithauer.
\newblock Effects of rotation on the helium burning shell source in accreting
  white dwarfs.
\newblock {\em Astron. Astroph.}, 425:217, 2004.

\bibitem{Yoon05}
S.-C. Yoon and N.~Langer.
\newblock On the evolution of rapidly rotating massive white dwarfs towards
  supernovae or collapses.
\newblock {\em Astron. Astroph.}, 435:967, 2005.

\bibitem{Yoon07}
S.-C. Yoon, P.~Podsiadlowski, and S.~\&~Rosswog.
\newblock Remnant evolution after a carbon-oxygen white dwarf merger.
\newblock {\em Mon. Not. R. Astron. Soc.}, 380:933, 2007.

\bibitem{fryer09}
C.L. et~al. Fryer.
\newblock {Spectra and Light Curves of Failed Supernovae}.
\newblock {\em Astrophys. J.}, 707:193, 2009.

\bibitem{dessart:06aic}
L.~{Dessart}, A.~{Burrows}, C.~D. {Ott}, E.~{Livne}, S.-Y. {Yoon}, and
  N.~{Langer}.
\newblock {Multidimensional Simulations of the Accretion-induced Collapse of
  White Dwarfs to Neutron Stars}.
\newblock {\em \apj}, 644:1063, 2006.

\bibitem{abdikamalov:10}
E.~B. {Abdikamalov}, C.~D. {Ott}, L.~{Rezzolla}, L.~{Dessart},
  H.~{Dimmelmeier}, A.~{Marek}, and {H.-T.} {Janka}.
\newblock {Axisymmetric general relativistic simulations of the
  accretion-induced collapse of white dwarfs}.
\newblock {\em \prd}, 81:044012, 2010.

\bibitem{dimonte04}
G.~Dimonte et~al.
\newblock {A comparative study of the turbulent Rayleigh-Taylor instability
  using high-resolution three-dimensional numerical simulations: The
  Alpha-Group collaboration}.
\newblock {\em Astrophys. J.}, 16:1668, 2004.

\bibitem{porter06}
D.H. Porter and P.R. Woodward.
\newblock Using ppm to model turbulent stellar convection.
\newblock In F.~Grinstein, L.~Margolin, and W.~Rider, editors, {\em Implicit
  Large Eddy Simulation: Computing Turbulent Fluid Dynamics}, Los Alamos, NM,
  2006. Cambridge University Press.

\bibitem{2001PhRvL..87x1101A}
N.~{Andersson} and G.~L. {Comer}.
\newblock {Probing Neutron-Star Superfluidity with Gravitational-Wave Data}.
\newblock {\em Physical Review Letters}, 87(24):241101, 2001.

\bibitem{mcwilliams2012}
S.~T. {McWilliams}, J.~P. {Ostriker}, and F.~{Pretorius}.
\newblock {The imminent detection of gravitational waves from massive
  black-hole binaries with pulsar timing arrays}, 2012.
\newblock {arXiv:1211.4590}.

\bibitem{sesana2012}
A.~{Sesana}.
\newblock {Systematic investigation of the expected gravitational wave signal
  from supermassive black hole binaries in the pulsar timing band}, 2012.
\newblock {arXiv:1211.5375}.

\end{thebibliography}

\end{document}